\documentclass[pra,aps,superscriptaddress,showpacs,floatfix,tightenlines,twocolumn]{revtex4-1}
\usepackage{graphicx}
\usepackage{dcolumn}
\usepackage{bm}
\usepackage{amssymb}
\usepackage{amsmath}
\usepackage{url}
\usepackage{amsthm}
\usepackage{layout}
\usepackage{epsfig}
\usepackage{graphicx}
\usepackage{epstopdf}
\usepackage{booktabs}
\usepackage{float}
\usepackage{subfigure}
\usepackage[hyperindex,breaklinks]{hyperref}
\usepackage{xcolor}

\begin{document}

\title{Witnessing nonlocality in quantum network of continuous-variable systems  by generalized quasiprobability functions}
\author{Taotao Yan}
\affiliation{College of Mathematics, Taiyuan University of Technology, Taiyuan, 030024, China}
\author{Jinchuan Hou}
\email{
houjinchuan@tyut.edu.cn, jinchuanhou@aliyun.com}
\affiliation{College of Mathematics, Taiyuan University of Technology, Taiyuan, 030024, China}
\author{Xiaofei Qi}
\email{
qixf1981@126.com}
\affiliation{School of Mathematics and Statistics, Shanxi University,
Taiyuan, 030006, China}
\author{Kan He}
\email{hekanquantum@163.com}
\affiliation{College of Mathematics, Taiyuan University of Technology, Taiyuan, 030024, China}

\date{\today }
\begin{abstract}
Gaussian measurements can not be used to witness nonlocality in Gaussian states as well as the network nonlocality in networks of continuous-variable (CV) systems. Thus special non-Gaussian measurements have to be utilized.
 In the present paper,  we first propose a kind of  nonlinear Bell-type inequality that is applicable to  quantum networks of both finite or infinite dimensional systems.  Violation of the inequality will witness the network nonlocality. This inequality allows us to propose a method  of the supremum strategy for detecting network nonlocality in CV systems with source states being  any multipartite multimode Gaussian states according to the configurations of the networks by utilizing non-Gaussian measurements  based on generalized quasiprobability functions.  The nonlinear Bell-type inequalities for CV networks, which depend solely on the generalized quasiprobability functions of Gaussian states, are straightforward to construct and implement.    As  illustrations,  we propose the corresponding nonlinear Bell-type inequalities for any chain, star, tree-shaped and cyclic networks  in CV systems with source states being $(1+1)$-mode Gaussian states. The examples show that this method works well.  Particularly, a thorough discussion is given for the entanglement swapping network. Our study provide a strong signature for the network nonlocality nature of CV systems and lead to precise recipes for its experimental verification.

\vspace{2mm} \noindent
\end{abstract}

\maketitle

%%%%%%%%%%%%%%%%%%%%%%%%%%%%%%%%%%%%%%%%%%%%%%%%%%%%%%%%%%%%%%%%%%%

\section{Introduction}
Bell nonlocality has emerged as a topic of significant interest over the past a few decades, with its implications recognized in various domains of quantum information processing \cite{NB}-\cite{SASA}, quantum communication \cite{AK}-\cite{DA} and quantum key distribution \cite{ANL,ANNS}. To detect whether a quantum state is nonlocality, Bell inequalities were proposed \cite{JSB}-\cite{BJS}. In \cite{BJS}, Bell derived inequalities
that any theory based on local hidden variables must satisfy. Consequently, the violation of these Bell inequalities is indicative of Bell nonlocality.

Recently, a generalization of the concept of Bell nonlocality was proposed to tackle the question of nonlocality phenomenon in quantum networks \cite{CNS}, which we call {\it network nonlocality} in the present paper to distinguish
from the usual nonlocality.
The nonlocality within various quantum network configurations has been extensively explored and studied, including chain networks \cite{CNS,RAK}, star networks \cite{FGL,APD}, tree-shaped networks \cite{YLH}-\cite{HFC}, cyclic networks \cite{TFR,AFR}, by establishing corresponding nonlinear Bell-type inequalities.  Research has demonstrated that network nonlocality plays an indispensable role across multiple domains, including secure communication, distributed quantum computing, cryptography, quantum key distribution, random number generation, and quantum sensing \cite{AJM}-\cite{PEM}. Detecting the nonlocality
in quantum networks becomes a basic task in this field.
The current approach is to establish nonlinear Bell-type
inequalities, violation any one of these inequalities will
witness the nonlocality in the involved quantum networks. { However, the majority of studies on network nonlocality have focused on finite-dimensional systems, and most known nonlinear Bell-type inequalities are confined to the two-input-two-output scenario.}

The continuous-variable (CV) systems are also fundamentally important from theoretical and experimental views. Specifically, Gaussian states have become important objects in the study of CV quantum information theory due to their ease of production and manipulation in laboratories, as well as their elegant mathematical structures. The Bell nonlocality of Gaussian states has been the subject of intensive study. It is noteworthy that Gaussian states cannot exhibit nonlocality through Gaussian measurements. This is because they inherently have a positive-definite Wigner function, which allows for the construction of a hidden-variable model \cite{Bell}. Therefore, the set of Gaussian nonlocal states is defined as the intersection of the set of Gaussian states and the set of nonlocal states. Additionally, it is essential to employ special non-Gaussian measurements that  are close to Gaussian measurements or can be implemented in the laboratory to detect the nonlocality inherent in Gaussian states.
In \cite{KKB}, Banaszek and W{\'{o}}dkiewicz first demonstrated the Bell nonlocality of the two-mode squeezed vacuum state using the Wigner functions. They highlighted that there is no direct correlation between the negativity of the Wigner function and Bell nonlocality. Following their initial work, several studies have expanded upon the concepts introduced in \cite{KKB}. Some of these works   considered more phase space displacements, as referenced in \cite{HWM}, while others   introduced new observables that allow for Bell inequality violations for Gaussian states, as seen in \cite{AAL}. These advancements contribute to a deeper understanding of the conditions under which Gaussian states can be shown to exhibit nonlocal properties, emphasizing the role of non-Gaussian measurements in detecting such nonlocality.

The CV quantum networks have garnered increasing attention due to their high transmission efficiency, low channel loss, and ease of implementation  \cite{JPV}-\cite{FCC}.  Moreover, experimental advancements have been at the forefront of developing the hybrid quantum optical paradigm \cite{PVL}-\cite{PM}, with a significant interest from the scientific community in studying hybrid networks that combine CV and discrete variable (DV) systems  \cite{SMP}-\cite{GTH}.  Given the practical advantages of Gaussian quantum information processing \cite{JM,GF}, such as deterministic entanglement generation \cite{SP}, it is both natural and crucial to delve into the study of quantum correlations within CV quantum networks with source states being Gaussian states. Currently, almost all known studies concerning  networks in CV systems, such as on the topics of entanglement swapping \cite{JPV,MSP,ZHH}, entanglement transmission \cite{MBF}, entanglement distillation \cite{FJ},  and hybrid quantum communication network \cite{NBD}, assume the source states to be $(1+1)$-mode Gaussian states. { Recently, Chakrabarty et al. \cite{SAA} detected the nonlocality of entanglement swapping network in CV systems via pseudospin measurements. However, compared with DV systems, research on network nonlocality in CV systems remains relatively limited. One of the main challenges is the lack of nonlinear Bell-type inequalities applicable to infinite-dimensional networks with infinitely many measurement outcomes.} Another challenge is that Gaussian measurements are not able
to witness the nonlocality in Gaussian states, and thus
one has to find and utilize suitable non-Gaussian measurements that can be realized in laboratory. The third
issue is that, when these non-Gaussian measurements are
performed, whether or not the measurement scheme is
sufficient to detect the network nonlocality contained in
the networks of CV systems.

Addressing these issues mentioned above, the purpose
of the present paper is to propose an approach for detecting the nonlocality of networks in CV systems with
source states being Gaussian states. To do this, we first
establish a new nonlinear Bell-type inequality for general networks of finite or infinite dimensional systems, in
the scenarios where all parties have two inputs but the
number of outputs is not limited (which may be infinite).
Secondly, we utilize this inequality and the non-Gaussian
measurement scheme based on the generalized quasiprobability functions, which are achievable in the laboratory,
to develop a general method of the supremum strategy for detecting network nonlocality in CV systems networks. To the third issue, several classes of networks are
discussed to illustrate that the measurements based on
generalized quasiprobability functions are effective to detect the network nonlocality of CV systems. { Our work
establishes a fundamental connection between Bell
nonlocality and quantum networks in CV systems.} This
scheme not only deepens the understanding of quantum
nonlocality, but also highlights their potential application
prospects in various quantum communication protocols.

This paper is organized as follows. In Section \ref{sec:2}, we establish a  nonlinear Bell-type inequality for a general network, which is applicable to the scenarios that all parties have two inputs and arbitrary outputs.  In Section \ref{sec:3}, we review some notions and notations
concerning Gaussian states and generalized quasiprobability functions. In addition, by applying the inequality established in Section \ref{sec:2}, we propose a general approach, that is, the supremum strategy, to detect network nonlocality in CV systems with any Gaussian states as source states. As illustrations of how to apply the supremum strategy,   Section \ref{sec:4}   derives a Bell-type inequality for arbitrary chain networks   with each source state being a $(1+1)$-mode Gaussian state. Particularly, the entanglement swapping network is discussed in more details. The corresponding nonlinear inequalities are established respectively for the star networks, the determinate $f$-forked tree-shaped networks and the cyclic networks in Section \ref{sec:5}, \ref{sec:6} and  \ref{sec:7}. A short conclusion and discussion is provided as Section \ref{sec:8}. All mathematical proofs are presented in Appendix.

\section{A nonlinear Bell-type inequality for quantum networks of any dimensional systems}\label{sec:2}
{ There are no known nonlinear Bell-type inequalities for quantum networks that are valid for networks of CV systems with infinitely many measurement outcomes.} So, to discuss the network nonlocality of CV systems, we need  to develop nonlinear Bell-type inequalities applicable to CV systems.  This is the purpose of the present section.

 Consider a finite-size network $\Xi(y,z)$ consisting of $y$ parties (nodes)  $A_{1}, A_{2},\cdots, A_{y}$, and $z$ independent sources $S_{1}, S_{2},\cdots, S_{z}$ that distribute hidden variables $\lambda_{1},\lambda_{2},\cdots,\lambda_{z}$, respectively.  Here, for any $j=1,2,\cdots, z$, $S_{j}$ usually emits a  multipartite quantum source state $\rho_j$, with the number of parties determined by the number of recipient nodes. Note that every $\rho_j$ is at least a bipartite state. The number $w=\max\{ \mbox{\rm partite of } \rho_j: j=1,2,\cdots, z  \}$ is called the {\it depth} of the network. Each party $A_{i}$ ($i=1,2,\cdots, y$) receives hidden variables $\Lambda_{i}=\{\lambda_{i_{1}}, \lambda_{i_{2}}, \cdots,\lambda_{i_{e_{i}}}\}$ from the corresponding sources $\mathbf{S}_{i}=\{S_{i_{1}}, S_{i_{2}}, \cdots,S_{i_{e_{i}}}\}$, where $\mathbf{S}_{1}, \mathbf{S}_{2},\cdots, \mathbf{S}_{y}$ satisfy $\bigcup_{i=1}^{y}\mathbf{S}_{j}=\{S_{1}, S_{2},\cdots,S_{z}\}$.  The {\it network locality} of $\Xi(y,z)$ suggests a joint conditional probability distribution of the measurement outcomes as
\begin{equation}\label{eq2.1}P(\mathbf{a}|\mathbf{x})=\int\cdot\cdot\cdot\int d\lambda_{1}\cdot\cdot\cdot d\lambda_{z} \prod_{j=1}^{z} \mu_{j}(\lambda_{j})\prod_{i=1}^{y}P(a_{i}|x_{i},\Lambda_{i}),\tag{2.1}\end{equation}
where $\mathbf{a}=(a_{1},a_{2},\cdots,a_{y})$ denotes the parties' outputs, $\mathbf{x}=(x_{1},x_{2},\cdots,x_{y})$ denotes their inputs. Additionally, $\mu_{j}(\lambda_{j})$ is the probability distribution of $\lambda_{j}$ with the normalization condition $\int d\mu_{j}(\lambda_{j})=1$  \cite{JP}-\cite{PNR}. Otherwise, $\Xi(y,z)$ is {\it network nonlocal}. It is clear that $\Xi(y,z)$ is network local if all its source states are separable.

For a  finite-size quantum network $\Xi(y,z)$ in systems of any dimensions,
recall that, a subset $\{B_1,B_2,\cdots, B_k\}\subset\{A_i\}_{i=1}^y$ of $k$ elements is an independent subset if any two elements of  $\{B_1,B_2,\cdots, B_k\}$ do not share source of the network. In the case that such an independent subset of $k$ parties exists, we say that the network  $\Xi(y,z)$ has $k$ independent parties \cite{LMX}. It is clear that, if $\Xi(y,z)$ has $k$ independent parties, then it has $l$ independent parties for any positive integer $l\leq k$.

\if false Let  $M_{x_{i}}$ be the local measurement operators of the party $A_{i}$, where $x_{i}\in\{0,1\}$, $i=1,2,\cdot\cdot\cdot, y$.\fi
The following is our main result in this section which establishes a nonlinear Bell-type inequality for general quantum networks  of any depth.

{\bf Theorem 2.1.} {\it Let $\Xi(y,z)$ be a finite-size network with $k$ independent parties. Suppose that  the inputs of all parties are dichotomic, while no restriction is placed on the number of possible measurement outcomes (which may be infinite when the system is infinite dimensional).  If $\Xi(y,z)$ is network local, then, for any local measurements $\{M_{x_i} : x_i\in\{0,1\}\}_{i=1}^y$ of countable outcomes,  the following inequality holds:

\begin{equation}\label{eq2.2}\mathcal{B}=|\mathcal I|^{\frac{1}{k}}+|\mathcal J|^{\frac{1}{k}}\leq1,\tag{2.2}\end{equation}
 where
\begin{equation}\label{eq2.3}\mathcal I=\frac{1}{2^{k}}\langle \prod_{i_{s}\in\mathcal K}(M_{x_{i_{s}}=0}+M_{x_{i_{s}=1}})\prod_{j\in\bar{\mathcal K}} M_{x_j=0}\rangle,\tag{2.3}\end{equation}
\begin{equation}\label{eq2.4}\mathcal J=\frac{1}{2^{k}}\langle \prod_{i_{s}\in\mathcal K}(M_{x_{i_{s}}=0}-M_{x_{i_{s}}=1})\prod_{j\in\bar{\mathcal K}} M_{x_j=1}\rangle,\tag{2.4}\end{equation}
$\mathcal{K}=\{i_{1},i_{2},\cdot\cdot\cdot,i_{k}\}$ is any fixed index set  so that $A_{i_{1}}$, $A_{i_{2}}$, $\cdot\cdot\cdot$, $A_{i_{k}}$ are $k$ independent parties and $\bar{\mathcal{K}}=\{1,2,\cdot\cdot\cdot,y\}\backslash\mathcal{K}$.

In addition,   for any $i=1,2,\cdot\cdot\cdot,y$,  with $\Upsilon_{x_i}=\{a_{h}: a_{h}\ \text{is the eigenvalue of}\  M_{x_{i}}\}$,  we have
$$\begin{array}{rl}
\langle M_{x_{1}}M_{x_{2}}\cdot\cdot\cdot M_{x_{y}}\rangle=&\displaystyle\sum_{a_{1}\in\Upsilon_{x_1}}
\cdot\cdot\cdot\displaystyle\sum_{a_{y}\in\Upsilon_{x_y}}a_{1}\cdot\cdot\cdot a_{y}\nonumber\\&\times P(a_{1},\cdot\cdot\cdot, a_{y}|x_{1},\cdot\cdot\cdot, x_{y})\nonumber,
\end{array}
$$
where $P(a_{1},\cdot\cdot\cdot, a_{y}|x_{1},\cdot\cdot\cdot, x_{y})$ is the same as defined in Eq.(\ref{eq2.1}).}

A proof of Theorem 2.1 is presented in Appendix A.

From Theorem 2.1, violation of the inequality (\ref{eq2.2}) demonstrates the network nonlocality of $\Xi(y,z)$.  Thus $\mathcal I$ and $\mathcal J$ are important quantities to characterize the network nonlocality. For different network configurations, $\mathcal I$ and $\mathcal J$ have different expressions. In addition, $k$ is another important quantity representing the network nonlocality. Roughly speaking, the larger $k$ is, the more network nonlocalities are involved by the inequality (\ref{eq2.2}). Therefore, it is optimal to find the largest such $k$, denoted by $k_{\max}$, and the corresponding   index set $\mathcal K_{\max}$ for the maximal set of independent parties. However, for a general network, its index set $\mathcal K_{\max}$ for the maximal set of independent parties and consequently, its  maximal independent  number  $k_{\max}=\sharp (\mathcal K_{\max})$ (the cardinal  of $\mathcal K_{\max}$) are difficult to be found out. \if failse
 is related to the configuration of the network and it requires us to check the independence of any subsets of all parties, it is difficult to obtain the $k_{\max}$ for a general network.\fi  Nevertheless, analytical methods to determine $k_{\max}$ exist  for the chain, tree-shaped, star, cyclic networks, and even some more complex networks \cite{RAK,FGL}.

Theorem 2.1 is valid for finite-size quantum networks of any depth in  both  DV systems and  CV systems. However,
in the present paper, we mainly consider its applications to  quantum networks in CV systems.

\section{A general approach to detect network nonlocality in CV  systems by generalized quasiprobability functions}\label{sec:3}

In this section, we first briefly review some notions and notations concerning Gaussian states
and the generalized quasiprobability functions.                                                                                                                                                                                                                                                                                                                                                                                                                                                                                                                                                                                                                                                                                                                                                                                                                                                                                                                                                                                                                                                                                                                                                                                                        Subsequently, we propose a general approach to detect network nonlocality in CV systems, applicable to any {\it Gaussian networks}. Here a quantum network in CV system is called a Gaussian network if all sources of it emit Gaussian states.

\subsection{ Gaussian states}
 Every state $\rho$
in an $n$-mode CV system with state space $H=H_1\otimes H_2\otimes\cdots\otimes H_n$ has a characteristic function
$\chi_{\rho}(\bm{\alpha})$  defined as
$$\chi_{\rho}(\bm{\alpha})={\rm Tr}(\rho V(\bm{\alpha})),$$
where $\bm{\alpha}=(x_{1},y_{1},\cdots,x_{n},y_{n})^{\rm T}\in\mathbb{R}^{2n}$, $V(\bm{\alpha})=\exp(i{\bm{R}^{\rm T}}\bm{\alpha})$ is the Weyl displacement operator,  $\bm{R}=(\hat{R}_1,\hat{R}_2,\cdots,\hat{R}_{2n})=(\hat{Q}_1,\hat{P}_1,\cdots,\hat{Q}_n,\hat{P}_n)$. As usual, $\hat{Q}_g=(\hat{a}_g+\hat{a}_g^\dag)/\sqrt{2}$ and
$\hat{P}_g=-i(\hat{a}_g-\hat{a}_g^\dag)/\sqrt{2}$ ($g=1,2,\cdots,n$)
are respectively the position and the momentum operators of $g$th mode, where
$\hat{a}_g^\dag$ and $\hat{a}_g$ are respectively the creation and the annihilation
operators of the $g$th mode  satisfying the Canonical Commmutation Relation:
$$[\hat{a}_g, \hat{a}_l^\dag]=\delta_{gl}I, [\hat{a}_g^\dag, \hat{a}_l^\dag]=[\hat{a}_g, \hat{a}_l]=0, \ \forall \ g,l=1,2,\cdot\cdot\cdot, n.$$
Particularly, $\rho$
is called a Gaussian state  if its characteristic function
$\chi_{\rho}(\bm{\alpha})$ has the form as
\begin{equation}\label{eq3.1}\chi_{\rho}(\bm{\alpha})=\exp[-\frac{1}{4}\bm{\alpha}^{\rm T}\Gamma \bm{\alpha}+i{\mathbf d}^{\rm T}\bm{\alpha}],\tag{3.1}\end{equation}
where
$$\begin{array}{rl}
{\mathbf d}=&(\langle\hat R_1 \rangle, \langle\hat R_2
\rangle, \cdots,\langle\hat R_{2n} \rangle)^{\rm T}\nonumber\\=&({\rm Tr}(\rho
\hat{R}_1), {\rm Tr}(\rho \hat{R}_2), \cdots, {\rm Tr}(\rho \hat{R}_{2n}))^{\rm
T}\in{\mathbb R}^{2n}\nonumber
\end{array}
$$
 is called the mean vector of $\rho$ and
$\Gamma=(\gamma_{kl})\in \mathcal{M}_{2n}(\mathbb R)$, the algebra of all $2n\times 2n$ matrices over the real field $\mathbb R$, is the covariance matrix of $\rho$ defined by $$\gamma_{kl}={\rm Tr}[\rho
(\Delta\hat{R}_k\Delta\hat{R}_l+\Delta\hat{R}_l\Delta\hat{R}_k)],$$
in which, $\Delta\hat{R}_k=\hat{R}_k-\langle\hat{R}_k\rangle$,
$\langle\hat{R}_k\rangle={\rm Tr}[\rho\hat{R}_k]$ \cite{SP,GJ}.  Note that a matrix $\Gamma$ is a  covariance matrix for some Gaussian state if and only if $\Gamma$ is real symmetric and
satisfies the condition $\Gamma +i\Omega_{n}\geq 0$, where
$\Omega_n= \underbrace{\Omega\oplus \Omega\cdots\oplus\Omega}_n\in{\mathcal M}_{2n}(\mathbb R)$ with $\Omega=\left(\begin{array}{cc} 0 & 1\\ -1 & 0 \end{array}\right)$. By (\ref{eq3.1}), every Gaussian state $\rho$ is uniquely determined by its covariance matrix $\Gamma$ and mean vector ${\mathbf d}$, and thus, one can write $\rho=\rho(\Gamma, {\mathbf d})$.

Besides the characteristic function, quantum
states can also be represented by the Wigner function \cite{SP}. The Wigner function of an $n$-mode state $\rho$ is defined as the symplectic Fourier transform of its characteristic function, that is,
$$W_\rho(\bm{\alpha})=\frac{1}{(2\pi)^{2n}}\int e^{-i\bm{\alpha}^{\rm T}\bm{\eta}}\chi_\rho(\bm{\eta})d^{2n}\bm{\eta},$$
where $\bm{\alpha}=(x_{1},y_{1},\cdots,x_{n},y_{n})^{\rm T}\in\mathbb{R}^{2n}$.

\subsection{The generalized quasiprobability function}

For a single-mode quantum state $\rho$ in CV system, its generalized quasiprobability function  is
\begin{equation}\tag{3.2}\label{eq3.2}
\begin{aligned}
Q_{\rho}^{(1)}(\alpha;s)= & \frac{2}{\pi(1-s)}{\rm Tr}[\rho\hat{\Pi}(\alpha;s)]\\=&\frac{2}{\pi(1-s)}\sum_{n=0}^{\infty}(\frac{s+1}{s-1})^{n}\langle\alpha,n|\rho|\alpha,n\rangle,
\end{aligned}
\end{equation}
where $\hat{\Pi}(\alpha;s)=\sum_{n=0}^{\infty}(\frac{s+1}{s-1})^{n}|\alpha,n\rangle\langle\alpha,n|$ with $|\alpha,n\rangle=D(\alpha)|n\rangle$. Here, as usual, $\{|n\rangle\}_{n=0}^\infty$ is the Fock basis and $D(\alpha)={\rm exp}(\alpha\hat{a}^{\dagger}+\alpha^{\ast}\hat{a})$ for each $\alpha\in\mathbb{C}$ \cite{KER,HPL}. Since the eigenvalues of $\hat{\Pi}(\alpha;s)$ are unbounded when $s>0$, we always assume $s\leq 0$  in this paper unless otherwise specified. It should be noted that some literature refers to the generalized quasiprobability function $Q_{\rho}^{(1)}(\alpha;s)$ as the $s$-parameterized quasiprobability function. \if false but for the purpose of consistency, we will use generalized quasiprobability function here.\fi

Generally, for any $n$-mode Gaussian state $\rho$, its generalized quasiprobability function  is defined as
$$
Q_{\rho}^{(n)}(\bm{\alpha};s)=\frac{2^{n}}{\pi^{n}(1-s)^{n}}{\rm Tr}[\rho(\otimes_{i=1}^{n}\hat{\Pi}(\alpha_{i};s)]
$$
with $\bm{\alpha}=(\alpha_1,\alpha_2,\cdots,\alpha_n)\in\mathbb{C}^{n}$.

Then, for any bipartite $(n_{1}+n_{2})$-mode Gaussian state $\rho$, its generalized quasiprobability function is
\begin{equation}\tag{3.3}\label{eq3.3}
\begin{aligned}
&Q_{\rho}^{(n_1,n_2)}(\bm{\alpha},\bm{\beta};s)\\=&\frac{2^{n_1+n_2}}{\pi^{n_1+n_2}(1-s)^{n_1+n_2}}{\rm Tr}[\rho(\otimes_{i=1}^{n_{1}}\hat{\Pi}(\alpha_{i};s)\otimes_{j=1}^{n_{2}}\hat{\Pi}(\beta_{j};s))],
\end{aligned}
\end{equation}
where $\bm{\alpha}=(\alpha_{1},\alpha_{2},\cdots,\alpha_{n_{1}})\in\mathbb{C}^{n_{1}}$ and $\bm{\beta}=(\beta_{1},\beta_{2},\cdots,\beta_{n_{2}})\in\mathbb{C}^{n_{2}}$. In addition, the marginal distribution of $Q_{\rho}^{(n_1,n_2)}(\bm{\alpha},\bm{\beta};s)$ is
\begin{equation}\label{eq3.4}Q_{\rho}^{(n_{1},\Box)}(\bm{\alpha};s)=\int Q_{\rho}^{(n_1+n_2)}(\bm{\alpha},\bm{\beta};s)d\bm{\beta}\tag{3.4}\end{equation}
and $Q_{\rho}^{(\Box, n_2)}(\bm{\beta};s)$ is defined similarly.

For any $r$-partite $(n_1+n_2+\cdots +n_r)$-mode Gaussian state $\rho$, its generalized quasiprobability function can be defined similarly.

Particularly, if $\rho(\Gamma_{\rho},0)$ is a (1+1)-mode Gaussian state with zero mean, its Wigner function can be represented as
\begin{equation}\label{eq3.5}W_{\rho}(\alpha,\beta)=\frac{4}{\pi^{2}\sqrt{{\rm det}\Gamma_{\rho}}}{\rm exp}\{-\frac{1}{2}X^{\rm T}\Gamma_{\rho}^{-1}X\},\tag{3.5}\end{equation}
where $X^{\rm T}$ is the real row vector $[\alpha+\alpha^{\ast}, -i(\alpha-\alpha^{\ast}), \beta+\beta^{\ast}, -i(\beta-\beta^{\ast})]$ and $\alpha,\beta\in\mathbb{C}$. Then the generalized quasiprobability function of $\rho$ is
\begin{equation}\tag{3.6}\label{eq3.6}
\begin{aligned}
&Q_\rho^{(1,1)}(\alpha,\beta;s)\\=&\frac{1}{4s^{2}\pi^{2}}\int d^{4}Y W_{\rho}(Y){\rm exp}\{-\frac{1}{2}(Y-X)^{\rm T}|s|^{-1}(Y-X)\}.
\end{aligned}
\end{equation}
Using the Gaussian convolutional ground property and  replacing
$\Gamma_{\rho}$ by $ \Gamma_{\rho}^{s}=\Gamma_{\rho}+|s|I_{4}$ in Eq.(\ref{eq3.5}),
 one sees that the Wigner function of $\rho$ becomes the generalized  quasiprobability function $Q_{\rho}^{(1,1)}(\alpha,\beta;s)$, here $I_{4}$ is the $4\times4$ unit matrix \cite{AAL,UL}.

 Below, we list the generalized quasiprobability functions as well as their marginal distributions of two common types of $(1+1)$-mode Gaussian states, namely EPR state and the squeezed thermal state (STS), which are used repeatedly in this paper.

{\bf Example 3.1.} The $(1+1)$-mode EPR state is a pure Gaussian state of the form $\rho^{r}=|\psi(r)\rangle\langle\psi(r)|$, where
\begin{equation}\label{eq3.7}|\psi(r)\rangle={\rm sech}r\sum_{n=0}^{\infty}{\rm tanh}^{n}r|n,n\rangle\tag{3.7}\end{equation}
and $r\geq0$ is the squeezing parameter. For a non-positive real number $s$, the generalized  quasiprobability function of the EPR state $\rho^{r}$ is given by
\begin{equation}\tag{3.8}\label{eq3.8}
\begin{aligned}
&Q_{\rho^{r}}^{(1,1)}(\alpha,\beta;s)\\=&\frac{4}{\pi^{2}R(s)}{\rm exp}\{-\frac{2}{R(s)}[S(s)(|\alpha|^{2}+|\beta|^{2})\\&-{\rm sinh}2r(\alpha\beta+\alpha^{\ast}\beta^{\ast})]\}
\end{aligned}
\end{equation}
and its marginal single-mode distribution is
\begin{equation}\label{eq3.9}Q_{\rho^{r}}^{(1,\Box)}(\alpha;s)=\frac{2}{\pi S(s)}{\rm exp}(-\frac{2|\alpha|^{2}}{S(s)}),\tag{3.9}
\end{equation}
where $R(s)=s^{2}-2s{\rm cosh}(2r)+1$, $S(s)={\rm cosh}(2r)-s$, and $\alpha,\beta\in\mathbb{C}$ \cite{SHD}. Obviously, $\rho^{r}$ is   $(1+1)$-mode entangled   if only and if $r>0$.

{\bf Example 3.2.} The STS is a $(1+1)$-mode mixed Gaussian state defined as a special unitary transform of some $(1+1)$-mode thermal state, that is,
 \begin{equation}\label{eq3.10}\rho(v_{1}, v_{2}, r)=S_{12}(r)(\rho_{T_{1}}\otimes\rho_{T_{2}})S_{12}^{\dag}(r),\tag{3.10}
\end{equation}
where $S_{12}(r)=\exp[r(\hat{a}_{1}^{\dag}\hat{a}_{2}^{\dag}-\hat{a}_{1}\hat{a}_{2})]$ is a two-mode squeeze operator with the squeeze factor $r>0$ and
$$\rho_{T_{j}}=\frac{1}{\bar{n}_{j}+1}\sum_{n=0}^{\infty}(\frac{\bar{n}_{j}}{\bar{n}_{j}+1})^{n}|n\rangle\langle n|,\ (j=1,2)$$
is the density operator of the $j$th thermal field with $\bar{n}_{j}$  its average number of thermal photons. The covariance matrix for the $(1+1)$-mode STS $\rho$ is
$$\Gamma_\rho=\left(\begin{array}{cccc}a&0&c&0\\0&a&0&d\\c&0&b&0\\0&d&0&b
\end{array}\right),$$
where the matrix elements are given by
 $a=v_{1}{\rm cosh}^{2}(r)+v_{2}{\rm sinh}^{2}(r)$, $b=v_{1}{\rm sinh}^{2}(r)+v_{2}{\rm cosh}^{2}(r)$, $c=-d=\frac{1}{2}(v_{1}+v_{2}){\rm sinh}(2r)$
 and $v_{j}=2\bar{n}_{j}+1$ $(j=1,2)$. Thus a $(1+1)$-mode STS $\rho(v_{1}, v_{2}, r)$ is uniquely determined by parameters $v_1,v_2,r$. Recall that, $\rho(v_{1}, v_{2}, r)$ is separable if and only if $\cosh^2(r)\leq\frac{(v_1+1)(v_2+1)}{2(v_1+v_2)}$  \cite{PTH}.
 In the case $v_{1}=v_{2}=v$, we refer to $\rho(v , v , r)$ as the $(1+1)$-mode symmetric STS.

For a non-positive real number $s$, the generalized  quasiprobability function of the $(1+1)$-mode STS $\rho(v_{1}, v_{2}, r)$ is given by
\begin{equation}\label{eq3.11}\begin{array}{rl}
&Q_{\rho(v_{1}, v_{2}, r)}^{(1,1)}(\alpha,\beta;s)\nonumber\\=&\frac{4}{\pi^{2}A}\exp\{-\frac{2}{A}[A_{1}|\alpha|^{2}+A_{2}|\beta|^{2}\nonumber\\&-\frac{(v_{1}+v_{2}){\rm sinh}(2r)}{2}(\alpha\beta+\alpha^{\ast}\beta^{\ast})]\}\nonumber,
\end{array}\eqno(3.11)
\end{equation}
and its marginal single-mode distributions are
 \begin{equation}\label{eq3.12}Q_{\rho(v_{1}, v_{2}, r)}^{(1,\Box)}(\alpha;s)=\frac{2}{\pi A_{2}}\exp\{-\frac{2}{A_{2}}|\alpha|^{2}\}\tag{3.12}
\end{equation}
and
 \begin{equation}\label{eq3.13}Q_{\rho(v_{1}, v_{2}, r)}^{(\Box,1)}(\beta;s)=\frac{2}{\pi A_{1}}\exp\{-\frac{2}{A_{1}}|\beta|^{2}\},\tag{3.13}
\end{equation}
where $A=-2{\rm cosh}^{2}(r)s(v_{1}+v_{2})+(v_{1}+s)(v_{2}+s)$, $A_{1}=(v_{1}+v_{2}){\rm cosh}^{2}(r)-v_{1}-s$, $A_{2}=(v_{1}+v_{2}){\rm cosh}^{2}(r)-v_{2}-s$, and $\alpha,\beta\in\mathbb{C}$.

In the sequel, without causing confusion, we simply denote by $Q_\rho^{(n_1)}({\bm \alpha};s)=Q_\rho^{(n_1,\Box)}({\bm \alpha},{\bm \beta};s)$ and $Q_\rho^{(n_2)}({\bm \beta};s)=Q_\rho^{(\Box,n_2)}({\bm \alpha},{\bm \beta};s)$ with ${\bm \alpha}=(\alpha_1,\cdots,\alpha_{n_1})\in\mathbb C^{n_1}$ and ${\bm \beta}=(\beta_1,\cdots,\beta_{n_2})\in\mathbb C^{n_2}$.

\subsection{A general approach to detect network nonlocality in CV systems}

Though our approach is valid for any Gaussian networks with arbitrary depth, for the sake of simplicity, we mainly present our approach for  the most common class of finite-size Gaussian networks $\Xi(y,z)$, that is, those networks with depth 2, or equivalently, those networks of which all sources emit   bipartite multimode Gaussian states.  Also, we assume that all  measurement operators  are of the form
\begin{equation}\tag{3.14}\label{eq3.14}
\begin{aligned}
&M_{x_{i}}(\bm{\alpha}_{x_{i}}^{\lambda_{i_{1}}},\bm{\alpha}_{x_{i}}^{\lambda_{i_{2}}}, \cdots, \bm{\alpha}_{x_{i}}^{\lambda_{i_{e_{i}}}};s)\\ = & \hat{O}^{(n_{i_{1}})}(\bm{\alpha}_{x_{i}}^{\lambda_{i_{1}}};s) \otimes \hat{O}^{(n_{i_{2}})}(\bm{\alpha}_{x_{i}}^{\lambda_{i_{2}}};s) \\
            & \otimes \cdots \otimes \hat{O}^{(n_{i_{e_{i}}})}(\bm{\alpha}_{x_{i}}^{\lambda_{i_{e_{i}}}};s), \quad x_{i}\in\{0,1\}
\end{aligned}
\end{equation}
$i=1,2,\cdot\cdot\cdot,y$,  where
 $\Lambda _i=\{\lambda_{i_1}, \lambda_{i_2},\cdots, \lambda_{i_{e_i}}\}$  denotes the set of local variables associated with the sources $\mathbf{S}_{i}=\{S_{i_{1}}, S_{i_{2}}, \cdots, S_{i_{e_{i}}}\}$ that
connect to party $A_i$. For each $j=1,2,\cdots,e_{i}$, $\bm{\alpha}_{x_{i}}^{\lambda_{i_{j}}}=(\alpha_{x_{i},1}^{\lambda_{i_{j}}},\alpha_{x_{i},2}^{\lambda_{i_{j}}}, \cdots,\alpha_{x_{i},n_{i_{j}}}^{\lambda_{i_{j}}})\in\mathbb{C}^{n_{i_{j}}}$.
Thus each source $S_{i_{j}}$ emits an $n_{i_{j}}$-mode  Gaussian state to $A_{i}$, and  the party $A_{i}$ shares an $n_{i}(=\sum_{j=1}^{e_{i}}n_{i_{j}})$-mode $e_i$-partite Gaussian state. In addition,
\begin{equation}\tag{3.15}\label{eq3.15}
\begin{aligned}&\hat{O}^{(n_{i_{j}})}(\bm{\alpha}_{x_{i}}^{\lambda_{i_{j}}};s)\\=&\begin{cases}
(1-s)\otimes_{t=1}^{n_{i_{j}}}\hat{\Pi}(\alpha_{x_{i},t}^{\lambda_{i_{j}}};s)+sI, & \mbox{\rm if }-1<s\leq0,\\
2\otimes_{t=1}^{n_{i_{j}}}\hat{\Pi}(\alpha_{x_{i},t}^{\lambda_{i_{j}}};s)-I, & \mbox{\rm if }s\leq-1
\end{cases}
\end{aligned}
\end{equation}
is a Hermitian operator parameterized by a non-positive real number $s$ and some arbitrary complex variables $\alpha_{x_{i},1}^{\lambda_{i_{j}}},\alpha_{x_{i},2}^{\lambda_{i_{j}}},\cdots,\alpha_{x_{i},n_{i_{j}}}^{\lambda_{i_{j}}}$ \cite{UL}.
As usual, $I$ stands for the identity operator. The possible measurement outcomes of $\hat{O}^{(n_{i_{j}})}(\bm{\alpha}_{x_{i}}^{\lambda_{i_{j}}};s)$ are given by its eigenvalues
\[ \xi_{m}=\begin{cases}
(1-s)(\frac{s+1}{s-1})^{m}+s, & \mbox{\rm if }-1<s\leq0,\\
2(\frac{s+1}{s-1})^{m}-1, &\mbox{\rm if }s\leq-1,
\end{cases}
\]
where $m=0,1,2,\cdot\cdot\cdot$. Clearly, for any non-positive $s$, $|\xi_{m}|\leq1$.  Ref.\cite{SHD} shows that when  all sources emit $(1+1)$-mode Gaussian states in the quantum network, the experimental design of the measurement scheme is extremely simple, only requiring the beam-splitter connected to the photondetector. In particular, when the parameter $s=-1$, the measurement scheme only needs to couple the beam-splitter with a strong coherent state, and then use a photodetector to distinguish between vacuum and non-vacuum states.

Consider  a finite-size CV network $\Xi(y,z)$ where each source $S_{j}(j=1,2,\cdots,z)$ emits arbitrary bipartite multimode Gaussian state $\rho_{j}$. Then  the whole network state is a Gaussian state $\rho=\otimes_{j=1}^{z}\rho_{j}$. More clearly, we denote the network $\Xi(y,z)$ with network state $\rho$ by $\Xi(y,z;\rho)$.  Each  party $A_{i}$ performs two measurements of the form given in Eq.(\ref{eq3.14}) for any $i=1,2,\cdots,y$. If $\Xi(y,z)$ has  $k$ independent parties,  then, according to Theorem 2.1,   $\Xi(y,z;\rho)$ is network local implies that the nonlinear Bell-type inequality
 \begin{equation}\label{eq3.161}
 \begin{array}{rl} \mathcal{B}_{\rho}^{\Xi}(\bm{ \alpha}_{0},\bm{ \alpha}_{1};s)= |\mathcal I|^{\frac{1}{k}}+|\mathcal J|^{\frac{1}{k}}\leq   1
 \end{array}\tag{3.16}
\end{equation}
holds for any complex variables $$\bm{ \alpha}_{t}=(\bm{ \alpha}_{x_{1}=t}^{\lambda_{1_{1}}}, \cdots, \bm{\alpha}_{x_{1}=t}^{\lambda_{1_{e_{1}}}}, \cdots, \bm{\alpha}_{x_{y}=t}^{\lambda_{y_{1}}}, \cdots, \bm{\alpha}_{x_{y}=t}^{\lambda_{y_{e_{y}}}})$$ with $t\in\{0,1\}$ and real $s\leq 0$. Furthermore, Eqs.(\ref{eq3.3})-(\ref{eq3.4}) and Eqs.(\ref{eq3.14})-(\ref{eq3.15}) allow us to transform the nonlinear inequality in Eq.(\ref{eq3.161}) into an expression involving {\it the generalized  quasiprobability functions of   Gaussian states}.

 Also note that, by Eqs.(\ref{eq3.3}) and (\ref{eq3.4}),  for any fixed $s\leq0$, the generalized quasiprobability function $Q_{\sigma}^{(n)}(\bm{\alpha},\bm{\beta};s)$ and their marginal distribution $Q_{\sigma}^{(n_{1})}(\bm{\alpha};s)$ of any $(n_{1}+n_{2})$-mode Gaussian state $\sigma$ are bounded functions, where $n=n_{1}+n_{2}$, $\bm{\alpha}\in\mathbb{C}^{n_{1}}$ and $\bm{\beta}\in\mathbb{C}^{n_{2}}$. Then, for every fixed $s\leq0$, $\mathcal{B}_{\rho}^{\Xi}(\bm{ \alpha}_{0},\bm{ \alpha}_{1};s)$ in Eq.(\ref{eq3.161}) is a bounded function of $\bm{ \alpha}_{0}$ and $\bm{ \alpha}_{1}$. By Theorem 2.1, the network $\Xi(y,z;\rho)$  is network local will imply
 \begin{equation}\label{eq3.171}B^{\Xi}(s,\rho)=\mathop{\sup}\limits_{\bm{\alpha}_{0},\bm{\alpha}_{1}}
 \mathcal{B}_{\rho}^{\Xi}(\bm{\alpha}_{0}, \bm{\alpha}_{1};s)\leq 1 \tag{3.17}
\end{equation}
holds for all $s\leq 0$. Therefore, we have proved the following

{\bf Theorem 3.1.} {\it The Gaussian network $\Xi(y,z;\rho)$ is network nonlocal if $B^\Xi(s,\rho)>1$ for some $s\leq 0$. }

We refer to this general approach described above as {\it the supremum strategy}.

Note that $B^\Xi(s,\rho)$ can be calculated by Eqs.(\ref{eq3.3})-(\ref{eq3.4}). Although it is difficult to write out the analytic formula of $B^\Xi(s,\rho)$ in general,   it may still be utilized as a tool to detect the network nonlocality in $\Xi(y,z;\rho)$ by numerical method.

 In the following four sections, we applying Theorems 2.1 and 3.1 to the chain networks, the star networks, the tree-shaped networks and the cyclic networks, where all source states are arbitrary bipartite multimode Gaussian states.  Our results
 show that the measurements $M_{x_i}$s in Eq.(\ref{eq3.14}) will work well for witnessing the network nonlocality of Gaussian networks.
 To avoid cumbersome presentation of the results, we in fact focus on $(1+1)$-mode Gaussian states as source states in Sections \ref{sec:4}-\ref{sec:7}. Also, for convenience, we omit the superscript (1+1) of the generalized quasiprobability function  of the $(1+1)$-mode Gaussian state $\rho$ and the superscript (1) of its marginal distribution, denoting both of them with $Q_{\rho}$. Similarly, we omit the superscript of $\hat{O}$ in Eq.(\ref{eq3.14}). For any $p$, $q$,  $\alpha_{p}^{q}$ always stands for a complex number. Moreover, we introduce some functions for the generalized  quasiprobability functions of the $(1+1)$-mode Gaussian states to facilitate subsequent discussion in next four sections.

 Let $\rho_{12}$ be a $(1+1)$-mode Gaussian state. We denote by
 \begin{widetext}
  \begin{equation}\label{eq3.16}C_{\rho_{12}}^{+}(\alpha_{1},\alpha_{2},\alpha_{1}^{'},\alpha_{2}^{'};s)=Q_{\rho_{12}}(\alpha_{1},\alpha_{2};s)
+Q_{\rho_{12}}(\alpha_{1}^{'},\alpha_{2}^{'};s),\tag{3.18}
\end{equation}
\begin{equation}\label{eq3.17}C_{\rho_{12}}^{-}(\alpha_{1},\alpha_{2},\alpha_{1}^{'},\alpha_{2}^{'};s)=Q_{\rho_{12}}(\alpha_{1},\alpha_{2};s)
-Q_{\rho_{12}}(\alpha_{1}^{'},\alpha_{2}^{'};s),\tag{3.19}
\end{equation}
\begin{equation}\label{eq3.18}D_{\rho_{12}}^{+}(\alpha_{1},\alpha_{2},\alpha_{1}^{'},\alpha_{2}^{'};s)=Q_{\rho_{12}}(\alpha_{1};s)+Q_{\rho_{12}}(\alpha_{2};s)
+Q_{\rho}(\alpha_{1}^{'};s)+Q_{\rho}(\alpha_{2}^{'};s),\tag{3.20}
\end{equation}
\begin{equation}\label{eq3.19}D_{\rho_{12}}^{-}(\alpha_{1},\alpha_{2},\alpha_{1}^{'},\alpha_{2}^{'};s)=Q_{\rho_{12}}(\alpha_{1};s)
+Q_{\rho_{12}}(\alpha_{2};s)-Q_{\rho_{12}}(\alpha_{1}^{'};s)-Q_{\rho_{12}}(\alpha_{2}^{'};s),\tag{3.21}
\end{equation}
\end{widetext}
where $Q_{\rho_{12}}(\alpha_{1},\alpha_{2};s)$ and $Q_{\rho_{12}}(\alpha_{1}^{'},\alpha_{2}^{'};s)$ are the generalized  quasiprobability functions of the $(1+1)$-mode Gaussian state $\rho_{12}$ defined
in Eq.(\ref{eq3.3}), $Q_{\rho_{12}}(\alpha_{1};s)$ is the marginal function of $Q_{\rho_{12}}(\alpha_{1},\alpha_{2};s)$  defined as in Eq.(\ref{eq3.4}), that is,
$$Q_{\rho_{12}}(\alpha_1;s)=\int Q_{\rho_{12}}(\alpha_1,\alpha_2;s)d\alpha_2. $$
 $Q_{\rho_{12}}(\alpha_{1}^{'};s)$, $Q_{\rho_{12}}(\alpha_{2};s)$ and $Q_{\rho_{12}}(\alpha_{2}^{'};s)$ are defined similarly.

\section{Witnessing nonlocality in  chain networks of CV systems}\label{sec:4}

\begin{figure}[]
\centering
\includegraphics[width=9cm,height=1.5cm]{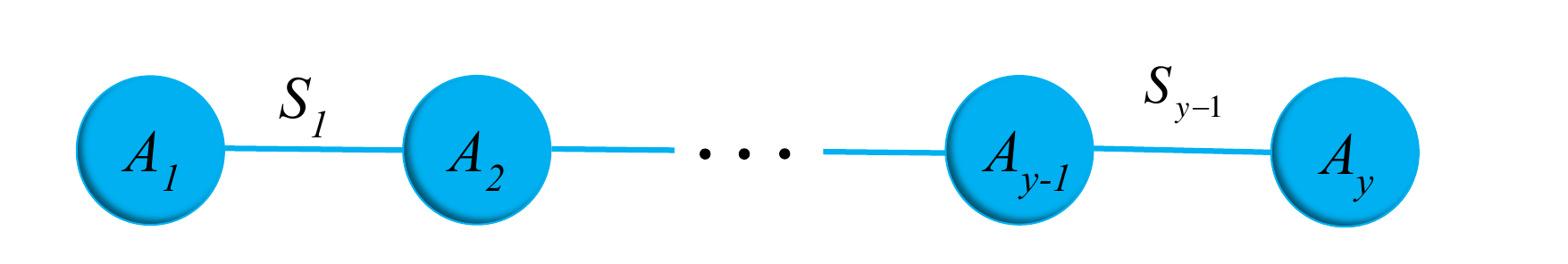}
\caption{\quad \small  A chain network ${\mathcal Cha}(y)$ consists of $y$ parties  $A_{1}$, $A_{2}$, $\cdot\cdot\cdot$, $A_{y}$  and $y-1$ independent sources $S_{1}$, $S_{2}$, $\cdot\cdot\cdot$, $S_{y-1}$. Each  source $S_{j}$ emits $(1+1)$-mode Gaussian state $\rho^{A_{j}A_{j+1}}$.}
 \label{fig1}
\end{figure}

Consider an arbitrary chain network ${\mathcal Cha}(y)$ as described in Figure \ref{fig1}. This is a network consisting of $y$ parties $A_{1}$, $A_{2}$, $\cdot\cdot\cdot$, $A_{y}$,  and $y-1$ independent sources $S_{1}$, $S_{2}$, $\cdot\cdot\cdot$, $S_{y-1}$. The parties and the sources are arranged linearly such that any two neighboring parties share a common source. We assume that each $S_{j}$ emits a $(1+1)$-mode Gaussian state for   $j=1,2,\cdot\cdot\cdot,y-1$, and each $A_{i}$ performs the measurements of the form in Eq.(\ref{eq3.14}) for $i=1,2,\cdot\cdot\cdot,y$. We apply Theorem 2.1 to establish corresponding Bell-type inequality for arbitrary chain network ${\mathcal Cha}(y)$ in Figure \ref{fig1}.

For such ${\mathcal Cha}(y)$, it is clear that (1) if $y$ is odd, then $k_{\rm max}=\frac{y+1}{2}$ with the index set of  maximal set of independent parties ${\mathcal K}=\mathcal K_{\max}=\{1,3,\cdots, y\}$; (2) if $y$ is even, then $k_{\rm max}=\frac{y}{2}$ with the index set of  maximal set of independent parties ${\mathcal K}=\mathcal K_{\max}=\{2,4,\cdots, y\}$. Denote by $\overline{\mathcal K}=\{1,2,\cdots, y\}\setminus {\mathcal K}$ and ${\mathcal K}'$  the set of indexes from ${\mathcal K}$ with the biggest index excluded. For example, when $y$ is odd, ${\mathcal K}'=\{1,3,\cdots, y-2\}$. Also, $C^{+}$, $C^{-}$, $D^{+}$, $D^{-}$ are defined as in Eqs.(\ref{eq3.16})-(\ref{eq3.19}). Then, by Theorem 2.1, we have

{\bf Theorem 4.1.} {\it Let ${\mathcal Cha}(y)$ be a chain network as in Figure \ref{fig1}. Assume that each independent source $S_{j}$ emits $(1+1)$-mode Gaussian state $\rho^{A_{j}A_{j+1}}$ and denote its generalized  quasiprobability function by $Q_{\rho^{A_{j}A_{j+1}}}(\alpha_{x_j}^{\lambda_{j}},\alpha_{x_{j+1}}^{\lambda_{j}};s)$, $x_{j},x_{y}\in\{0,1\}$, $j=1,2,\cdot\cdot\cdot,y-1$. If ${\mathcal C}ha(y)$ is network local, then the following nonlinear Bell-type inequality holds:}
\begin{widetext}
\begin{equation}\label{eq4.1}\mathcal{B}_{\rho}^{ch}(\alpha_{x_1=0}^{\lambda_{1}},\alpha_{x_2=0}^{\lambda_{1}},\cdots, \alpha_{x_{y-1}=0}^{\lambda_{y-1}},\alpha_{x_y=0}^{\lambda_{y-1}}, \alpha_{x_1=1}^{\lambda_{1}},\alpha_{x_2=1}^{\lambda_{1}},\cdots, \alpha_{x_{y-1}=1}^{\lambda_{y-1}},\alpha_{x_y=1}^{\lambda_{y-1}};s)=|\mathcal I_{s}|^{\frac{1}{k_{\rm max}}}+|\mathcal J_{s}|^{\frac{1}{k_{\rm max}}}\leq 1 ,\tag{4.1}
\end{equation}
{\it where $ \rho=\rho^{A_{1}A_{2}}\otimes\cdots\otimes\rho^{A_{y}A_{y-1}}$, $k_{\rm max}=\frac{y+1}{2}$ when $y$ is odd, $k_{\rm max}=\frac{y}{2}$ when $y$ is even; for $-1<s\leq0$,
\begin{equation}
\begin{aligned}
\mathcal I_{s}=&\frac{1}{2^{k_{\rm max}}}\mathop{\prod}\limits_{i\in\mathcal{K}^{'}}\{\frac{\pi^{2}(1-s)^{4}}{4}C_{\rho^{A_{i}A_{i+1}}}^{+}(\alpha_{x_{i}=0}^{\lambda_{i}},\alpha_{x_{i+1}=0}^{\lambda_{i}},\alpha_{x_{i}=1}^{\lambda_{i}},\alpha_{x_{i+1}=0}^{\lambda_{i}};s)+\frac{\pi s(1-s)^{2}}{2}D_{\rho^{A_{i}A_{i+1}}}^{+}\nonumber\\&(\alpha_{x_{i}=0}^{\lambda_{i}},
\alpha_{x_{i+1}=0}^{\lambda_{i}},\alpha_{x_{i}=1}^{\lambda_{i}},\alpha_{x_{i+1}=0}^{\lambda_{i}};s)
+2s^{2}\}\times\mathop{\prod}\limits_{j\in\overline{\mathcal K}}\{\frac{\pi^{2}(1-s)^{4}}{4}C_{\rho^{A_{j}A_{j+1}}}^{+}(\alpha_{x_{j}=0}^{\lambda_{j}},\alpha_{x_{j+1}=0}^{\lambda_{j}},\nonumber\\&\alpha_{x_{j}=0}^{\lambda_{j}},\alpha_{x_{j+1}=1}^{\lambda_{j}};s)
+\frac{\pi s(1-s)^{2}}{2}D_{\rho^{A_{j}A_{j+1}}}^{+}(\alpha_{x_{j}=0}^{\lambda_{j}},
\alpha_{x_{j+1}=0}^{\lambda_{j}},\alpha_{x_{j}=0}^{\lambda_{j}},\alpha_{x_{j+1}=1}^{\lambda_{j}};s)
+2s^{2}\}\nonumber
\end{aligned}
\end{equation}
and
\begin{equation}
\begin{aligned}
\mathcal J_{s}=&\frac{1}{2^{k_{\rm max}}}\mathop{\prod}\limits_{i\in\mathcal{K}^{'}}\{\frac{\pi^{2}(1-s)^{4}}{4}C_{\rho^{A_{i}A_{i+1}}}^{-}(\alpha_{x_{i}=0}^{\lambda_{i}},\alpha_{x_{i+1}=1}^{\lambda_{i}},\alpha_{x_{i}=1}^{\lambda_{i}},\alpha_{x_{i+1}=1}^{\lambda_{i}};s)+\frac{\pi s(1-s)^{2}}{2}D_{\rho^{A_{i}A_{i+1}}}^{-}\nonumber\\&(\alpha_{x_{i}=0}^{\lambda_{i}},
\alpha_{x_{i+1}=1}^{\lambda_{i}},\alpha_{x_{i}=1}^{\lambda_{i}},\alpha_{x_{i+1}=1}^{\lambda_{i}};s)\}
\times\mathop{\prod}\limits_{j\in\overline{\mathcal K}}\{\frac{\pi^{2}(1-s)^{4}}{4}C_{\rho^{A_{j}A_{j+1}}}^{-}(\alpha_{x_{j}=1}^{\lambda_{j}},\alpha_{x_{j+1}=0}^{\lambda_{j}},\nonumber\\&\alpha_{x_{j}=1}^{\lambda_{j}},\alpha_{x_{j+1}=1}^{\lambda_{j}};s)
+\frac{\pi s(1-s)^{2}}{2}D_{\rho^{A_{j}A_{j+1}}}^{-}(\alpha_{x_{j}=1}^{\lambda_{j}},\alpha_{x_{j+1}=0}^{\lambda_{j}},
\alpha_{x_{j}=1}^{\lambda_{j}},\alpha_{x_{j+1}=1}^{\lambda_{j}};s)\}\nonumber;
\end{aligned}
\end{equation}
while for $s\leq-1$,
\begin{equation}
\begin{aligned}
\mathcal I_{s}=&\frac{1}{2^{k_{\rm max}}}\mathop{\prod}\limits_{i\in\mathcal{K}^{'}}\{\pi^{2}(1-s)^{2}C_{\rho^{A_{i}A_{i+1}}}^{+}(\alpha_{x_{i}=0}^{\lambda_{i}},\alpha_{x_{i+1}=0}^{\lambda_{i}},\alpha_{x_{i}=1}^{\lambda_{i}},\alpha_{x_{i+1}=0}^{\lambda_{i}};s)-\pi(1-s)
D_{\rho^{A_{i}A_{i+1}}}^{+}\nonumber\\&(\alpha_{x_{i}=0}^{\lambda_{i}},\alpha_{x_{i+1}=0}^{\lambda_{i}},
\alpha_{x_{i}=1}^{\lambda_{i}},\alpha_{x_{i+1}=0}^{\lambda_{i}};s)+2\}\times\mathop{\prod}\limits_{j\in\overline{\mathcal K}}\{\pi^{2}(1-s)^{2}C_{\rho^{A_{j}A_{j+1}}}^{+}(\alpha_{x_{j}=0}^{\lambda_{j}},\alpha_{x_{j+1}=0}^{\lambda_{j}},\nonumber\\&\alpha_{x_{j}=0}^{\lambda_{j}},\alpha_{x_{j+1}=1}^{\lambda_{j}};s)
-\pi(1-s)D_{\rho^{A_{j}A_{j+1}}}^{+}(\alpha_{x_{j}=0}^{\lambda_{j}},\alpha_{x_{j+1}=0}^{\lambda_{j}},
\alpha_{x_{j}=0}^{\lambda_{j}},\alpha_{x_{j+1}=1}^{\lambda_{j}};s)+2\}\nonumber
\end{aligned}
\end{equation}
and
\begin{equation}
\begin{aligned}
\mathcal J_{s}=&\frac{1}{2^{k_{\rm max}}}\mathop{\prod}\limits_{i\in\mathcal{K}^{'}}\{\pi^{2}(1-s)^{2}C_{\rho^{A_{i}A_{i+1}}}^{-}(\alpha_{x_{i}=0}^{\lambda_{i}},\alpha_{x_{i+1}=1}^{\lambda_{i}},\alpha_{x_{i}=1}^{\lambda_{i}},\alpha_{x_{i+1}=1}^{\lambda_{i}};s) -\pi(1-s)D_{\rho^{A_{i}A_{i+1}}}^{-}\nonumber\\&(\alpha_{x_{i}=0}^{\lambda_{i}},
\alpha_{x_{i+1}=1}^{\lambda_{i}},\alpha_{x_{i}=1}^{\lambda_{i}},\alpha_{x_{i+1}=1}^{\lambda_{i}};s)\}
\times\mathop{\prod}\limits_{j\in\overline{\mathcal K}}\{\pi^{2}(1-s)^{2}C_{\rho^{A_{j}A_{j+1}}}^{-}(\alpha_{x_{j}=1}^{\lambda_{j}},\alpha_{x_{j+1}=0}^{\lambda_{j}},\nonumber\\&\alpha_{x_{j}=1}^{\lambda_{j}},\alpha_{x_{j+1}=1}^{\lambda_{j}};s)
-\pi(1-s)D_{\rho^{A_{j}A_{j+1}}}^{-}(\alpha_{x_{j}=1}^{\lambda_{j}},\alpha_{x_{j+1}=0}^{\lambda_{j}},
\alpha_{x_{j}=1}^{\lambda_{j}},\alpha_{x_{j+1}=1}^{\lambda_{j}};s)\}\nonumber.
\end{aligned}
\end{equation}
Here, $C^{+}$, $C^{-}$, $D^{+}$, $D^{-}$ are defined as in Eqs.(\ref{eq3.16})-(\ref{eq3.19}).}
\end{widetext}

A proof of Theorem 4.1 is provided in Appendix B.

By the above result, if the inequality (\ref{eq4.1}) is violated by some choice of parameters $s$, $\alpha_{x_{i}=0}^{\lambda_{i}}$, $\alpha_{x_{i}=1}^{\lambda_{i}}$, $\alpha_{x_{i+1}=0}^{\lambda_{i}}$ and $\alpha_{x_{i+1}=1}^{\lambda_{i}}$ for $i=1,2,\cdots, y-1$,
then it demonstrates the nonlocality  in the chain network ${\mathcal C}ha(y)$ of CV system (Figure \ref{fig1}).  Applying the supremum strategy described by  Eq.(\ref{eq3.171}) and Theorem 3.1, one can readily show that if the chain network ${\mathcal C}ha(y)$ is network local, then
\begin{equation}\tag{4.2}\label{eq4.2}
B^{ch}(s,\rho)=\mathop{\sup}\limits_{\bm{\alpha}_0,\bm{\alpha}_1}
\mathcal{B}_{\rho}^{ch}(\bm{\alpha}_0,\bm{\alpha}_1;s) \leq 1,
\end{equation}
where $\bm{\alpha}_t=(\alpha_{x_1=t}^{\lambda_{1}},\alpha_{x_2=t}^{\lambda_{1}},\cdots, \alpha_{x_{y-1}=t}^{\lambda_{y-1}},\alpha_{x_y=t}^{\lambda_{y-1}})$ for $t=0,1$.  Therefore,  ${\mathcal C}ha(y)$ is network nonlocal if $B^{ch}(s,\rho)>1$.

 As a particular category of ${\mathcal C}ha(y)$, the entanglement swapping network ${\mathcal E}(3)={\mathcal C}ha(3)$ is a standard tool in quantum
information processing. Next, for the scenario of all quantum source states consisting of either  pure Gaussian states or STSs, we  conduct a detailed  analysis  of how to use the inequalities (\ref{eq4.1}) and (\ref{eq4.2}) to witness the nonlocality in ${\mathcal E}(3)$ of CV system.

\begin{figure*}[]
\center
\subfigure [$s=-0.5$]
{\includegraphics[width=4cm,height=4cm]{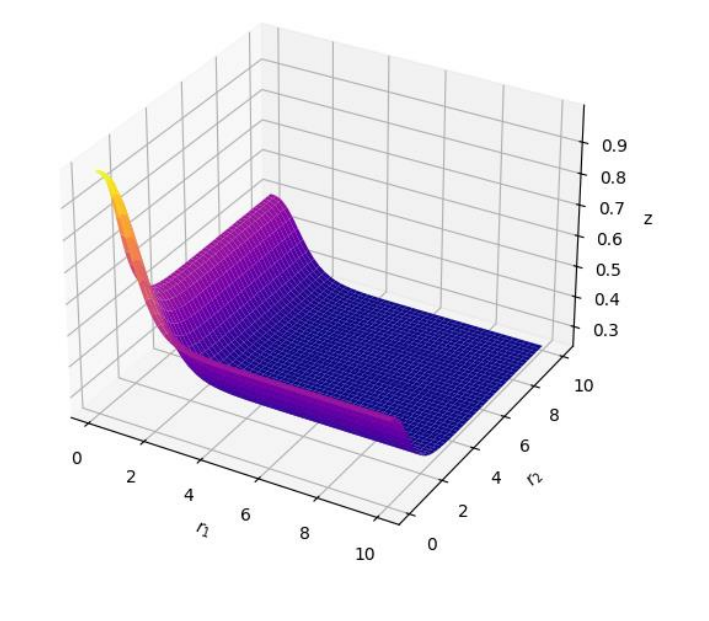}}
\subfigure [$s=-1$]
{\includegraphics[width=4cm,height=4cm]{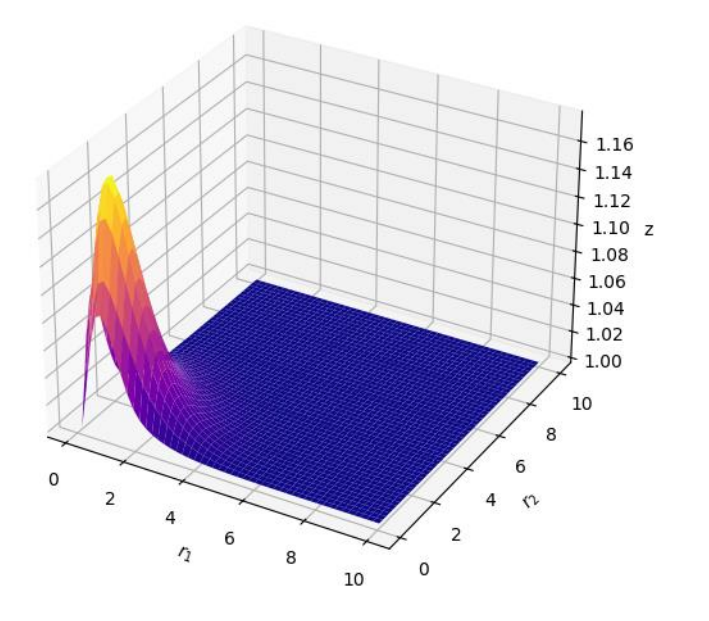}}
\subfigure [$s=-2$]
{\includegraphics[width=4cm,height=4cm]{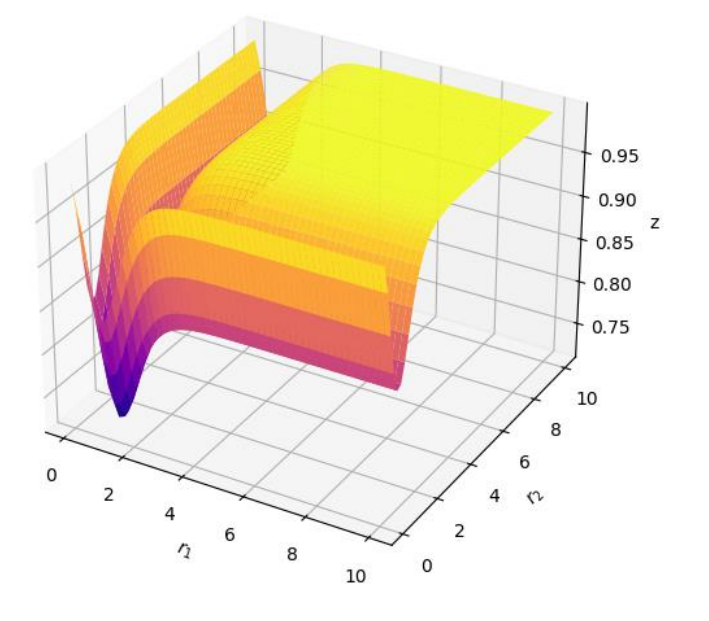}}

\caption{\small\quad  Values of $z=B^{en}(s,r_{1},r_{2})$ in ${\mathcal E}(3)$  for $\rho^{A_{1}A_{2}}=\rho^{r_{1}}$ and $\rho^{A_{2}A_{3}}=\rho^{r_{2}}$, as a function of the parameters $r_{1}$ and $r_{2}$ for fixed $s$. The subfigures (a), (b) and (c) exhibit respectively the cases of $s=-0.5$, $s=-1$, $s=-2$.}
\centering
\label{fig21}
\end{figure*}

\begin{figure}[]
\centering
\includegraphics[width=7cm,height=5cm]{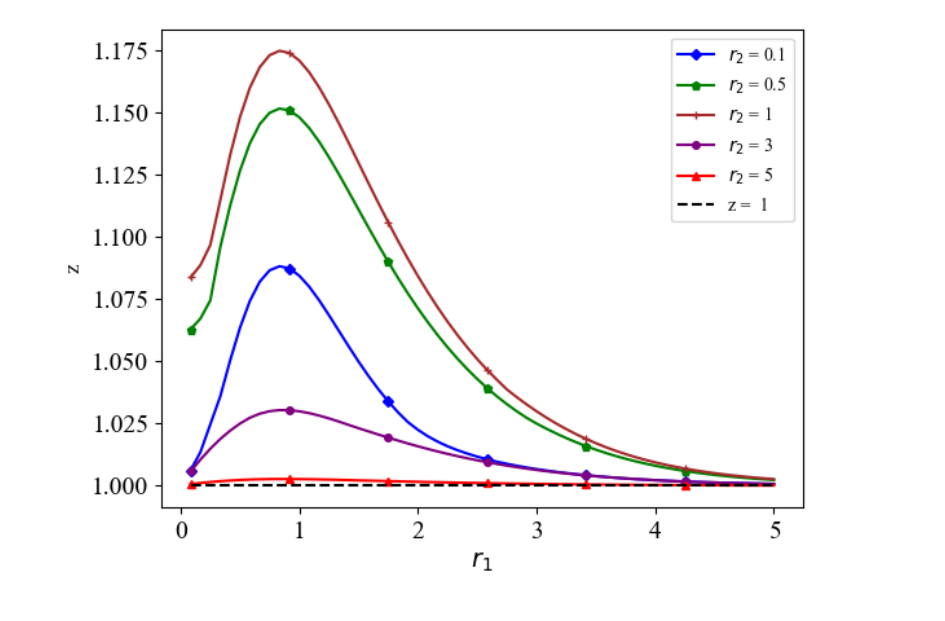}
\caption{\quad \small  Values of $z=B^{en}(-1,r_{1},r_{2})$ in ${\mathcal E}(3)$ for $\rho^{A_{1}A_{2}}=\rho^{r_{1}}$ and $\rho^{A_{2}A_{3}}=\rho^{r_{2}}$, as a function of the parameter $r_{1}$ for fixed  $r_{2}=0.1,0.5,1,3,5$.}
\label{fig22}
\end{figure}

{\bf Example 4.2}. {\it The scenario that source states are arbitrary  $(1+1)$-mode pure Gaussian states in $\mathcal E(3)$.} We consider the situation where both  source states of entanglement swapping network ${\mathcal E}(3)$ are $(1+1)$-mode pure Gaussian states. As  every  $(1+1)$-mode Gaussian pure state can be transformed into an EPR state via a local Gaussian unitary transformation  and local unitary operations do not change the network locality, it suffices to consider the scenario where both  source states are (may be different) $(1+1)$-mode EPR states in ${\mathcal E}(3)$.

Assume two sources emit both (1+1)-mode EPR states, namely $\rho^{r_{1}}$ and $\rho^{r_{2}}$. By  inequalities (\ref{eq4.1}) and (\ref{eq4.2}),  $B^{en}(s,\rho)=B^{ch}(s,\rho)$  is a function of $r_{1}$, $r_{2}$ and $s$, denoted it as $B^{en}(s,r_{1},r_{2})$. Then, $B^{en}(s,r_{1},r_{2})>1$ witnesses the network nonlocality in ${\mathcal E}(3)$. $B^{en}(s,r_{1},r_{2})$ can be calculated by Eqs.(\ref{eq3.8})-(\ref{eq3.9}). \if false Although it is difficult to write out the analytic formula of $B^{en}(s,r_{1},r_{2})$,   it may still be utilized as a tool to detect the network nonlocality in ${\mathcal E}(3)$ by numerical method.\fi

Figures \ref{fig21}(a), \ref{fig21}(b) and \ref{fig21}(c) respectively exhibit  the behaviour of $B^{en}(s,r_{1},r_{2})$ as a function of  $r_{1}$ and $r_{2}$  for specific, predetermined values of $s$. To illustrate more
intuitively the behavior of $B^{en}(s,r_{1},r_{2})$ when $s=-1$, we present Figure \ref{fig22}.

By comparing Figures \ref{fig21}(a), \ref{fig21}(b) and \ref{fig21}(c), we see that, if $s=-0.5,-2$, the generalized quasiprobability functions cannot witness the network nonlocality.
Additionally, by Figures \ref{fig21}(b) and \ref{fig22},  when taking $s=-1$, the inequality $B^{en}(-1,r_{1},r_{2})>1$ holds for all $r_1>0$ and $r_2>0$ and hence,  ${\mathcal E}(3)$ demonstrates the network nonlocality nature if the two sources emit  $(1+1)$-mode entangled EPR states. Furthermore, since any Gaussian pure state can be transformed into an EPR state through a local Gaussian unitary transformation, {\it the generalized quasiprobability function with $s=-1$ alone can witness all network nonlocality in the entanglement swapping network ${\mathcal E}(3)$ if the two sources  emit $(1+1)$-mode entangled Gaussian pure states}. {\color{red} This result is consistent with the result in \cite{SAA}.} In addition, it is also analogous to the result  for DV systems obtained in \cite{NQA}, where it was  shown that if two source states are entangled pure states, then the entanglement swapping network of DV system is network nonlocal. Figure \ref{fig22} further reveals that the entanglement swapping network ${\mathcal E}(3)$ may be network nonlocal even if one of the source states is separable as $B^{en}(-1,0,0.5)>1$. Another behavior of  $B^{en}(-1,r_{1},r_{2})$ is that $\lim_{r_1,r_2\to\infty}B^{en}(-1,r_{1},r_{2})=1^+$, which reveals that the generalized quasiprobability functions are not very powerful to detect the network nonlocality for larger $r_1,r_2$.  Anyway, {\it our result  suggests that measurements based solely on the generalized quasiprobability functions with $s=-1$ are sufficient to detect the network nonlocality of the entanglement swapping network ${\mathcal E}(3)$ in CV systems  if the source states are pure Gaussian states}.

\begin{figure*}[]
\center
\subfigure [$s=-0.5$]
{\includegraphics[width=4cm,height=3.8cm]{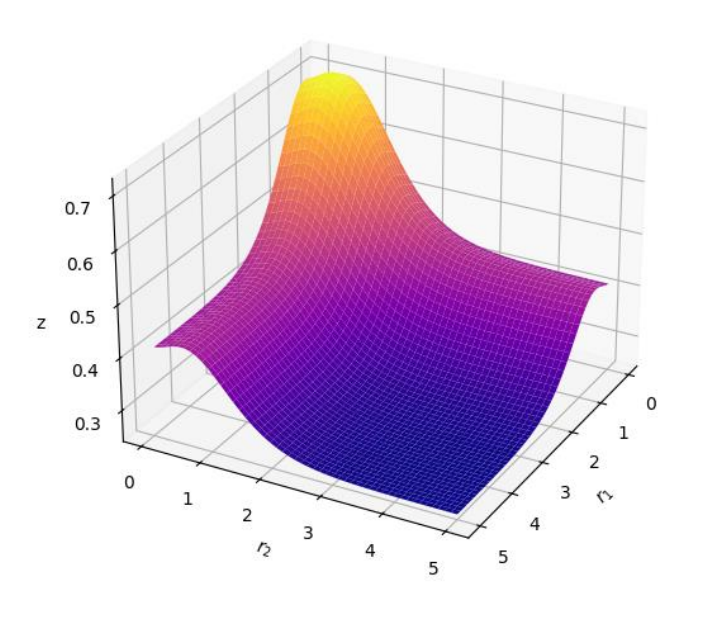}}
\subfigure [$s=-1$]
{\includegraphics[width=4cm,height=3.8cm]{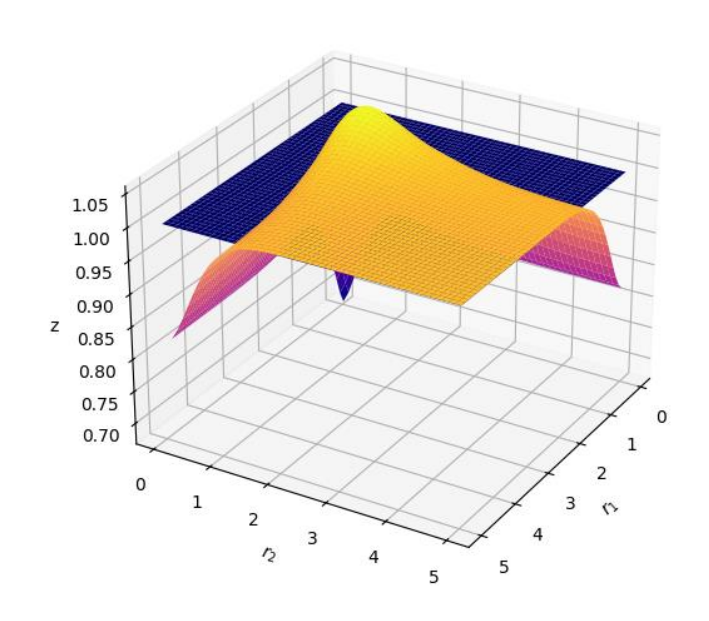}}
\subfigure [$s=-2$]
{\includegraphics[width=4cm,height=3.8cm]{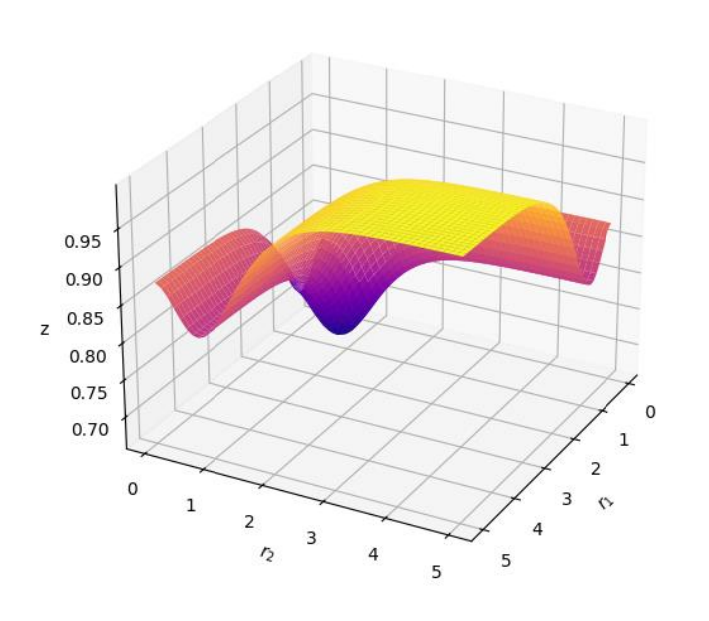}}

\caption{\small\quad  Values of  $z=B^{en}(s,r_{1},r_{2})$ in ${\mathcal E}(3)$ for $\rho^{A_{1}A_{2}}=\rho(1.2, 1.2, r_{1})$ and $\rho^{A_{2}A_{3}}=\rho(1.2, 1.2, r_{2})$, as a function of the parameters $r_{1}$ and $r_{2}$  for fixed $s$. The subfigures (a), (b) and (c) exhibit respectively the cases of $s=-0.5$, $s=-1$, $s=-2$. The blue part represents the plane $z=1$ in subfigure (b), from which one can easily see the region that $z>1$.}
\centering
\label{fig23}
\end{figure*}

\begin{figure*}[]
\center
\subfigure [$0<r_{2}\leq1.2$]
{\includegraphics[width=7cm,height=5cm]{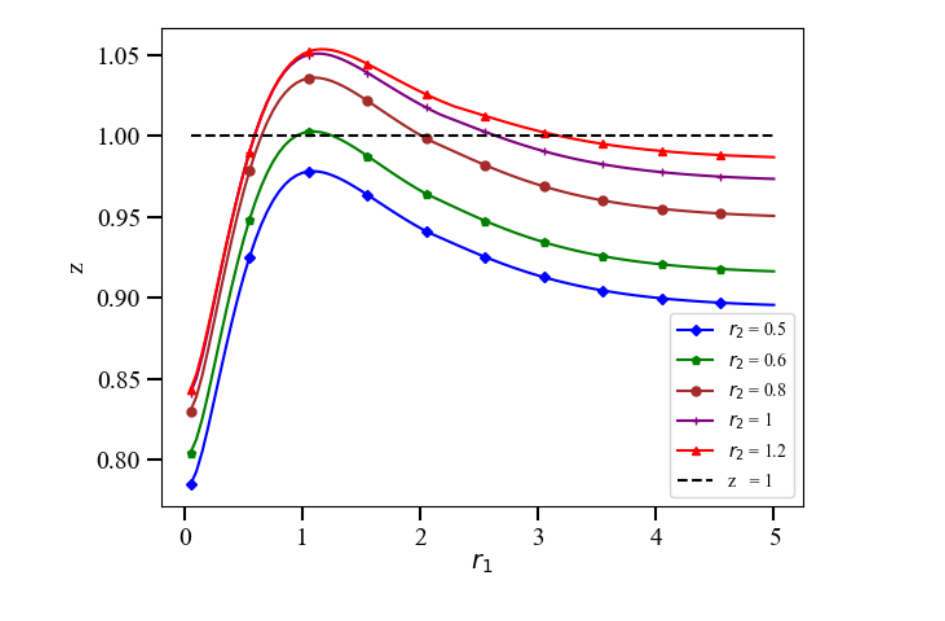}}
\subfigure [$ r_{2}>1.2$]
{\includegraphics[width=7cm,height=5cm]{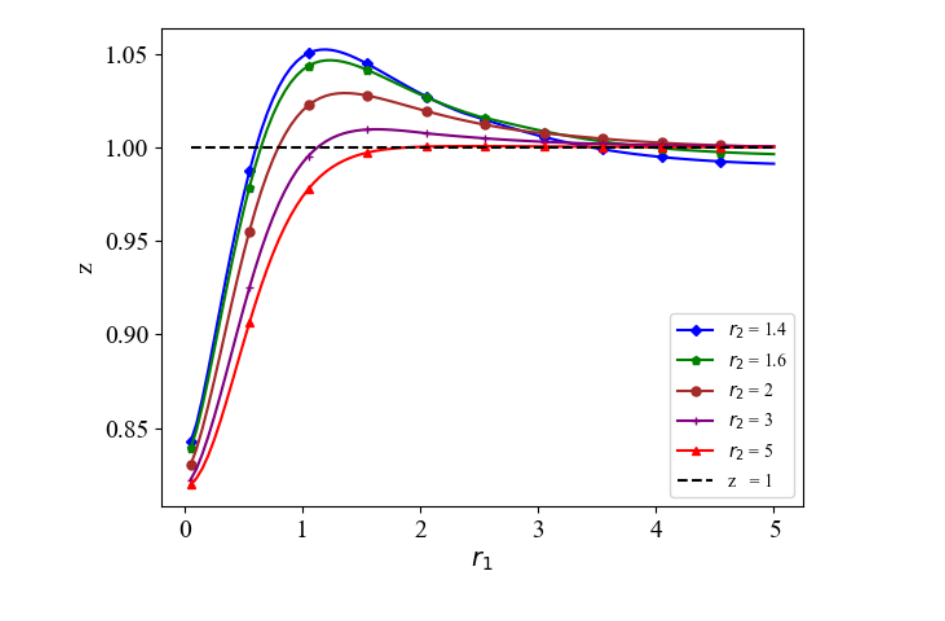}}

\caption{\small\quad  Values of  $z=B^{en}(-1,r_{1},r_{2})$ in ${\mathcal E}(3)$ for $\rho^{A_{1}A_{2}}=\rho(1.2, 1.2, r_{1})$ and $\rho^{A_{2}A_{3}}=\rho(1.2, 1.2, r_{2})$, as a function of the parameter $r_{1}$ for fixed $r_{2}$.  The subfigure (a) shows the case of $0<r_{2}\leq1.2$, while the subfigure (b) corresponds to the case of $r_{2}>1.2$.}
\centering
\label{fig24}
\end{figure*}

{\bf Example 4.3}. {\it The scenario that source states are  $(1+1)$-mode symmetric STSs in $\mathcal E(3)$.} For the situation where two quantum source states of ${\mathcal E}(3)$ are mixed Gaussian states, we consider the scenario that the two quantum source states specifically constitute $(1+1)$-mode symmetric STSs.

For simplicity's sake, assume that the two quantum source states in ${\mathcal E}(3)$ are  $(1+1)$-mode mixed symmetric STSs, specifically $\rho(1.2, 1.2, r_{1})$ and $\rho(1.2, 1.2, r_{2})$ as defined in Eq.(\ref{eq3.10}).
Then  $B^{en}(s,r_{1},r_{2})=B^{ch}(s,\rho)$ in the inequality (\ref{eq4.2}) is expressed as a function that depends on $s$, $r_{1}$ and $r_{2}$. Thus, $B^{en}(s,r_{1},r_{2})>1$ will witness the network nonlocality in ${\mathcal E}(3)$.
Considering $B^{en}(s,r_{1},r_{2})$ as a function of the parameters $r_{1}$ and $r_{2}$, with
$s$ as constants, the behavior of $B^{en}(s,r_{1},r_{2})$ is shown in Figure \ref{fig23}. To more intuitively illustrate the behavior of $B^{en}(s,r_{1},r_{2})$ when $s=-1$, we present Figure \ref{fig24}. As $\rho(1.2, 1.2, r_{j})$ is separable if and only if $\cosh(r_j)\leq\sqrt{\frac{121}{120}}$, i.e., $0\leq r_j\leq r_s={\rm arcosh}(\sqrt{\frac{121}{120}})\in (0.09116,0.09117)$, $j=1,2$, we see that  ${\mathcal E}(3)$ is network local if both $r_1,r_2\in[0,r_s]$, and thus $B^{en}(s,r_{1},r_{2})\leq 1$ for $(r_1,r_2)\in [0,r_s]\times [0,r_s]$.

By comparing Figures \ref{fig23}(a), \ref{fig23}(b) and \ref{fig23}(c), we see that the inequality $B^{en}(s,r_{1},r_{2})>1$ may hold only when $s=-1$. And from Figure \ref{fig23}(b), it is known that when both $r_{1}$ and $r_{2}$ are relatively large, the condition $B^{en}(-1,r_{1},r_{2})>1$ is satisfied, demonstrating the network nonlocality in
${\mathcal E}(3)$. Specifically, by Figure \ref{fig24}(a), with $0.8\leq r_{2}\leq1.2$,  $B^{en}(-1,r_{1},r_{2})>1$ holds true for $0.6\leq r_{1}\leq 2.2$,  while $B^{en}(-1,r_{1},r_{2})>1$ if $r_2=1.2$ and $0.5\leq r_1\leq 3.2$, demonstrating the network nonlocality. For the situation $s=-1$ and $r_{2}>1.2$,
 Figure \ref{fig24}(b) reveals that ${\mathcal E}(3)$    demonstrates the network nonlocality if $r_2=1.4$ and $0.6\leq r_1\leq 3.3$. Additionally, if $2\leq r_{2}\leq5$,  $B^{en}(-1,r_{1},r_{2})>1$ holds for $r_{1}\geq2$, demonstrating the network nonlocality. Similar conclusions may be obtained when considering $B^{en}(s,r_{1},r_{2})$ as a function of $r_2$.

%From Fig.5(b), if $s=-1$ and $r_{2}\leq1$, for a fixed $r_{2}$, the value of $B(-1,r_{1},r_{2})$ first increases and then decreases with the increase of $r_{1}$, eventually converging to a fixed value, and the maximum violation of inequality (4.1) is achieved when $r_{1}\approx1.1$. In Fig.5(c), if $s=-1$ and $r_{2}>1$,  fixing $r_{2}$, $B(-1,r_{1},r_{2})$ gradually approaches 1 as $r_{1}$ increases.
%Combining Fig.5(b) and Fig.5(c), we conclude that violating the network local inequality (4.1) implies that at least one of the two symmetrical squeezed thermal states distributed within the network exhibits network nonlocality.

In summary, {\it if the two sources in the entanglement swapping network ${\mathcal E}(3)$ emit $(1+1)$-mode mixed symmetric STSs, say, for example, $\rho(1.2, 1.2, r_{1})$ and $\rho(1.2, 1.2, r_{2})$, using the generalized quasiprobability functions with $s=-1$ enables the detection of network nonlocality for many values of  $r_1, r_2$}.

 \if false  which is determined by appropriately selecting the parameters $\alpha_{x_{i}=0}^{\lambda_{i}}$, $\alpha_{x_{i}=1}^{\lambda_{i}}$, $\alpha_{x_{i+1}=0}^{\lambda_{i}}$ and $\alpha_{x_{i+1}=1}^{\lambda_{i}}$ for $i=1,2,\cdot\cdot\cdot, y-1$.\fi

 For a chain  network ${\mathcal C}ha(y)$ with $y\geq3$, due to its complexity, we only consider cases where all sources emit the same $(1+1)$-mode EPR states (see Example 4.4) and the same $(1+1)$-mode symmetric STSs (see Example 4.5).

 \begin{figure*}[]
\center
\subfigure [$-1<s\leq0$]
{\includegraphics[width=7cm,height=5cm]{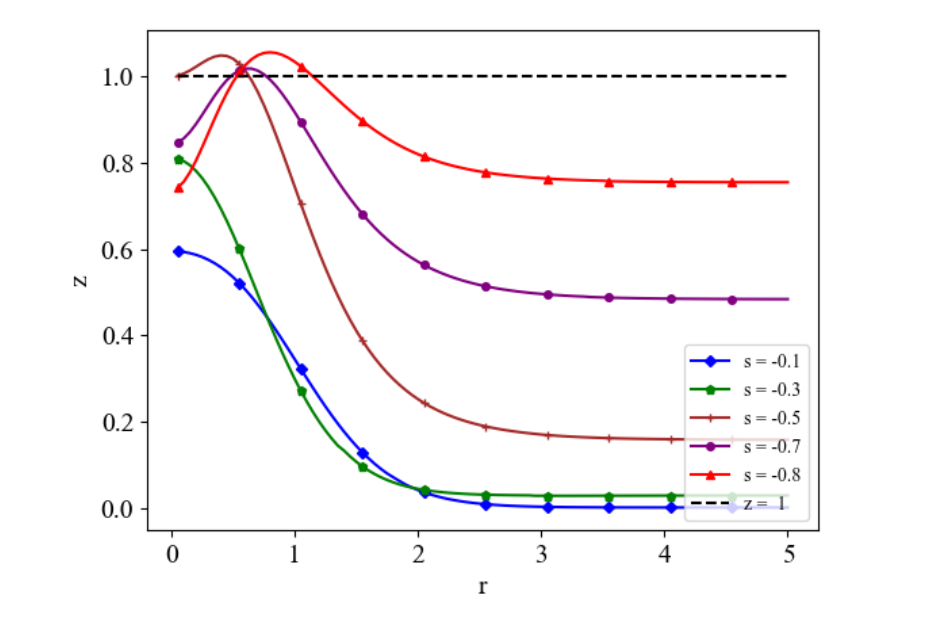}}
\subfigure [$s\leq-1$]
{\includegraphics[width=7cm,height=5cm]{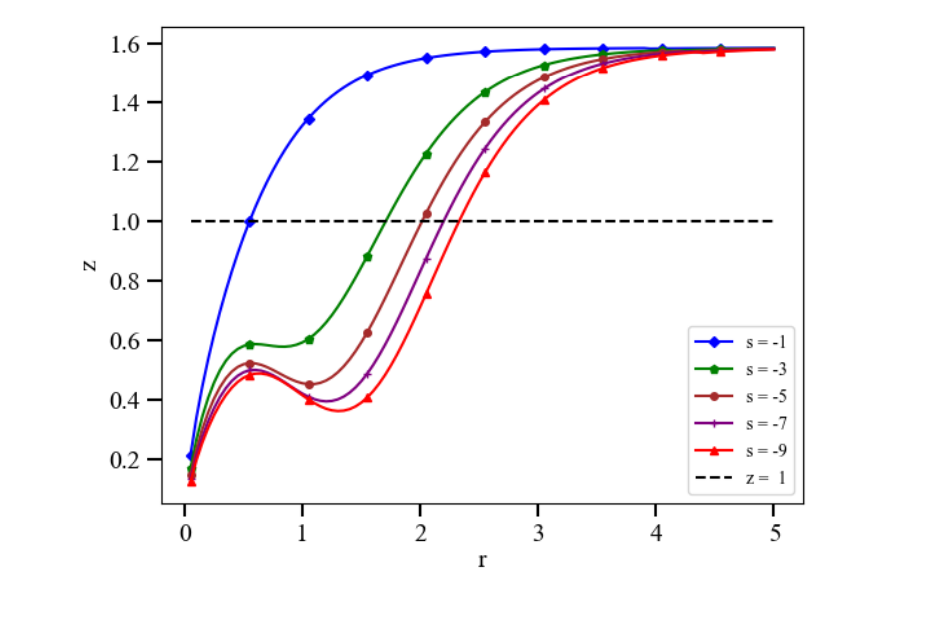}}

\caption{\small\quad Values of $z=B^{ch}(s,r)$ in ${\mathcal C}ha(6)$ for $(1+1)$-mode Gaussian state $\rho^{A_{i}A_{i+1}}=\rho^{r}(i=1,2,3,4,5)$, as a function of the parameter $r$ for fixed $s$. The figure (a) displays the behavior  of $z=B^{ch}(s,r)$ when $-1<s\leq0$, whereas the figure (b) illustrates the behavior  of $z=B^{ch}(s,r)$ when $s\leq-1$.}
\centering
\label{fig2}
\end{figure*}

\begin{figure*}[]
\center
\subfigure [$-1<s\leq0$]
{\includegraphics[width=7cm,height=5.5cm]{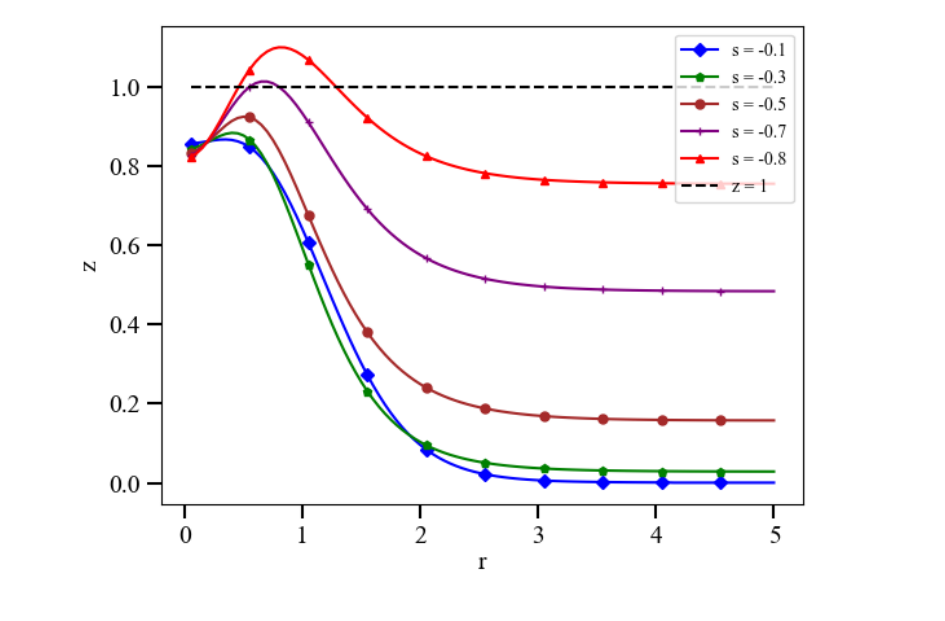}}
\subfigure [$s\leq-1$]
{\includegraphics[width=7cm,height=5.5cm]{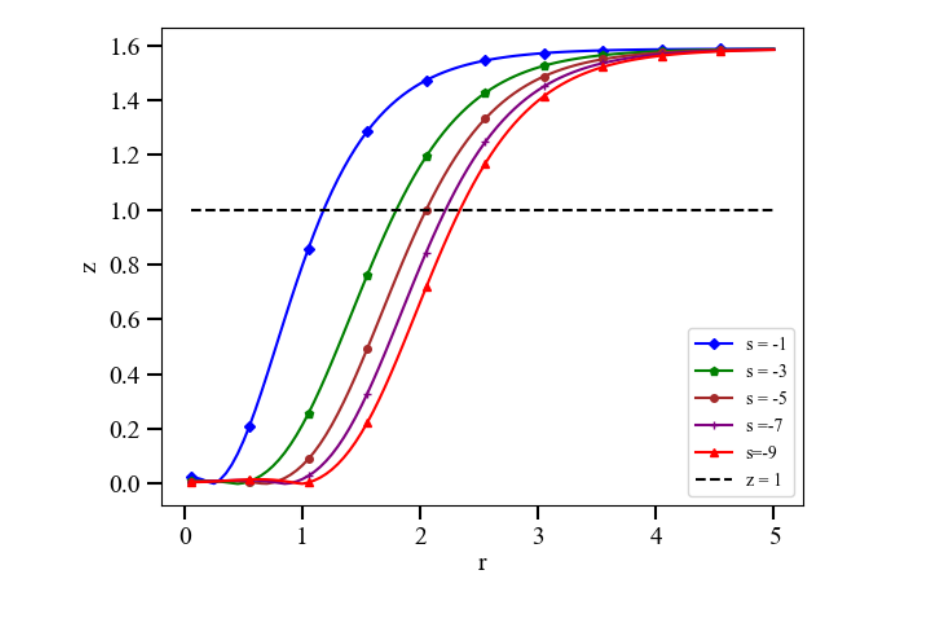}}

\caption{\small\quad  Values of $z=B^{ch}(s,r)$ in ${\mathcal C}ha(6)$ for $(1+1)$-mode Gaussian state $\rho^{A_{i}A_{i+1}}=\rho(1.2,1.2,r)\ (i=1,2,3,4,5)$, as a function of the parameter $r$ for fixed $s$. The figure (a) displays the behavior  of $z=B^{ch}(s,r)$ when $-1<s\leq0$, whereas the figure (b) illustrates the behavior  of $z=B^{ch}(s,r)$ when $s\leq-1$.}
\centering
\label{fig3}
\end{figure*}

{\bf Example 4.4.} Consider  the chain network ${\mathcal C}ha(6)$. Assume that all source states are the same  EPR state $\rho^{r}$ as defined in Eq.(\ref{eq3.7}), that is, $\rho^{A_{i}A_{i+1}}= \rho^{r}$ for each $i=1,2,3,4,5$. Then $B^{ch}(s,\rho)$ in the inequality (\ref{eq4.2}) is a function $B^{ch}(s,r)$ of  $s$ and $r$, and $B^{ch}(s,r)>1$ witnesses the network nonlocality in ${\mathcal C}ha(6)$.

 For certain fixed values of  $s$, Figure \ref{fig2} is derived by considering $B^{ch}(s,r)$ as a function of the parameter $r$. It reveals that, as a function of $r$, $B^{ch}(s,r)$ exhibits entirely different behaviors for $s\in(-1,0]$ and $s\in(-\infty, -1]$. If $s\in(-1,0]$, $B^{ch}(s,r)$ has advantage to witness the network nonlocality for those $r$ near $0$, while, if $s\in(-\infty, -1]$, $B^{ch}(s,r)$ has advantage to witness the network nonlocality for those $r$ far from $0$. \if false
As observed in Figure 2(a), if $s\in(-1,0]$, the condition $B^{ch}(s,r)>1$ is satisfied only when $r$ is relatively small. However, as shown in Figure 2(b), for sufficiently large values of $r$,  it becomes necessary to use generalized quasiprobability functions with $s\in(-\infty, -1]$ to achieve $B^{ch}(s,r)>1$.\fi  Observe also that, for $s\in(-\infty, -1]$, as $s$ increases, more values of $r$ make the inequality $B^{ch}(s,r) > 1$ holding true. When $s=-1$, a larger set of $r$ values results in  $B^{ch}(s,r)>1$, that is, using the generalized quasiprobability function with $s=-1$ can witness the network nonlocality for more EPR states.

Furthermore, as depicted in Figure \ref{fig2}(a), when $0<r\leq0.63$, we observe that $B^{ch}(-0.5,r)>1$. Similarly, in Figure \ref{fig2}(b), when $r\geq0.55$, we have $B^{ch}(-1,r)>1$. Consequently, for any $r>0$, the inequality $B^{ch}(s,r)>1$  holds for some choice of $s$, demonstrating the network nonlocality nature. Therefore, we conclude that {\it as long as the five sources in the chain network ${\mathcal C}ha(6)$ emit any identical entangled Gaussian pure states, ${\mathcal C}ha(6)$ will generate network nonlocality, which can be witnessed by the generalized quasiprobability functions}. This result is true for any chain network ${\mathcal C}ha(y)$ of CV systems.

{\bf Example 4.5.} Assume that all source states of the chain network ${\mathcal C}ha(6)$ are the same symmetric STSs $\rho(v,v,r)$ as given in Eq.(\ref{eq3.10})  with $v=1.2$, i.e., $\rho^{A_{i}A_{i+1}}=\rho(1.2,1.2,r)$   for each $i=1,2,3,4,5$. Note that $\rho(1.2,1.2,r)$ is a mixed  Gaussian state and it is entangled if and only if $r> r_s$, where $0.09116<r_s<0.09117$. It follows that ${\mathcal C}ha(6)$ is network local if $0\leq r\leq r_s$. To demonstrate the network nonlocality,  we focus our attention on the  $B^{ch}(s,\rho)$ in the inequality (\ref{eq4.2}), which is a function $B^{ch}(s,r)$ of $s$ and $r$. \if false This function is obtained by selecting the appropriate parameters $\alpha_{x_{i}=0}^{\lambda_{i}}$, $\alpha_{x_{i}=1}^{\lambda_{i}}$, $\alpha_{x_{i+1}=0}^{\lambda_{i}}$ and $\alpha_{x_{i+1}=1}^{\lambda_{i}}$ for $i=1,2,3,4,5$. Subsequently,\fi By Theorem 4.1, $B^{ch}(s,r)>1$ witnesses the network nonlocality in ${\mathcal C}ha(6)$. Figure \ref{fig3} is derived by considering
$B^{ch}(s,r)$ as a function of the parameter $r$, with
$s$ held constant.

From Figure \ref{fig3}(a), within the range $-1<s\leq0$, it is observed that only when $s\leq-0.7$ does a small subset of symmetric STSs lead to $B^{ch}(s,r)>1$. However, as depicted in Figure \ref{fig3}(b), when $s\leq-1$, $B^{ch}(s,r)$ can detect network nonlocality in a larger number of STSs. Moreover, within the interval $s\leq-1$, as $s$ increases, an increasing number of STSs contributes to the condition $B^{ch}(s,r)>1$. Therefore, when all sources emit symmetric STSs in ${\mathcal C}ha(6)$, employing the  generalized quasiprobability functions with $s=-1$ allows for the detection of network nonlocality in a greater number of instances.

In addition, from Figure \ref{fig3}(a), the inequality $B^{ch}(-0.8,r)>1$ holds  when $0.5\leq r\leq1.25$. And in Figure \ref{fig3}(b), the inequality $B^{ch}(-1,r)>1$ is satisfied for $r\geq1.18$. Thus we conclude that {\it if all sources emit symmetric STSs $\rho^{A_{i}A_{i+1}}=\rho(1.2,1.2,r)$ $(i=1,2,3,4,5)$  in ${\mathcal Cha}(6)$ and $r\geq0.5$, then ${\mathcal Cha}(6)$ exhibits network nonlocality, which can be witnessed by the generalized quasiprobability functions}. However, for this scenario,  we still do not know whether ${\mathcal Cha}(6)$ is network local or not for $r_{s}<r<0.5$. We believe that, there exists $r_{1.2}$ with $0<r_s\leq r_{1.2}<0.5$ such that ${\mathcal Cha}(6)$ is network local if and only if $0\leq r\leq r_{1.2}$. Generally, if all source states are the same symmetric STSs $\rho(v,v,r)$, there exists $r_v>0$ such that ${\mathcal Cha}(6)$ is network local if and only if $0\leq r\leq r_{v}$.

\begin{figure}[]
\centering
\includegraphics[width=5.5cm,height=5.3cm]{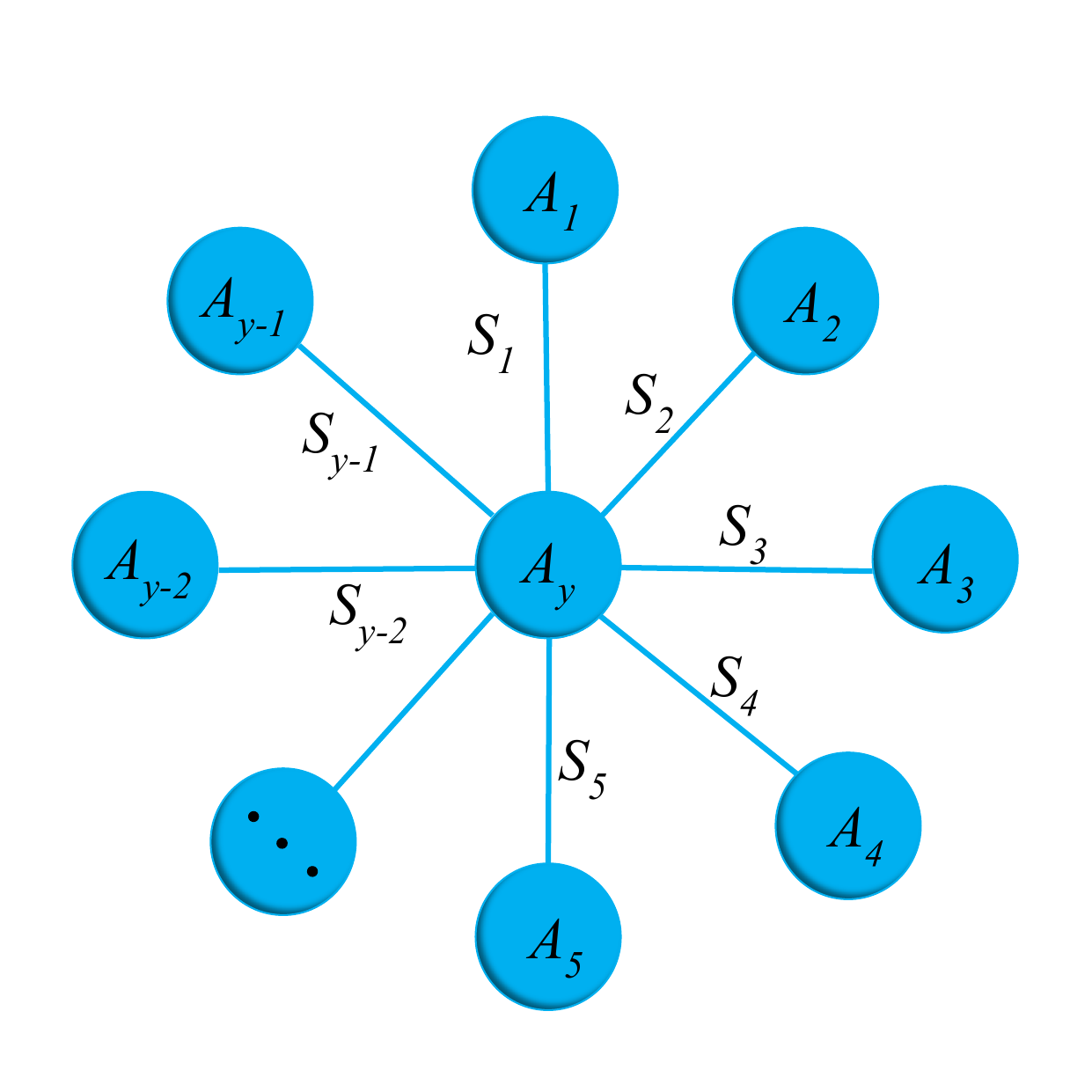}
\caption{\small\quad   A star network ${\mathcal S}(y)$ consists of $y$ parties  $ A_{1}, A_{2}, \cdot\cdot\cdot, A_{y}$, and $y-1$ independent sources $S_{1}$, $S_{2}$, $\cdot\cdot\cdot$, $S_{y-1}$. And each independent source $S_{j}$ emits $(1+1)$-mode Gaussian state $\rho^{A_{j}A_{y}}$ for each $j=1,2,\cdot\cdot\cdot,y-1$.}
\label{fig4}
\end{figure}

\section{Witnessing nonlocality in  star networks of CV systems }\label{sec:5}

In this section, we consider the star network $\mathcal{S}(y)$ depicted in Figure \ref{fig4}, which consists of $y$
parties $A_{1}, A_{2}, \cdots, A_{y}$, and $y-1$ independent sources $S_{1}$, $S_{2}$, $\cdots$, $S_{y-1}$. It is clear that in star network $\mathcal{S}(y)$, the number of maximal independent parties  $k_{\rm max}=y-1$, and the corresponding index set of maximal set of  independent parties $\mathcal{K}=\mathcal K_{\max}=\{1,2,\cdots,y-1\}$. We assume that each $S_{j}$ emits  the $(1+1)$-mode Gaussian state for $j=1,2,\cdots,y-1$, and each $A_{i}$ performs the measurements of the form in Eq.(\ref{eq3.14}) for $i=1,2,\cdots,y$. Then, by applying Theorem 2.1, we obtain the following nonlinear Bell-type inequality for the star network $\mathcal{S}(y)$ in Figure \ref{fig4}.

{\bf Theorem 5.1.} {\it Let ${\mathcal S}(y)$ be a star network as in Figure \ref{fig4}. Assume that each independent source $S_{j}$ emits $(1+1)$-mode Gaussian state $\rho^{A_{j}A_{y}}$, and denote its generalized  quasiprobability function as $Q_{\rho^{A_{j}A_{y}}}(\alpha_{x_j}^{\lambda_{j}},\alpha_{x_y}^{\lambda_{j}};s)$ with $x_j,x_y\in\{0,1\}$, $j=1,2,\cdots,y-1$. If ${\mathcal S}(y)$ is network local, then the following nonlinear Bell-type inequality holds:}
\begin{widetext}
\begin{equation}\label{eq5.1}\mathcal{B}_{\rho}^{st}(\alpha_{x_1=0}^{\lambda_{1}},\alpha_{x_y=0}^{\lambda_{1}},\cdots,\alpha_{x_{y-1}=0}^{\lambda_{y-1}}, \alpha_{x_y=0}^{\lambda_{y-1}}, \alpha_{x_1=1}^{\lambda_{1}},\alpha_{x_y=1}^{\lambda_{1}},\cdots,\alpha_{x_{y-1}=1}^{\lambda_{y-1}}, \alpha_{x_y=1}^{\lambda_{y-1}};s)=|\mathcal I_{s}|^{\frac{1}{y-1}}+|\mathcal J_{s}|^{\frac{1}{y-1}}\leq1,\tag{5.1}
\end{equation}
{\it where  $\rho=\rho^{A_{1}A_{y}}\otimes \rho^{A_{2}A_{y}}\otimes\cdots\otimes\rho^{A_{y-1}A_{y}}$, for $-1<s\leq 0$,
\begin{equation}
\begin{aligned}
\mathcal I_{s}=&\frac{1}{2^{y-1}}\times\mathop{\prod}\limits_{j\in\mathcal{K}}\{\frac{\pi^{2}(1-s)^{4}}{4}C_{\rho^{A_{j}A_{y}}}^{+}(\alpha_{x_{j}=0}^{\lambda_{j}},\alpha_{x_{y}=0}^{\lambda_{j}},\alpha_{x_{j}=1}^{\lambda_{j}},\alpha_{x_{y}=0}^{\lambda_{j}};s)
\nonumber\\&+\frac{\pi s(1-s)^{2}}{2}D_{\rho^{A_{j}A_{y}}}^{+}(\alpha_{x_{j}=0}^{\lambda_{j}},\alpha_{x_{y}=0}^{\lambda_{j}},\alpha_{x_{j}=1}^{\lambda_{j}},\alpha_{x_{y}=0}^{\lambda_{j}};s)+2s^{2}\}\nonumber
\end{aligned}
\end{equation}
and
\begin{equation}
\begin{aligned}
\mathcal J_{s}=&\frac{1}{2^{y-1}}\times\mathop{\prod}\limits_{j\in\mathcal{K}}\{\frac{\pi^{2}(1-s)^{4}}{4}C_{\rho^{A_{j}A_{y}}}^{-}(\alpha_{x_{j}=0}^{\lambda_{j}},\alpha_{x_{y}=1}^{\lambda_{j}},\alpha_{x_{j}=1}^{\lambda_{j}},\alpha_{x_{y}=1}^{\lambda_{j}};s)
\nonumber\\&+\frac{\pi s(1-s)^{2}}{2}D_{\rho^{A_{j}A_{y}}}^{-}(\alpha_{x_{j}=0}^{\lambda_{j}},\alpha_{x_{y}=1}^{\lambda_{j}},\alpha_{x_{j}=1}^{\lambda_{j}},\alpha_{x_{y}=1}^{\lambda_{j}};s)\};\nonumber
\end{aligned}
\end{equation}
while for $s\leq-1$,
\begin{equation}
\begin{aligned}
\mathcal I_{s}=&\frac{1}{2^{y-1}}\times\mathop{\prod}\limits_{j\in\mathcal{K}}\{\pi^{2}(1-s)^{2}C_{\rho^{A_{j}A_{y}}}^{+}(\alpha_{x_{j}=0}^{\lambda_{j}},\alpha_{x_{y}=0}^{\lambda_{j}},\alpha_{x_{j}=1}^{\lambda_{j}},\alpha_{x_{y}=0}^{\lambda_{j}};s)
\nonumber\\&-\pi(1-s)D_{\rho^{A_{j}A_{y}}}^{+}(\alpha_{x_{j}=0}^{\lambda_{j}},\alpha_{x_{y}=0}^{\lambda_{j}},\alpha_{x_{j}=1}^{\lambda_{j}},\alpha_{x_{y}=0}^{\lambda_{j}};s)+2\}\nonumber
\end{aligned}
\end{equation}
and
\begin{equation}
\begin{aligned}
\mathcal \mathcal J_{s}=&\frac{1}{2^{y-1}}\times\mathop{\prod}\limits_{j\in\mathcal{K}}\{\pi^{2}(1-s)^{2}C_{\rho^{A_{j}A_{y}}}^{-}(\alpha_{x_{j}=0}^{\lambda_{j}},\alpha_{x_{y}=1}^{\lambda_{j}},\alpha_{x_{j}=1}^{\lambda_{j}},\alpha_{x_{y}=1}^{\lambda_{j}};s)
\nonumber\\&-\pi(1-s)D_{\rho^{A_{j}A_{y}}}^{-}(\alpha_{x_{j}=0}^{\lambda_{j}},\alpha_{x_{y}=1}^{\lambda_{j}},\alpha_{x_{j}=1}^{\lambda_{j}},\alpha_{x_{y}=1}^{\lambda_{j}};s)\}.\nonumber
\end{aligned}
\end{equation}
Here, $C^{+}$, $C^{-}$, $D^{+}$, $D^{-}$ are defined as in Eqs.(\ref{eq3.16})-(\ref{eq3.19}).
}
\end{widetext}

A proof of Theorem 5.1 can be found in Appendix C.

 According to Theorem 5.1, if we select appropriate parameters $s$, $\alpha_{x_{j}=0}^{\lambda_{j}}$, $\alpha_{x_{j}=1}^{\lambda_{j}}$, $\alpha_{x_{y}=0}^{\lambda_{j}}$ and $\alpha_{x_{y}=1}^{\lambda_{j}}$ for $j=1,2,\cdots,y-1$  such that the inequality (\ref{eq5.1}) is violated, then it demonstrates the network nonlocality in the star network ${\mathcal S}(y)$.   Similar to Eq.(\ref{eq3.171}), we use the supremum strategy described in Section \ref{sec:3} that
\begin{equation}\tag{5.2}\label{eq5.2}
B^{st}(s,\rho)=\mathop{\sup}\limits_{\bm \alpha }\mathcal{B}_{\rho}^{st}({\bm \alpha};s) \leq 1
\end{equation}
if the star network ${\mathcal S}(y)$ is network local, where
\begin{widetext}
$$ \bm \alpha=(\alpha_{x_1=0}^{\lambda_{1}},\alpha_{x_y=0}^{\lambda_{1}},\cdots,\alpha_{x_{y-1}=0}^{\lambda_{y-1}}, \alpha_{x_y=0}^{\lambda_{y-1}}, \alpha_{x_1=1}^{\lambda_{1}},\alpha_{x_y=1}^{\lambda_{1}},\cdots,\alpha_{x_{y-1}=1}^{\lambda_{y-1}}, \alpha_{x_y=1}^{\lambda_{y-1}}).$$
\end{widetext}
\if false to study the nonlocality of star networks ${\mathcal S}(y)$ consisting of $(1+1)$-mode EPR states and symmetric STSs, respectively, in Examples 5.2 and 5.3.\fi

\begin{figure*}[]
\center
\subfigure [$-1<s\leq0$]
{\includegraphics[width=7cm,height=5.5cm]{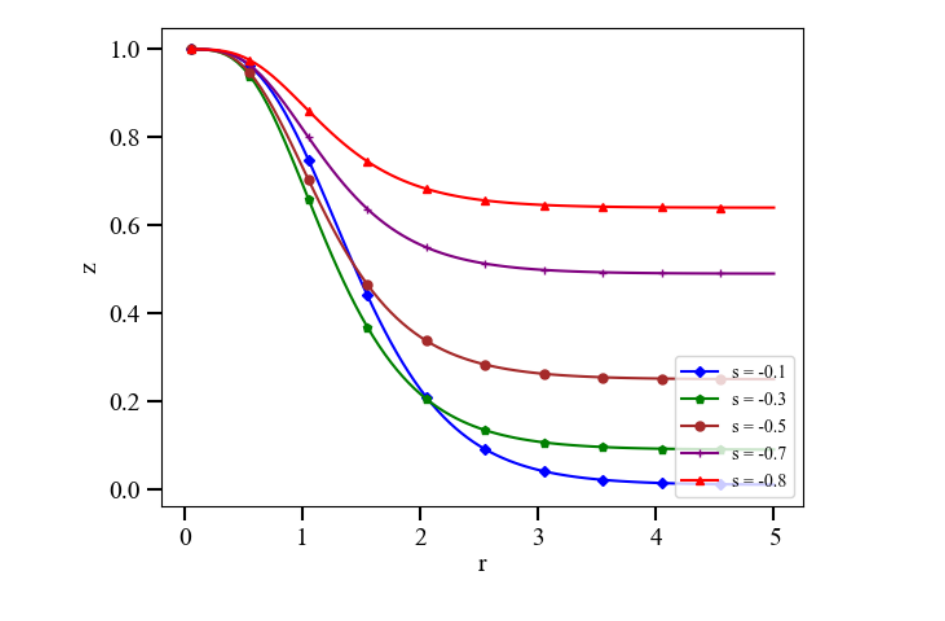}}
\subfigure [$s\leq-1$]
{\includegraphics[width=7cm,height=5.5cm]{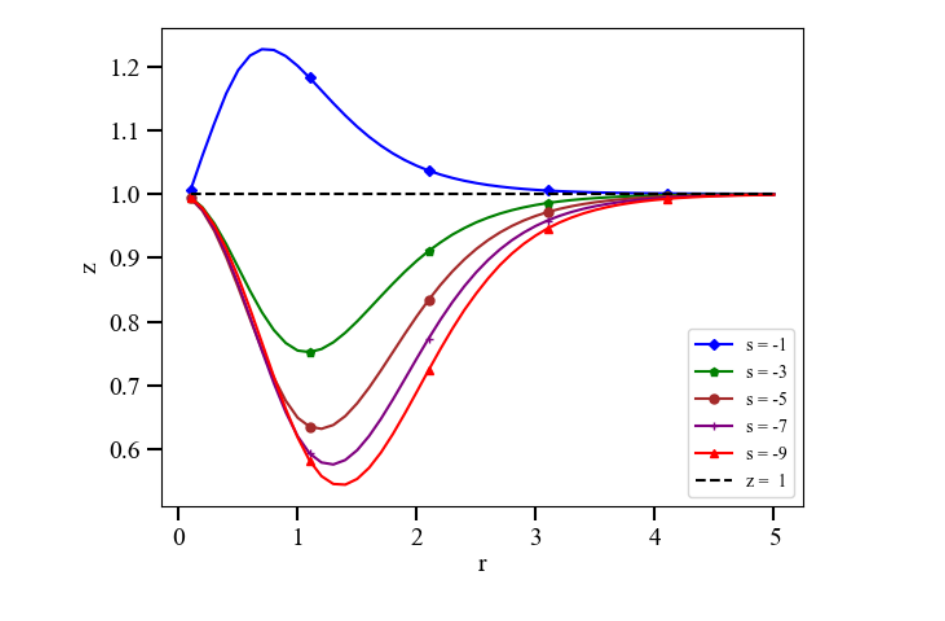}}
\caption{\quad\small Value of $z=B^{st}(s,r)$ in $\mathcal{S}(6)$ for $\rho^{A_{j}A_{6}}=\rho^{r}(j=1,2,3,4,5)$, as a function of the parameter $r$ for fixed $s$. The figure (a) shows the case of $-1<s\leq0$, while the figure (b) shows the case of $s\leq-1$.}
\centering
\label{fig5}
\end{figure*}

\begin{figure*}[]
\center
\subfigure [$-1<s\leq0$]
{\includegraphics[width=7cm,height=5.5cm]{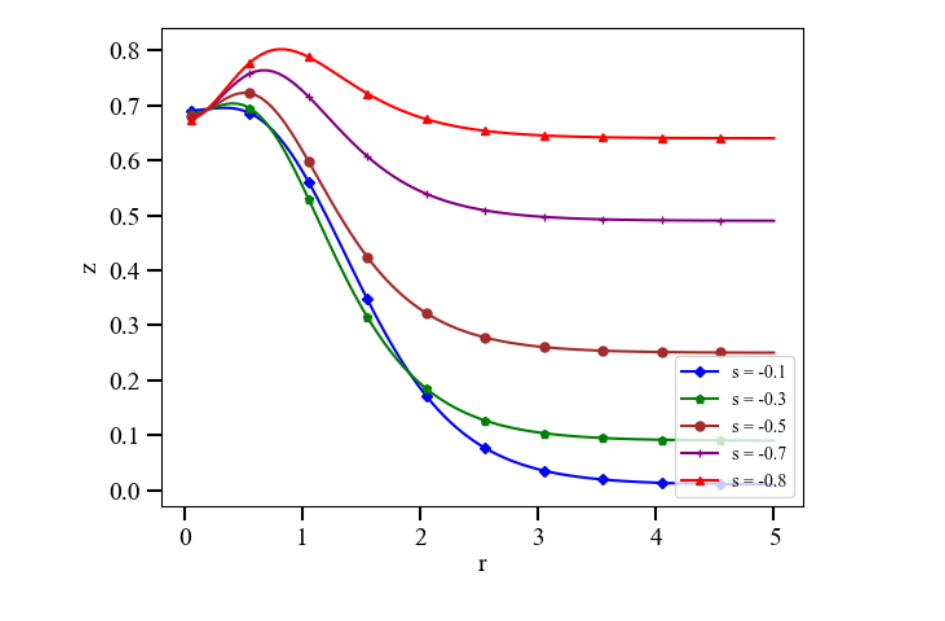}}
\subfigure [$s\leq-1$]
{\includegraphics[width=7cm,height=5.5cm]{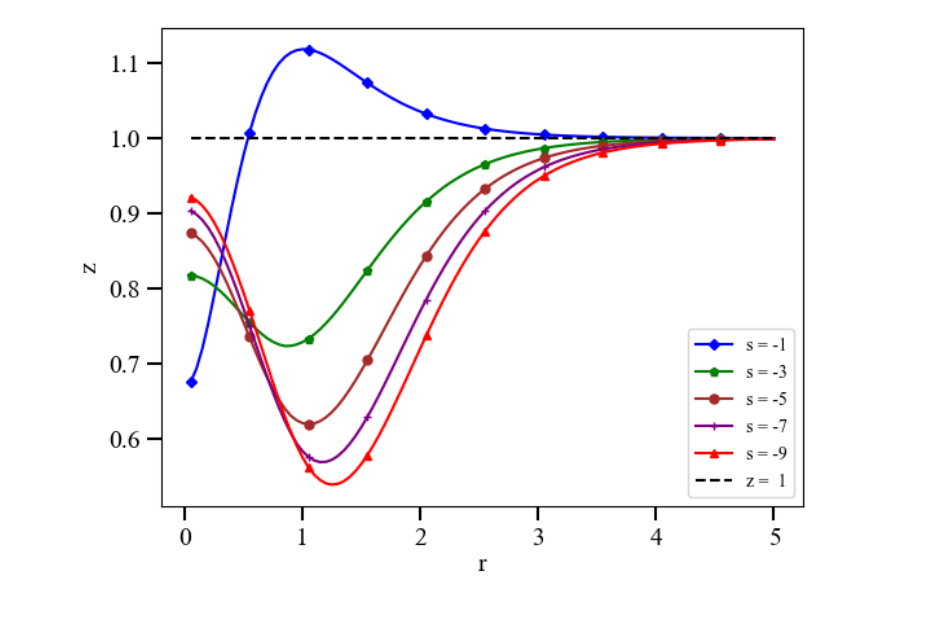}}
\caption{\small\quad Values of $z=B^{st}(s,r)$ in $\mathcal{S}(6)$ for $\rho^{A_{j}A_{6}}=\rho(1.2,1.2,r)(j=1,2,3,4,5)$, as a function of the parameter $r$ for fixed $s$. The figure (a) shows the case of $-1<s\leq0$, while the figure (b) shows the case of $s\leq-1$.}
\centering
\label{fig6}
\end{figure*}

{\bf Example 5.2.} Consider the scenario of the star network ${\mathcal S}(6)$ with the same pure Gaussian source states.  Without loss of generality, we assume that $\rho^{A_{j}A_{6}}=\rho^{r}$ is an EPR state for each $j=1,2,3,4,5$. Then $B^{st}(s,\rho)$ in the inequality (\ref{eq5.2}) is  a function $B^{st}(s,r)$ of $s$ and $r$. From Theorem 5.1,   $B^{st}(s,r)>1$  demonstrates the network nonlocality in ${\mathcal S}(6)$.   Figure \ref{fig5} illustrates the value of $B^{st}(s,r)$  when regarding $B^{st}(s,r)$  as a function of $r$ for some fixed $s$.

Figure \ref{fig5} reveals that the inequality $B^{st}(s,r)>1$ holds true only when $s=-1$. In this case, as $r$ increases, the value of $B^{st}(-1,r)$ initially rises and then declines. The largest violation   $B^{st}(-1,r)\approx1.227$ occurs at $r=0.75$. Furthermore, for any $r>0$, it holds that $B^{st}(-1,r)>1$, indicating that if the five quantum source states are identical entangled EPR states, they will exhibit network nonlocality in ${\mathcal S}(6)$. However, $B^{st}(-1,r)\to 1^+$ quickly  as $r\to\infty$. Since every Gaussian pure state can be transformed into an EPR state through a local Gaussian unitary transformation, we conclude that {\it if all five sources in star network ${\mathcal S}(6)$ emit any identical entangled Gaussian pure states, it will generate network nonlocality}.

{\bf Example 5.3.} Consider the scenario of the star network ${\mathcal S}(6)$ with the same mixed Gaussian states as the source states. Assume $\rho^{A_{j}A_{6}}=\rho(1.2,1.2,r)$  is a STS as defined in Eq.(\ref{eq3.10}) for each $j=1,2,3,4,5$. In this context,   $B^{st}(s,\rho)$ in the inequality (\ref{eq5.2}) is expressed as a function $B^{st}(s,r)$ that depends on both $s$ and $r$. According to Theorem 5.1, when  $B^{st}(s,r)>1$, it witnesses the network nonlocality in ${\mathcal S}(6)$. For certain specific, fixed values of $s$, we can obtain Figure \ref{fig6} by plotting $B^{st}(s,r)$ as a function of the parameter $r$.

From Figure \ref{fig6}, {\it it is evident that the inequality $B^{st}(s,r)>1$ holds exclusively when $s=-1$ and $r\geq0.55$, indicating the presence of the nonlocality in ${\mathcal S}(6)$ when $r$ meets or exceeds this threshold}. Additionally, as $r$ increases, the value of $B^{st}(-1,r)$ exhibits a trend of initial increase followed by a decrease, and ultimately approaches $1$. The maximum value $B^{st}(-1,r)\approx1.118$ is achieved at $r=1$.

\section{ Witnessing nonlocality in  tree-shaped  networks  of CV systems.}\label{sec:6}

\begin{figure*}[]
\centering
\includegraphics[width=9cm,height=5cm]{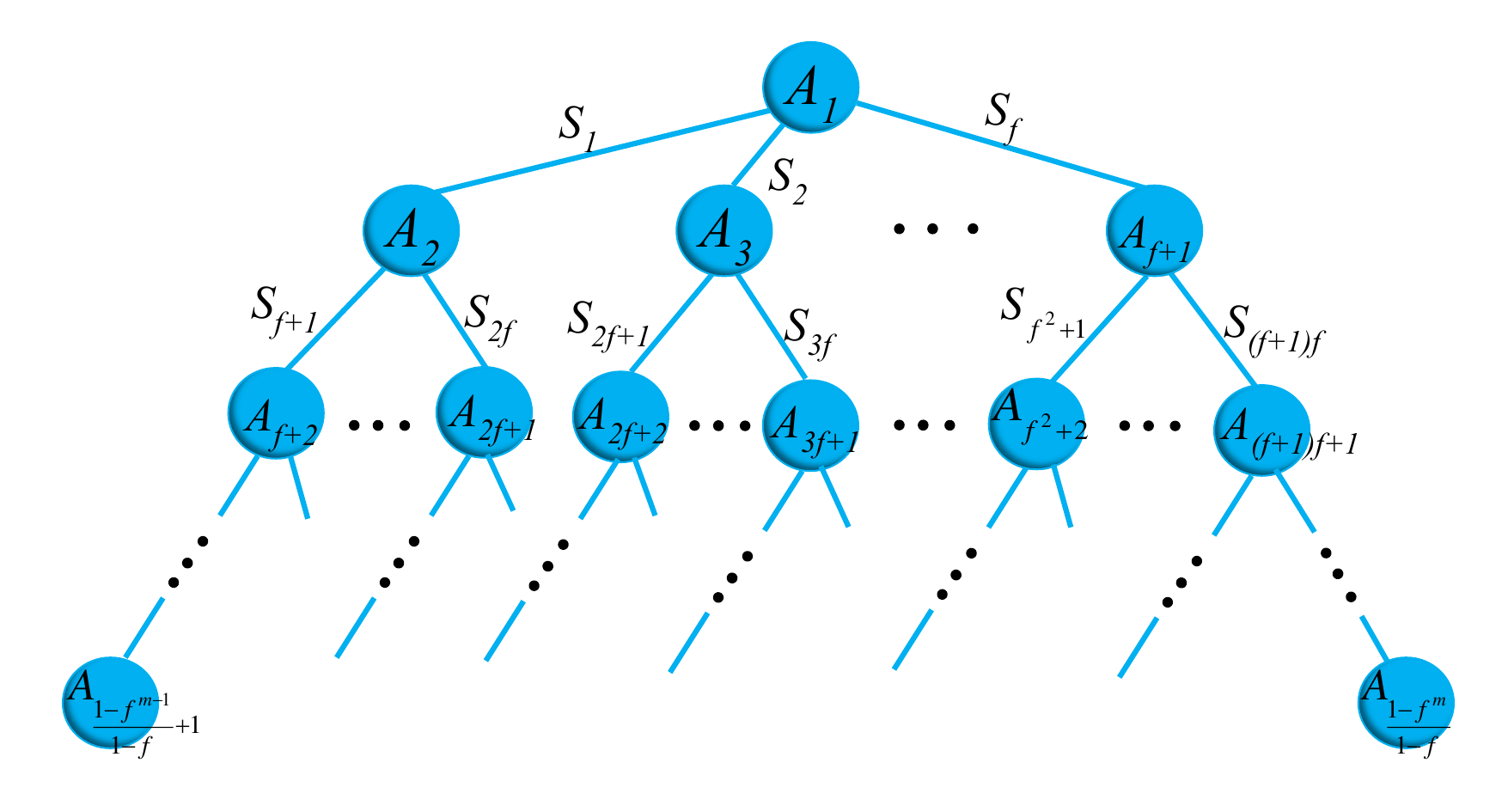}
\caption{\small\quad   A tree-shaped network $\mathcal{T}(m,f)$ consists of $\frac{1-f^{m}}{1-f}$ parties consisting of $ A_{1}, A_{2}, \cdot\cdot\cdot, A_{\frac{1-f^{m}}{1-f}}$ and $\frac{f-f^{m}}{1-f}$ independent sources $S_{1}$, $S_{2}$, $\cdot\cdot\cdot$, $S_{\frac{f-f^{m}}{1-f}}$. And each independent source $S_{i}$ emits $(1+1)$-mode Gaussian state $\rho^{A_{\lceil\frac{i}{f}\rceil}A_{i+1}}$ for each $i=1,2,\cdot\cdot\cdot,\frac{f-f^{m}}{1-f}$.}
\label{fig7}
\end{figure*}

In this section, we discuss the network nonlocality in  the deterministic $m$-layer $f$-forked tree-shaped network $\mathcal{T}(m,f)$, as depicted in Figure \ref{fig7}, which is a network consisting of $\frac{1-f^{m}}{1-f}$ parties $A_{1}, A_{2}, \cdots, A_{\frac{1-f^{m}}{1-f}}$ and $\frac{f-f^{m}}{1-f}$ sources $S_{1}, S_{2}, \cdots, S_{\frac{f-f^{m}}{1-f}}$. Employing the same methodology as in Sections \ref{sec:4} and \ref{sec:5}, we establish corresponding Bell-type inequality for tree-shaped network $\mathcal{T}(m,f)$ of CV systems.

In tree-shaped network $\mathcal{T}(m,f)$, it is clear that (1) if $m$ is odd, then $k_{\rm max}=\frac{f^{m+1}-1}{f^{2}-1}$ with the index set of  maximal set of independent parties ${\mathcal K}={\mathcal K}_{\rm max}=\{1,\frac{1-f^{t-1}}{1-f}+1, \frac{1-f^{t-1}}{1-f}+2, \cdots, \frac{1-f^{t}}{1-f}\}_{t=3,5,\cdots,m}$; (2) if $m$ is even, then $k_{\rm max}=\frac{f-f^{m+1}}{1-f^{2}}$ with the index set of  maximal set of independent parties ${\mathcal K}=\mathcal K_{\max}=\{\frac{1-f^{t-1}}{1-f}+1, \frac{1-f^{t-1}}{1-f}+2, \cdots, \frac{1-f^{t}}{1-f}\}_{t=2,4,\cdots,m}$. Denote by $\overline{\mathcal K}=\{1,2,\cdots, \frac{1-f^{m}}{1-f}\}\setminus {\mathcal K}$ and ${\mathcal K}'$   the set of indexes from ${\mathcal K}$ with the last layer of indexes excluded. For example, when $m$ is odd, ${\mathcal K}'=\{1,\frac{1-f^{t-1}}{1-f}+1, \frac{1-f^{t-1}}{1-f}+2, \cdots, \frac{1-f^{t}}{1-f}\}_{t=3,5,\cdots,m-2}$. Also, $C^{+}$, $C^{-}$, $D^{+}$, $D^{-}$ are defined as in Eqs.(\ref{eq3.16})-(\ref{eq3.19}). Then, by applying Theorem 2.1, we have

{\bf Theorem 6.1.}  {\it Let $\mathcal{T}(m,f)$ be a tree-shaped network as in Figure \ref{fig7}. Assume that each independent source $S_{i}$  emits $(1+1)$-mode Gaussian state $\rho^{A_{\lceil\frac{i}{f}\rceil}A_{i+1}}$ and denote its generalized  quasiprobability function by $Q_{\rho^{A_{\lceil\frac{i}{f}\rceil}A_{i+1}}}(\alpha_{x_{\lceil\frac{i}{f}\rceil}}^{\lambda_{i}},\alpha_{x_{i+1}}^{\lambda_{i}};s)$, where $ x_{\lceil\frac{i}{f}\rceil}, x_{i+1}\in\{0,1\}$, $i=1,2,\cdot\cdot\cdot,\frac{f-f^{m}}{1-f}$ and  $\lceil\cdot\rceil$ represents the ceiling function. If $\mathcal{T}(m,f)$ is network local, then the following nonlinear Bell-type inequality holds:
\begin{equation}\label{eq6.1}\mathcal{B}_{\rho}^{tr}(\bm{\alpha}_0,\bm{\alpha}_1;s)=|{\mathcal I}_{s}|^{\frac{1}{k_{\rm max}}}+|{\mathcal J}_{s}|^{\frac{1}{k_{\rm max}}}\leq1,\tag{6.1}
\end{equation}
 where $\rho=\otimes_{i=1}^{\frac{f-f^{m}}{1-f}}(\rho^{A_{\lceil\frac{i}{f}\rceil}A_{i+1}})$,
\begin{widetext}
 $$ \bm{\alpha}_t=(\alpha_{x_{1}=t}^{\lambda_{1}}, \alpha_{x_{2}=t}^{\lambda_{1}}, \cdots,  \alpha_{x_{\lceil\frac{i}{f}\rceil}=t}^{\lambda_{i}}, \alpha_{x_{i+1}=t}^{\lambda_{i}}, \cdots, \alpha_{x_{\frac{f-f^{m-1}}{1-f}}=t}^{\lambda_{\frac{f-f^{m}}{1-f}}}, \alpha_{x_{\frac{f-f^{m}}{1-f}+1}=t}^{\lambda_{\frac{f-f^{m}}{1-f}}})
$$
 for $t=0,1$, $k_{\rm max}=\frac{f^{m+1}-1}{f^{2}-1}$ when $m$ is odd, $k_{\rm max}=\frac{f-f^{m+1}}{1-f^{2}}$ when $m$ is even; for $-1<s\leq0$,
\begin{equation}
\begin{aligned}
\mathcal I_{s}=&\frac{1}{2^{k_{\rm max}}}\times\mathop{\prod}\limits_{j=(i-1)f+2}^{if+1}\mathop{\prod}\limits_{i\in\mathcal{K}^{'}}\{\frac{\pi^{2}(1-s)^{4}}{4}C_{\rho^{A_{i}A_{j}}}^{+}(\alpha_{x_{i}=0}^{\lambda_{j-1}},\alpha_{x_{j}=0}^{\lambda_{j-1}},\alpha_{x_{i}=1}^{\lambda_{j-1}},\alpha_{x_{j}=0}^{\lambda_{j-1}};s)
+\frac{\pi s(1-s)^{2}}{2}\nonumber\\&D_{\rho^{A_{i}A_{j}}}^{+}(\alpha_{x_{i}=0}^{\lambda_{j-1}},\alpha_{x_{j}=0}^{\lambda_{j-1}},\alpha_{x_{i}=1}^{\lambda_{j-1}},\alpha_{x_{j}=0}^{\lambda_{j-1}};s)+2s^{2}\}\times\mathop{\prod}\limits_{q=(p-1)f+2}^{pf+1}\mathop{\prod}\limits_{p\in\overline{\mathcal K}}\{\frac{\pi^{2}(1-s)^{4}}{4}C_{\rho^{A_{p}A_{q}}}^{+}(\alpha_{x_{p}=0}^{\lambda_{q-1}},\nonumber\\&\alpha_{x_{q}=0}^{\lambda_{q-1}},\alpha_{x_{p}=0}^{\lambda_{q-1}},\alpha_{x_{q}=1}^{\lambda_{q-1}};s)
+\frac{\pi s(1-s)^{2}}{2}D_{\rho^{A_{p}A_{q}}}^{+}(\alpha_{x_{p}=0}^{\lambda_{q-1}},\alpha_{x_{q}=0}^{\lambda_{q-1}},\alpha_{x_{p}=0}^{\lambda_{q-1}},\alpha_{x_{q}=1}^{\lambda_{q-1}};s)+2s^{2}\}\nonumber
\end{aligned}
\end{equation}
and
\begin{equation}
\begin{aligned}
\mathcal J_{s}=&\frac{1}{2^{k_{\rm max}}}\times\mathop{\prod}\limits_{j=(i-1)f+2}^{if+1}\mathop{\prod}\limits_{i\in\mathcal{K}^{'}}\{\frac{\pi^{2}(1-s)^{4}}{4}C_{\rho^{A_{i}A_{j}}}^{-}(\alpha_{x_{i}=0}^{\lambda_{j-1}},\alpha_{x_{j}=1}^{\lambda_{j-1}},\alpha_{x_{i}=1}^{\lambda_{j-1}},\alpha_{x_{j}=1}^{\lambda_{j-1}};s)
+\frac{\pi s(1-s)^{2}}{2}\nonumber\\&D_{\rho^{A_{i}A_{j}}}^{-}(\alpha_{x_{i}=0}^{\lambda_{j-1}},\alpha_{x_{j}=1}^{\lambda_{j-1}},\alpha_{x_{i}=1}^{\lambda_{j-1}},\alpha_{x_{j}=1}^{\lambda_{j-1}};s)\}\times\mathop{\prod}\limits_{q=(p-1)f+2}^{pf+1}\mathop{\prod}\limits_{p\in\overline{\mathcal K}}\{\frac{\pi^{2}(1-s)^{4}}{4}C_{\rho^{A_{p}A_{q}}}^{-}(\alpha_{x_{p}=1}^{\lambda_{q-1}},\nonumber\\&\alpha_{x_{q}=0}^{\lambda_{q-1}},\alpha_{x_{p}=1}^{\lambda_{q-1}},\alpha_{x_{q}=1}^{\lambda_{q-1}};s)
+\frac{\pi s(1-s)^{2}}{2}D_{\rho^{A_{p}A_{q}}}^{-}(\alpha_{x_{p}=1}^{\lambda_{q-1}},\alpha_{x_{q}=0}^{\lambda_{q-1}},\alpha_{x_{p}=1}^{\lambda_{q-1}},\alpha_{x_{q}=1}^{\lambda_{q-1}};s)\}\nonumber;
\end{aligned}
\end{equation}
while for $s\leq-1$,
\begin{equation}
\begin{aligned}
\mathcal I_{s}=&\frac{1}{2^{k_{\rm max}}}\times\mathop{\prod}\limits_{j=(i-1)f+2}^{if+1}\mathop{\prod}\limits_{i\in\mathcal{K}^{'}}\{\pi^{2}(1-s)^{2}C_{\rho^{A_{i}A_{j}}}^{+}(\alpha_{x_{i}=0}^{\lambda_{j-1}},\alpha_{x_{j}=0}^{\lambda_{j-1}},\alpha_{x_{i}=1}^{\lambda_{j-1}},\alpha_{x_{j}=0}^{\lambda_{j-1}};s)
-\pi(1-s)\nonumber\\&D_{\rho^{A_{i}A_{j}}}^{+}(\alpha_{x_{i}=0}^{\lambda_{j-1}},\alpha_{x_{j}=0}^{\lambda_{j-1}},\alpha_{x_{i}=1}^{\lambda_{j-1}},\alpha_{x_{j}=0}^{\lambda_{j-1}};s)+2\}
\times\mathop{\prod}\limits_{q=(p-1)f+2}^{pf+1}\mathop{\prod}\limits_{p\in\overline{\mathcal K}}\{\pi^{2}(1-s)^{2}C_{\rho^{A_{p}A_{q}}}^{+}(\alpha_{x_{p}=0}^{\lambda_{q-1}},\nonumber\\&\alpha_{x_{q}=0}^{\lambda_{q-1}},\alpha_{x_{p}=0}^{\lambda_{q-1}},\alpha_{x_{q}=1}^{\lambda_{q-1}};s)
-\pi(1-s)D_{\rho^{A_{p}A_{q}}}^{+}(\alpha_{x_{p}=0}^{\lambda_{q-1}},\alpha_{x_{q}=0}^{\lambda_{q-1}},\alpha_{x_{p}=0}^{\lambda_{q-1}},\alpha_{x_{q}=1}^{\lambda_{q-1}};s)+2\}\nonumber
\end{aligned}
\end{equation}
and
\begin{equation}
\begin{aligned}
\mathcal J_{s}=&\frac{1}{2^{k_{\rm max}}}\times\mathop{\prod}\limits_{j=(i-1)f+2}^{if+1}\mathop{\prod}\limits_{i\in\mathcal{K}^{'}}\{\pi^{2}(1-s)^{2}C_{\rho^{A_{i}A_{j}}}^{-}(\alpha_{x_{i}=0}^{\lambda_{j-1}},\alpha_{x_{j}=1}^{\lambda_{j-1}},\alpha_{x_{i}=1}^{\lambda_{j-1}},\alpha_{x_{j}=1}^{\lambda_{j-1}};s)
-\pi (1-s)\nonumber\\&D_{\rho^{A_{i}A_{j}}}^{-}(\alpha_{x_{i}=0}^{\lambda_{j-1}},\alpha_{x_{j}=1}^{\lambda_{j-1}},\alpha_{x_{i}=1}^{\lambda_{j-1}},\alpha_{x_{j}=1}^{\lambda_{j-1}};s)\}
\times\mathop{\prod}\limits_{q=(p-1)f+2}^{pf+1}\mathop{\prod}\limits_{p\in\overline{\mathcal K}}\{\pi^{2}(1-s)^{2}C_{\rho^{A_{p}A_{q}}}^{-}(\alpha_{x_{p}=1}^{\lambda_{q-1}},\nonumber\\&\alpha_{x_{q}=0}^{\lambda_{q-1}},\alpha_{x_{p}=1}^{\lambda_{q-1}},\alpha_{x_{q}=1}^{\lambda_{q-1}};s)
-\pi(1-s)D_{\rho^{A_{p}A_{q}}}^{-}(\alpha_{x_{p}=1}^{\lambda_{q-1}},\alpha_{x_{q}=0}^{\lambda_{q-1}},\alpha_{x_{p}=1}^{\lambda_{q-1}},\alpha_{x_{q}=1}^{\lambda_{q-1}};s)\}\nonumber.
\end{aligned}
\end{equation}

\end{widetext}
}
A proof of Theorem 6.1 is provided in Appendix D.

By the above theorem, a violation of inequality (\ref{eq6.1}) witnesses the nonlocality in the tree-shaped network $\mathcal{T}(m,f)$ of CV system.

\if false Clearly, if   $\mathcal{T}(m,f)$ is network local, then
$$\mathcal{B}_{s}^{tr}=\sup_{\alpha_{x_1}^{\lambda_{1}},\alpha_{x_2}^{\lambda_{2}},\ldots, \alpha_{x_y}^{\lambda_y}}
\mathcal{B}_{s}^{tr}(\alpha_{x_1}^{\lambda_{1}},\alpha_{x_2}^{\lambda_{2}},\ldots, \alpha_{x_y}^{\lambda_y})\leq 1. \eqno(6.2)
$$
\fi

The following two examples illustrate how to use Theorem 6.1  and  the supremum strategy to specifically analyze the network nonlocality  in $\mathcal{T}(m,f)$.

\begin{figure*}[]
\center
\subfigure [$-1<s\leq0$]
{\includegraphics[width=7cm,height=5.5cm]{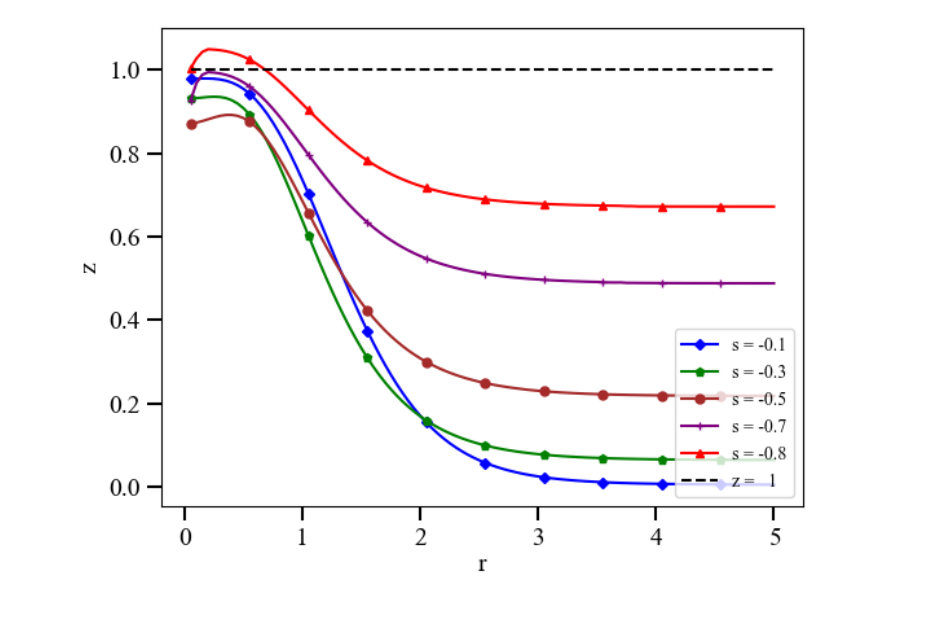}}
\subfigure [$s\leq-1$]
{\includegraphics[width=7cm,height=5.5cm]{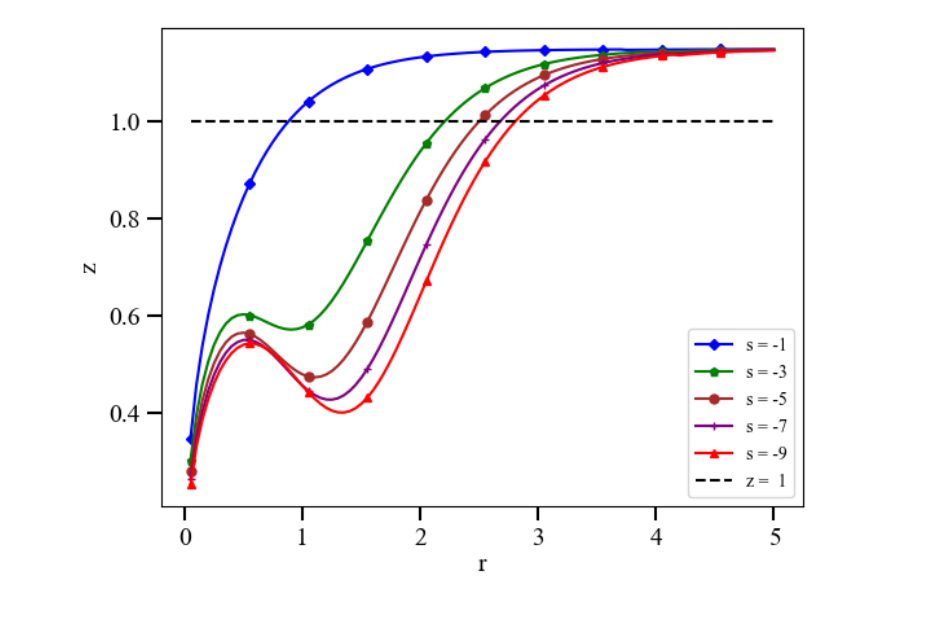}}
\caption{\small\quad Value of $z=B^{tr}(s,r)$ in $\mathcal{T}(3,2)$ for $(1+1)$-mode Gaussian state $\rho^{A_{1}A_{2}}=\rho^{A_{1}A_{3}}=\rho^{A_{2}A_{4}}=\rho^{A_{2}A_{5}}=\rho^{A_{3}A_{6}}=\rho^{A_{3}A_{7}}=\rho^{r}$, as a function of the parameter $r$ for fixed $s$. The figure (a) shows the case of $-1<s\leq0$, while the figure (b) shows the case of $s\leq-1$.}
\centering
\label{fig8}
\end{figure*}

\begin{figure*}[]
\center
\subfigure [$-1<s\leq0$]
{\includegraphics[width=7cm,height=5.5cm]{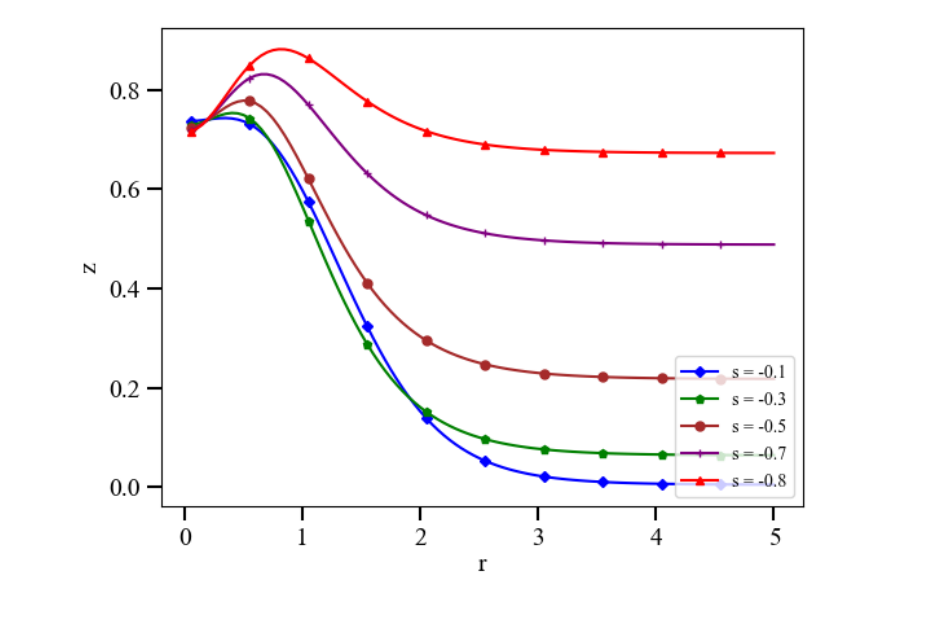}}
\subfigure [$s\leq-1$]
{\includegraphics[width=7cm,height=5.5cm]{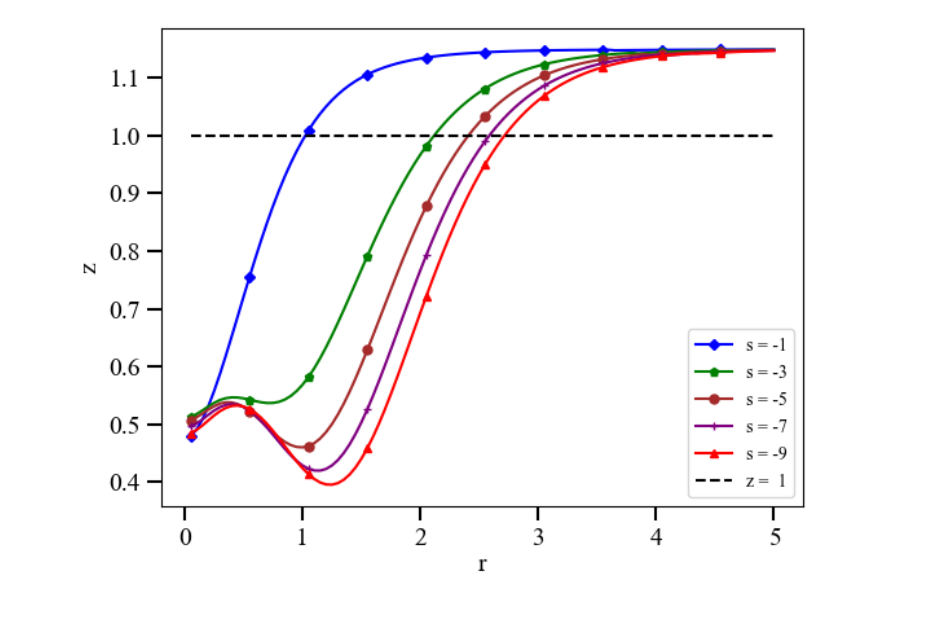}}
\caption{\small\quad Value of $z=B^{tr}(s,r)$ in $\mathcal{T}(3,2)$ for $(1+1)$-mode Gaussian state $\rho^{A_{1}A_{2}}=\rho^{A_{1}A_{3}}=\rho^{A_{2}A_{4}}=\rho^{A_{2}A_{5}}=\rho^{A_{3}A_{6}}=\rho^{A_{3}A_{7}}=\rho(1.2,1.2,r)$, as a function of the parameter $r$ for fixed $s$. The figure (a) shows the case of $-1<s\leq0$, while the figure (b) shows the case of $s\leq-1$.}
\centering
\label{fig9}
\end{figure*}

{\bf Example 6.2.} Consider the tree-shaped network $\mathcal{T}(3,2)$ with   sources states  that are  the same EPR  states. In this scenario,  $\rho^{A_{1}A_{2}}=\rho^{A_{1}A_{3}}=\rho^{A_{2}A_{4}}=\rho^{A_{2}A_{5}}=\rho^{A_{3}A_{6}}=\rho^{A_{3}A_{7}}=\rho^{r}$, an EPR state as defined in Eq.(\ref{eq3.7}). Let
$$\begin{array}{rl}
{\bm \alpha_{t}}=&(\alpha_{x_{1}=t}^{\lambda_{1}},\alpha_{x_{2}=t}^{\lambda_{1}},
\alpha_{x_{1}=t}^{\lambda_{2}},
\alpha_{x_{3}=t}^{\lambda_{2}},
\alpha_{x_{2}=t}^{\lambda_{3}},\alpha_{x_{4}=t}^{\lambda_{3}},\nonumber\\
&\alpha_{x_{2}=t}^{\lambda_{4}},\alpha_{x_{5}=t}^{\lambda_{4}},
\alpha_{x_{3}=t}^{\lambda_{5}},\alpha_{x_{6}=t}^{\lambda_{5}},
\alpha_{x_{3}=t}^{\lambda_{6}},\alpha_{x_{7}=t}^{\lambda_{6}})\in\mathbb{C}^{12},
\end{array}
$$
where $t\in\{0,1\}$. Then
$$ B^{tr}(s,r)=\sup_{\bm{\alpha}_{0},\bm{\alpha}_{1}}{\mathcal B}_{\rho}^{tr}(\bm{\alpha}_{0},\bm{\alpha}_{1};s)
$$
is a function of $s$ and $r$.  According to Theorem 6.1,  $B^{tr}(s,r)>1$ will demonstrate the network nonlocality in $\mathcal{T}(3,2)$.   Figure \ref{fig8} shows the image of  $B^{tr}(s,r)$ when considering $B^{tr}(s,r)$  as a function of $r$ for the cases $-1<s\leq0$ and $s\leq-1$, respectively.

From Figure \ref{fig8}(a), it is evident that the inequality $B^{tr}(-0.8,r)>1$  holds  when $0<r\leq0.75$. Additionally, in Figure \ref{fig8}(b), the inequality $B^{tr}(-1,r)>1$ is satisfied for $r\geq0.7$. Consequently, taking $s=-0.8$ and  $-1$ will detect the network nonlocality in $\mathcal{T}(3,2)$ for any $r>0$. \if false indicating that all six sources in the tree-shaped network $\mathcal{T}(3,2)$ emit any identical entangled EPR states, it will generate the network nonlocality.\fi Furthermore, since any Gaussian pure state can be transformed into an EPR state through a local Gaussian unitary transformation, we conclude that {\it if the six sources in tree-shaped network $\mathcal{T}(3,2)$ emit the same entangled Gaussian pure states, then the network is    network nonlocal.}

{\bf Example 6.3.} Now let us consider the tree-shaped network $\mathcal{T}(3,2)$ with all source states  being the same mixed Gaussian  states, say a symmetric STS $\rho(v,v,r)$ as defined in Eq.(\ref{eq3.10}).  For simplicity,  let $\rho^{A_{1}A_{2}}=\rho^{A_{1}A_{3}}=\rho^{A_{2}A_{4}}=\rho^{A_{2}A_{5}}=\rho^{A_{3}A_{6}}=\rho^{A_{3}A_{7}}
=\rho(1.2,1.2,r)$. Just like what we have done in Example 6.2, by taking the supremum of  $\mathcal{B}_{\rho}^{tr}(\bm{\alpha}_{0},\bm{\alpha}_{1};s)$ in the inequality (\ref{eq6.1}) over the parameters   $\bm{\alpha}_0$, $\bm{\alpha}_{1}$, one gets a  function $B^{tr}(s,r)$ of $s$ and $r$. Applying Theorem 6.1, if $B^{tr}(s,r)>1$, then it demonstrates network nonlocality. For some fixed values of $s$,   Figure \ref{fig9} is obtained  by treating $B^{tr}(s,r)$ as a function of the parameter $r$.

From Figures \ref{fig9}(a) and \ref{fig9}(b), we can see that the inequality $B^{tr}(s,r)>1$ holds for some $r$ only when $s\leq-1$, and as $s$ increases, more symmetric STSs $\rho(1.2,1.2,r)$ satisfy the inequality $B^{tr}(s,r)>1$. Therefore, using the generalized quasiprobability function with $s=-1$ allows for witnessing network nonlocality of a larger number of STSs in $\mathcal{T}(3,2)$. Furthermore, Figure \ref{fig9}(b) demonstrates that the inequality $B^{tr}(-1,r)>1$  whenever $r\geq1.025$. Then, we conclude that, {\it with every source state being symmetric STS $\rho(1.2,1.2,r)$,  if $r\geq1.025$, the tree-shaped network $\mathcal{T}(3,2)$  exhibits network nonlocality}.

\section{ Witnessing nonlocality in cyclic networks of CV systems}\label{sec:7}
\begin{figure*}[]
\centering
\includegraphics[width=4.5cm,height=4.5cm]{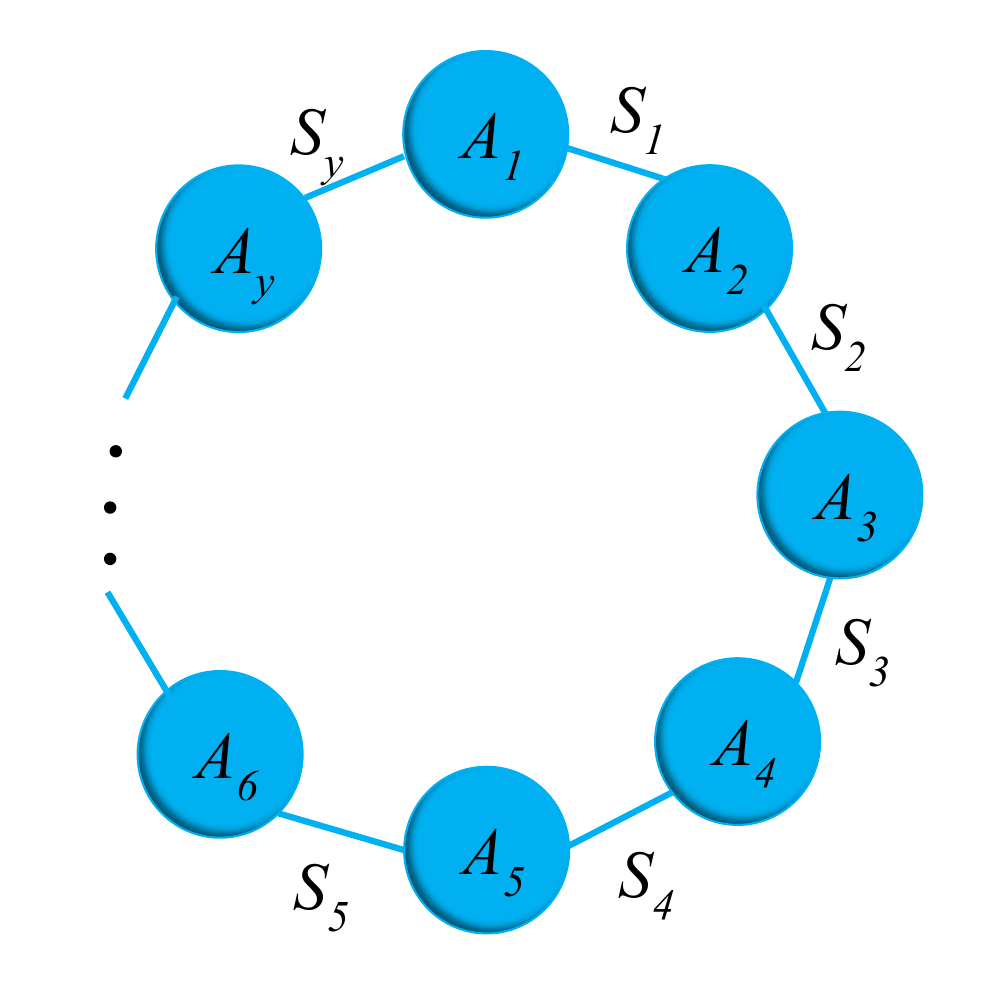}
\caption{\small\quad   A cyclic network $\mathcal{C}yc(y)$ consists of $y$ parties consisting of $ A_{1}, A_{2}, \cdot\cdot\cdot, A_{y}$ and $y$ independent sources $S_{1}$, $S_{2}$, $\cdot\cdot\cdot$, $S_{y}$. And each independent source $S_{j}$ emits $(1+1)$-mode Gaussian state $\rho^{A_{j}A_{j+1}}$ for each $j=1,2,\cdot\cdot\cdot,y-1$ and $S_{y}$ emits $(1+1)$-mode Gaussian states $\rho^{A_{y}A_{1}}$.}
\label{fig10}
\end{figure*}

 In a cyclic network $\mathcal{C}yc(y)$ (see Figure \ref{fig10}), there are $y$ parties
$A_{1}$, $A_{2}$, $\cdots$, $A_{y}$ with $A_{i}$ and $A_{i+1}$ share a source $S_{i}$ for $i=1,2, \cdots, y-1$, $A_{y}$ and $A_{1}$ share a source $S_{y}$. It is clear that (1) if $y$ is odd, then $k_{\rm max}=\frac{y-1}{2}$ with the index set of  maximal set of independent parties ${\mathcal K}=\mathcal K_{\max}=\{1,3, 5, \cdots, y-2\}$; (2) if $y$ is even, then $k_{\rm max}=\frac{y}{2}$ with the index set of  maximal set of independent parties ${\mathcal K}=\mathcal K_{\max}=\{1,3,5,\cdots, y-1\}$ or $\{2,4,6,\cdots, y\}$. Denote by $\overline{\mathcal K}=\{1,2,\cdots, y\}\setminus {\mathcal K}$. Using a measurement scheme based on generalized quasiprobability functions and applying Theorem 2.1, we can establish the following nonlinear Bell-type inequalities for cyclic network $\mathcal{C}yc(y)$ of CV systems. Here,  we have to address the scenarios where $y$ is odd and where $y$ is even as two distinct cases.

{\bf Theorem 7.1.} {\it Let $\mathcal{C}yc(y)$ be a cyclic network as in Figure \ref{fig10} with $y$  odd. Assume that each independent source $S_{j}$ emits $(1+1)$-mode Gaussian state $\rho^{A_{j}A_{j+1}}$ and denote its generalized  quasiprobability function as $Q_{\rho^{A_{j}A_{{j+1}}}}(\alpha_{x_j}^{\lambda_{j}},\alpha_{x_{j+1}}^{\lambda_{j}};s)$  for $j=1,2,\cdots,y-1$;  $S_{y}$ emits $(1+1)$-mode Gaussian state $\rho^{A_{y}A_{1}}$ and denote its generalized  quasiprobability function as $Q_{\rho^{A_{y}A_{1}}}(\alpha_{x_y}^{\lambda_{y}},\alpha_{x_1}^{\lambda_{y}};s)$, where $x_j, x_{y}\in\{0,1\}$, $j=1,2,\cdots,y-1$. Then $\mathcal{C}yc(y)$ is network local implies that the following nonlinear Bell-type inequality holds:
\begin{equation}\label{eq7.1}\mathcal{B}_{\rho}^{cy}(\bm{\alpha}_0,\bm{\alpha}_1;s )=|\mathcal I_{s}|^{\frac{2}{y-1}}+|\mathcal J_{s}|^{\frac{2}{y-1}}\leq1,\tag{7.1}
\end{equation}
 where
 $\rho=\rho^{A_{1}A_{2}}\otimes\rho^{A_{2}A_{3}}\otimes\cdots\otimes\rho^{A_{y-1}A_{y}}\otimes\rho^{A_{y}A_{1}}$, \begin{widetext}
$$ \bm{\alpha}_t=(\alpha_{x_1=t}^{\lambda_{1}},\alpha_{x_{2}=t}^{\lambda_{1}}, \cdots, \alpha_{x_{y-1}=t}^{\lambda_{y-1}},\alpha_{x_{y}=t}^{\lambda_{y-1}}, \alpha_{x_y=t}^{\lambda_{y}},\alpha_{x_1=t}^{\lambda_{y}}), \quad t\in\{0,1\},
$$
 and, for $-1<s\leq 0$,
\begin{equation}
\begin{aligned}
\mathcal I_{s}=&\frac{1}{2^{\frac{y-1}{2}}}\times\mathop{\prod}\limits_{i\in\mathcal{K}}\{\frac{\pi^{2}(1-s)^{4}}{4}C_{\rho^{A_{i}A_{i+1}}}^{+}(\alpha_{x_{i}=0}^{\lambda_{i}},\alpha_{x_{i+1}=0}^{\lambda_{i}},\alpha_{x_{i}=1}^{\lambda_{i}},\alpha_{x_{i+1}=0}^{\lambda_{i}};s)+\frac{\pi s(1-s)^{2}}{2}D_{\rho^{A_{i}A_{i+1}}}^{+}\nonumber\\&(\alpha_{x_{i}=0}^{\lambda_{i}},
\alpha_{x_{i+1}=0}^{\lambda_{i}},\alpha_{x_{i}=1}^{\lambda_{i}},\alpha_{x_{i+1}=0}^{\lambda_{i}};s)
+2s^{2}\}\times\mathop{\prod}\limits_{j\in\overline{\mathcal K}'}\{\frac{\pi^{2}(1-s)^{4}}{4}C_{\rho^{A_{j}A_{j+1}}}^{+}(\alpha_{x_{j}=0}^{\lambda_{j}},\alpha_{x_{j+1}=0}^{\lambda_{j}},\nonumber\\&\alpha_{x_{j}=0}^{\lambda_{j}},\alpha_{x_{j+1}=1}^{\lambda_{j}};s)
+\frac{\pi s(1-s)^{2}}{2}D_{\rho^{A_{j}A_{j+1}}}^{+}(\alpha_{x_{j}=0}^{\lambda_{j}},
\alpha_{x_{j+1}=0}^{\lambda_{j}},\alpha_{x_{j}=0}^{\lambda_{j}},\alpha_{x_{j+1}=1}^{\lambda_{j}};s)
+2s^{2}\}\nonumber\\&\times\{\frac{\pi^{2}(1-s)^{4}}{8}C_{\rho^{A_{y-1}A_{y}}}^{+}(\alpha_{x_{y-1}=0}^{\lambda_{y-1}},\alpha_{x_{y}=0}^{\lambda_{y-1}},\alpha_{x_{y-1}=0}^{\lambda_{y-1}},\alpha_{x_{y}=0}^{\lambda_{y-1}};s)
+\frac{\pi s(1-s)^{2}}{4}D_{\rho^{A_{y-1}A_{y}}}^{+}\nonumber\\&(\alpha_{x_{y-1}=0}^{\lambda_{y-1}},\alpha_{x_{y}=0}^{\lambda_{y-1}},\alpha_{x_{y-1}=0}^{\lambda_{y-1}},\alpha_{x_{y}=0}^{\lambda_{y-1}};s)
+s^{2}\}\nonumber
\end{aligned}
\end{equation}
and
\begin{equation}
\begin{aligned}
\mathcal J_{s}=&\frac{1}{2^{\frac{y-1}{2}}}\times\mathop{\prod}\limits_{i\in\mathcal{K}}\{\frac{\pi^{2}(1-s)^{4}}{4}C_{\rho^{A_{i}A_{i+1}}}^{-}(\alpha_{x_{i}=0}^{\lambda_{i}},\alpha_{x_{i+1}=1}^{\lambda_{i}},\alpha_{x_{i}=1}^{\lambda_{i}},\alpha_{x_{i+1}=1}^{\lambda_{i}};s)+\frac{\pi s(1-s)^{2}}{2}D_{\rho^{A_{i}A_{i+1}}}^{-}\nonumber\\&(\alpha_{x_{i}=0}^{\lambda_{i}},
\alpha_{x_{i+1}=1}^{\lambda_{i}},\alpha_{x_{i}=1}^{\lambda_{i}},\alpha_{x_{i+1}=1}^{\lambda_{i}};s)\}
\times\mathop{\prod}\limits_{j\in\overline{\mathcal K}'}\{\frac{\pi^{2}(1-s)^{4}}{4}C_{\rho^{A_{j}A_{j+1}}}^{-}(\alpha_{x_{j}=1}^{\lambda_{j}},\alpha_{x_{j+1}=0}^{\lambda_{j}},\nonumber\\&\alpha_{x_{j}=1}^{\lambda_{j}},\alpha_{x_{j+1}=1}^{\lambda_{j}};s)
+\frac{\pi s(1-s)^{2}}{2}D_{\rho^{A_{j}A_{j+1}}}^{-}(\alpha_{x_{j}=1}^{\lambda_{j}},\alpha_{x_{j+1}=0}^{\lambda_{j}},
\alpha_{x_{j}=1}^{\lambda_{j}},\alpha_{x_{j+1}=1}^{\lambda_{j}};s)\}\nonumber\\&\times\{\frac{\pi^{2}(1-s)^{4}}{8}C_{\rho^{A_{y-1}A_{y}}}^{+}(\alpha_{x_{y-1}=1}^{\lambda_{y-1}},\alpha_{x_{y}=1}^{\lambda_{y-1}},\alpha_{x_{y-1}=1}^{\lambda_{y-1}},\alpha_{x_{y}=1}^{\lambda_{y-1}};s)
+\frac{\pi s(1-s)^{2}}{4}D_{\rho^{A_{y-1}A_{y}}}^{+}\nonumber\\&(\alpha_{x_{y-1}=1}^{\lambda_{y-1}},\alpha_{x_{y}=1}^{\lambda_{y-1}},\alpha_{x_{y-1}=1}^{\lambda_{y-1}},\alpha_{x_{y}=1}^{\lambda_{y-1}};s)
+s^{2}\}\nonumber;
\end{aligned}
\end{equation}
while for $s\leq-1$,
\begin{equation}
\begin{aligned}
\mathcal I_{s}=&\frac{1}{2^{\frac{y-1}{2}}}\times\mathop{\prod}\limits_{i\in\mathcal{K}}\{\pi^{2}(1-s)^{2}C_{\rho^{A_{i}A_{i+1}}}^{+}(\alpha_{x_{i}=0}^{\lambda_{i}},\alpha_{x_{i+1}=0}^{\lambda_{i}},\alpha_{x_{i}=1}^{\lambda_{i}},\alpha_{x_{i+1}=0}^{\lambda_{i}};s)-\pi(1-s)
D_{\rho^{A_{i}A_{i+1}}}^{+}\nonumber\\&(\alpha_{x_{i}=0}^{\lambda_{i}},\alpha_{x_{i+1}=0}^{\lambda_{i}},
\alpha_{x_{i}=1}^{\lambda_{i}},\alpha_{x_{i+1}=0}^{\lambda_{i}};s)+2\}\times\mathop{\prod}\limits_{j\in\overline{\mathcal K}'}\{\pi^{2}(1-s)^{2}C_{\rho^{A_{j}A_{j+1}}}^{+}(\alpha_{x_{j}=0}^{\lambda_{j}},\alpha_{x_{j+1}=0}^{\lambda_{j}},\nonumber\\&\alpha_{x_{j}=0}^{\lambda_{j}},\alpha_{x_{j+1}=1}^{\lambda_{j}};s)
-\pi(1-s)D_{\rho^{A_{j}A_{j+1}}}^{+}(\alpha_{x_{j}=0}^{\lambda_{j}},\alpha_{x_{j+1}=0}^{\lambda_{j}},
\alpha_{x_{j}=0}^{\lambda_{j}},\alpha_{x_{j+1}=1}^{\lambda_{j}};s)+2\}\nonumber\\&\times\{\frac{\pi^{2}(1-s)^{2}}{2}C_{\rho^{A_{y-1}A_{y}}}^{+}(\alpha_{x_{y-1}=0}^{\lambda_{y-1}},\alpha_{x_{y}=0}^{\lambda_{y-1}},\alpha_{x_{y-1}=0}^{\lambda_{y-1}},\alpha_{x_{y}=0}^{\lambda_{y-1}};s)
-\frac{\pi(1-s)}{2}D_{\rho^{A_{y-1}A_{y}}}^{+}\nonumber\\&(\alpha_{x_{y-1}=0}^{\lambda_{y-1}},\alpha_{x_{y}=0}^{\lambda_{y-1}},\alpha_{x_{y-1}=0}^{\lambda_{y-1}},\alpha_{x_{y}=0}^{\lambda_{y-1}};s)
+1\}\nonumber
\end{aligned}
\end{equation}
and
\begin{equation}
\begin{aligned}
\mathcal J_{s}=&\frac{1}{2^{\frac{y-1}{2}}}\times\mathop{\prod}\limits_{i\in\mathcal{K}}\{\pi^{2}(1-s)^{2}C_{\rho^{A_{i}A_{i+1}}}^{-}(\alpha_{x_{i}=0}^{\lambda_{i}},\alpha_{x_{i+1}=1}^{\lambda_{i}},\alpha_{x_{i}=1}^{\lambda_{i}},\alpha_{x_{i+1}=1}^{\lambda_{i}};s) -\pi(1-s)D_{\rho^{A_{i}A_{i+1}}}^{-}\nonumber\\&(\alpha_{x_{i}=0}^{\lambda_{i}},
\alpha_{x_{i+1}=1}^{\lambda_{i}},\alpha_{x_{i}=1}^{\lambda_{i}},\alpha_{x_{i+1}=1}^{\lambda_{i}};s)\}
\times\mathop{\prod}\limits_{j\in\overline{\mathcal K}'}\{\pi^{2}(1-s)^{2}C_{\rho^{A_{j}A_{j+1}}}^{-}(\alpha_{x_{j}=1}^{\lambda_{j}},\alpha_{x_{j+1}=0}^{\lambda_{j}},\nonumber\\&\alpha_{x_{j}=1}^{\lambda_{j}},\alpha_{x_{j+1}=1}^{\lambda_{j}};s)
-\pi(1-s)D_{\rho^{A_{j}A_{j+1}}}^{-}(\alpha_{x_{j}=1}^{\lambda_{j}},\alpha_{x_{j+1}=0}^{\lambda_{j}},
\alpha_{x_{j}=1}^{\lambda_{j}},\alpha_{x_{j+1}=1}^{\lambda_{j}};s)\}\nonumber\\&\times\{\frac{\pi^{2}(1-s)^{2}}{2}C_{\rho^{A_{y-1}A_{y}}}^{+}(\alpha_{x_{y-1}=1}^{\lambda_{y-1}},\alpha_{x_{y}=1}^{\lambda_{y-1}},\alpha_{x_{y-1}=1}^{\lambda_{y-1}},\alpha_{x_{y}=1}^{\lambda_{y-1}};s)
-\frac{\pi(1-s)}{2}D_{\rho^{A_{y-1}A_{y}}}^{+}\nonumber\\&(\alpha_{x_{y-1}=1}^{\lambda_{y-1}},\alpha_{x_{y}=1}^{\lambda_{y-1}},\alpha_{x_{y-1}=1}^{\lambda_{y-1}},\alpha_{x_{y}=1}^{\lambda_{y-1}};s)
+1\}\nonumber.
\end{aligned}
\end{equation}
Here, $\mathcal K=\{1,3,5,\cdots, y-2\}$, $\overline{{\mathcal K}}'=\overline{{\mathcal K}}\backslash \{y-1\}=\{2,4, 6, \cdots, y-3,y\}$ and $C^{+}$, $C^{-}$, $D^{+}$, $D^{-}$ are defined as Eqs.(\ref{eq3.16})-(\ref{eq3.19}). Additionally, $y+1$ actually represents 1 due to the cyclic nature of the network.
\end{widetext}
}

The following result is for the case when $y$ is even.  With the same symbol $\bm{\alpha}_t$ for $t=0,1$ as in Theorem 7.1, we have

{\bf Theorem 7.2.} {\it Let $\mathcal{C}yc(y)$ be a cyclic network as in Figure \ref{fig10} with $y$  even. Assume each independent source $S_{j}$ emits $(1+1)$-mode Gaussian state $\rho^{A_{j}A_{j+1}}$ and denote its generalized  quasiprobability function as $Q_{\rho^{A_{j}A_{j+1}}}(\alpha_{x_j}^{\lambda_{j}},\alpha_{x_{j+1}}^{\lambda_{j}};s)$, where $j=1,2,\cdots,y-1$;   $S_{y}$ emits $(1+1)$-mode Gaussian state $\rho^{A_{y}A_{1}}$ and denote its generalized  quasiprobability function as $Q_{\rho^{A_{y}A_{1}}}(\alpha_{x_y}^{\lambda_{y}},\alpha_{x_1}^{\lambda_{y}};s)$, where $x_j, x_{y}\in\{0,1\}$, $j=1,2,\cdots,y-1$. If $\mathcal{C}yc(y)$ is network local, then the following nonlinear Bell-type inequality holds:
\begin{equation}\label{eq7.2}\mathcal{B}_{\rho}^{cy}(\bm{\alpha}_0,\bm{\alpha}_1;s)=|\mathcal I_{s}|^{\frac{2}{y}}+|\mathcal J_{s}|^{\frac{2}{y}}\leq1,\tag{7.2}
\end{equation}
 where $\rho=\rho^{A_{1}A_{2}}\otimes\rho^{A_{2}A_{3}}\otimes\cdots\otimes\rho^{A_{y-1}A_{y}}\otimes\rho^{A_{y}A_{1}}$, for $-1<s\leq 0$,
\begin{widetext}
\begin{equation}
\begin{aligned}
\mathcal I_{s}=&\frac{1}{2^{\frac{y}{2}}}\times\mathop{\prod}\limits_{i\in\mathcal{K}}\{\frac{\pi^{2}(1-s)^{4}}{4}C_{\rho^{A_{i}A_{i+1}}}^{+}(\alpha_{x_{i}=0}^{\lambda_{i}},\alpha_{x_{i+1}=0}^{\lambda_{i}},\alpha_{x_{i}=1}^{\lambda_{i}},\alpha_{x_{i+1}=0}^{\lambda_{i}};s)+\frac{\pi s(1-s)^{2}}{2}D_{\rho^{A_{i}A_{i+1}}}^{+}\nonumber\\&(\alpha_{x_{i}=0}^{\lambda_{i}},
\alpha_{x_{i+1}=0}^{\lambda_{i}},\alpha_{x_{i}=1}^{\lambda_{i}},\alpha_{x_{i+1}=0}^{\lambda_{i}};s)
+2s^{2}\}\times\mathop{\prod}\limits_{j\in\overline{\mathcal K}}\{\frac{\pi^{2}(1-s)^{4}}{4}C_{\rho^{A_{j}A_{j+1}}}^{+}(\alpha_{x_{j}=0}^{\lambda_{j}},\alpha_{x_{j+1}=0}^{\lambda_{j}},\nonumber\\&\alpha_{x_{j}=0}^{\lambda_{j}},\alpha_{x_{j+1}=1}^{\lambda_{j}};s)
+\frac{\pi s(1-s)^{2}}{2}D_{\rho^{A_{j}A_{j+1}}}^{+}(\alpha_{x_{j}=0}^{\lambda_{j}},
\alpha_{x_{j+1}=0}^{\lambda_{j}},\alpha_{x_{j}=0}^{\lambda_{j}},\alpha_{x_{j+1}=1}^{\lambda_{j}};s)
+2s^{2}\}\nonumber
\end{aligned}
\end{equation}
and
\begin{equation}
\begin{aligned}
\mathcal J_{s}=&\frac{1}{2^{\frac{y}{2}}}\times\mathop{\prod}\limits_{i\in\mathcal{K}}\{\frac{\pi^{2}(1-s)^{4}}{4}C_{\rho^{A_{i}A_{i+1}}}^{-}(\alpha_{x_{i}=0}^{\lambda_{i}},\alpha_{x_{i+1}=1}^{\lambda_{i}},\alpha_{x_{i}=1}^{\lambda_{i}},\alpha_{x_{i+1}=1}^{\lambda_{i}};s)+\frac{\pi s(1-s)^{2}}{2}D_{\rho^{A_{i}A_{i+1}}}^{-}\nonumber\\&(\alpha_{x_{i}=0}^{\lambda_{i}},
\alpha_{x_{i+1}=1}^{\lambda_{i}},\alpha_{x_{i}=1}^{\lambda_{i}},\alpha_{x_{i+1}=1}^{\lambda_{i}};s)\}
\times\mathop{\prod}\limits_{j\in\overline{\mathcal K}}\{\frac{\pi^{2}(1-s)^{4}}{4}C_{\rho^{A_{j}A_{j+1}}}^{-}(\alpha_{x_{j}=1}^{\lambda_{j}},\alpha_{x_{j+1}=0}^{\lambda_{j}},\nonumber\\&\alpha_{x_{j}=1}^{\lambda_{j}},\alpha_{x_{j+1}=1}^{\lambda_{j}};s)
+\frac{\pi s(1-s)^{2}}{2}D_{\rho^{A_{j}A_{j+1}}}^{-}(\alpha_{x_{j}=1}^{\lambda_{j}},\alpha_{x_{j+1}=0}^{\lambda_{j}},
\alpha_{x_{j}=1}^{\lambda_{j}},\alpha_{x_{j+1}=1}^{\lambda_{j}};s)\}\nonumber;
\end{aligned}
\end{equation}
while for $s\leq-1$,
\begin{equation}
\begin{aligned}
\mathcal I_{s}=&\frac{1}{2^{\frac{y}{2}}}\times\mathop{\prod}\limits_{i\in\mathcal{K}}\{\pi^{2}(1-s)^{2}C_{\rho^{A_{i}A_{i+1}}}^{+}(\alpha_{x_{i}=0}^{\lambda_{i}},\alpha_{x_{i+1}=0}^{\lambda_{i}},\alpha_{x_{i}=1}^{\lambda_{i}},\alpha_{x_{i+1}=0}^{\lambda_{i}};s)-\pi(1-s)
D_{\rho^{A_{i}A_{i+1}}}^{+}\nonumber\\&(\alpha_{x_{i}=0}^{\lambda_{i}},\alpha_{x_{i+1}=0}^{\lambda_{i}},
\alpha_{x_{i}=1}^{\lambda_{i}},\alpha_{x_{i+1}=0}^{\lambda_{i}};s)+2\}\times\mathop{\prod}\limits_{j\in\overline{\mathcal K}}\{\pi^{2}(1-s)^{2}C_{\rho^{A_{j}A_{j+1}}}^{+}(\alpha_{x_{j}=0}^{\lambda_{j}},\alpha_{x_{j+1}=0}^{\lambda_{j}},\nonumber\\&\alpha_{x_{j}=0}^{\lambda_{j}},\alpha_{x_{j+1}=1}^{\lambda_{j}};s)
-\pi(1-s)D_{\rho^{A_{j}A_{j+1}}}^{+}(\alpha_{x_{j}=0}^{\lambda_{j}},\alpha_{x_{j+1}=0}^{\lambda_{j}},
\alpha_{x_{j}=0}^{\lambda_{j}},\alpha_{x_{j+1}=1}^{\lambda_{j}};s)+2\}\nonumber
\end{aligned}
\end{equation}
and
\begin{equation}
\begin{aligned}
\mathcal J_{s}=&\frac{1}{2^{\frac{y}{2}}}\times\mathop{\prod}\limits_{i\in\mathcal{K}}\{\pi^{2}(1-s)^{2}C_{\rho^{A_{i}A_{i+1}}}^{-}(\alpha_{x_{i}=0}^{\lambda_{i}},\alpha_{x_{i+1}=1}^{\lambda_{i}},\alpha_{x_{i}=1}^{\lambda_{i}},\alpha_{x_{i+1}=1}^{\lambda_{i}};s) -\pi(1-s)D_{\rho^{A_{i}A_{i+1}}}^{-}\nonumber\\&(\alpha_{x_{i}=0}^{\lambda_{i}},
\alpha_{x_{i+1}=1}^{\lambda_{i}},\alpha_{x_{i}=1}^{\lambda_{i}},\alpha_{x_{i+1}=1}^{\lambda_{i}};s)\}
\times\mathop{\prod}\limits_{j\in\overline{\mathcal K}}\{\pi^{2}(1-s)^{2}C_{\rho^{A_{j}A_{j+1}}}^{-}(\alpha_{x_{j}=1}^{\lambda_{j}},\alpha_{x_{j+1}=0}^{\lambda_{j}},\nonumber\\&\alpha_{x_{j}=1}^{\lambda_{j}},\alpha_{x_{j+1}=1}^{\lambda_{j}};s)
-\pi(1-s)D_{\rho^{A_{j}A_{j+1}}}^{-}(\alpha_{x_{j}=1}^{\lambda_{j}},\alpha_{x_{j+1}=0}^{\lambda_{j}},
\alpha_{x_{j}=1}^{\lambda_{j}},\alpha_{x_{j+1}=1}^{\lambda_{j}};s)\}\nonumber.
\end{aligned}
\end{equation}
Here,  ${\mathcal K}=\{2,4,\cdots, y\}$, $\overline{{\mathcal K}}=\{1,3, \cdots, y-1\}$, $y+1$ actually represents 1 due to the cyclic nature of the network, and  $C^{+}$, $C^{-}$, $D^{+}$, $D^{-}$ are defined as Eqs.(\ref{eq3.16})-(\ref{eq3.19}).
\end{widetext}}

The proofs of Theorems 7.1 and 7.2 are provided in Appendix E.

According to Theorems 7.1 and 7.2, if the inequality (\ref{eq7.1}) when $y$ is odd or the inequality (\ref{eq7.2}) when $y$ is even is violated by any specific choice of parameters $s$, $\alpha_{x_{i}=0}^{\lambda_{i}}$, $\alpha_{x_{i}=1}^{\lambda_{i}}$, $\alpha_{x_{i+1}=0}^{\lambda_{i}}$, $\alpha_{x_{i+1}=1}^{\lambda_{i}}$ for $i=1,2,\cdots, y-1$ and $\alpha_{x_{y}=0}^{\lambda_{y}}$, $\alpha_{x_{y}=1}^{\lambda_{y}}$, $\alpha_{x_{1}=0}^{\lambda_{y}}$, $\alpha_{x_{1}=1}^{\lambda_{y}}$, then it demonstrates the nonlocality  in the cyclic network ${\mathcal C}yc(y)$ of CV system.  Moving forward, we will provide two specific examples to illustrate how to use the inequality (\ref{eq7.1}) to witness the nonlocality in cyclic network ${\mathcal C}yc(5)$ of CV systems.

\begin{figure*}[]
\center
\subfigure [$-1<s\leq0$]
{\includegraphics[width=7cm,height=5.5cm]{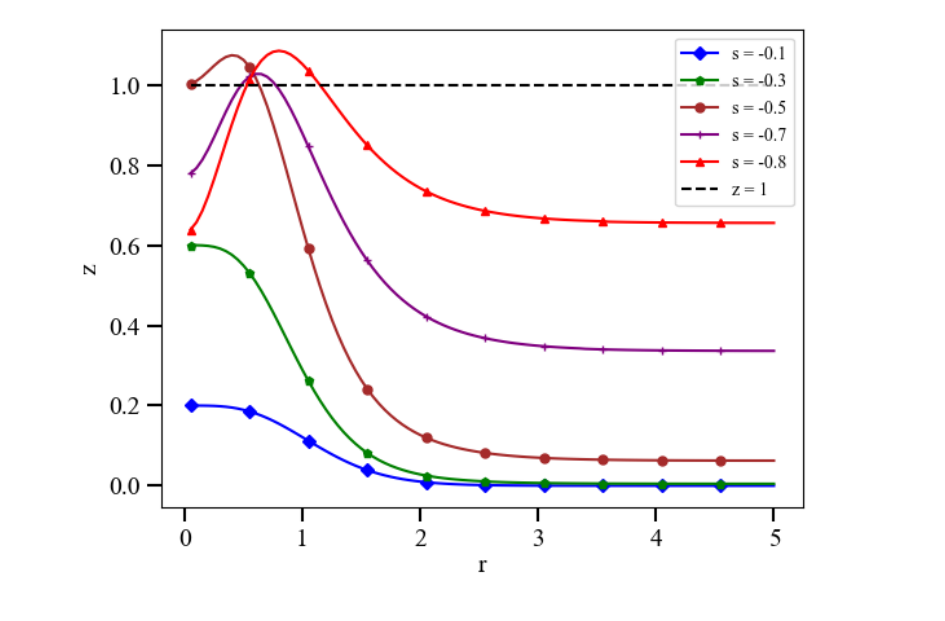}}
\subfigure [$s\leq-1$]
{\includegraphics[width=7cm,height=5.5cm]{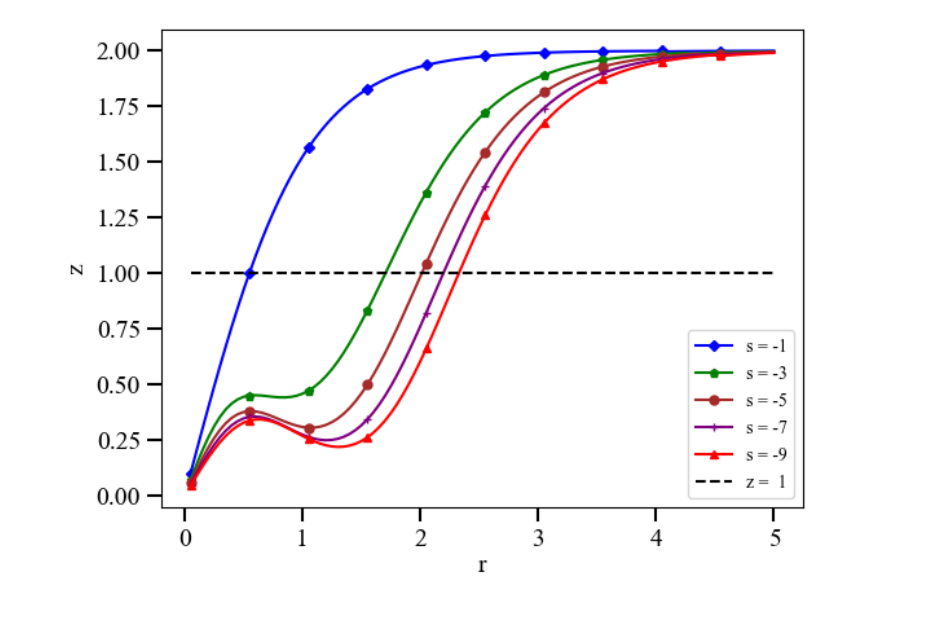}}
\caption{\small\quad Value of $z=B^{cy}(s,r)$ in $\mathcal{C}yc(5)$ for $(1+1)$-mode Gaussian state $\rho^{A_{j}A_{j+1}}=\rho^{A_{5}A_{1}}=\rho^{r}(j=1,2,3,4)$, as a function of the parameter $r$ for fixed $s$. The figure (a) shows the case of $-1<s\leq0$, while the figure (b) shows the case of $s\leq-1$.}
\centering
\label{fig11}
\end{figure*}

\begin{figure*}[]
\center
\subfigure [$-1<s\leq0$]
{\includegraphics[width=7cm,height=5.5cm]{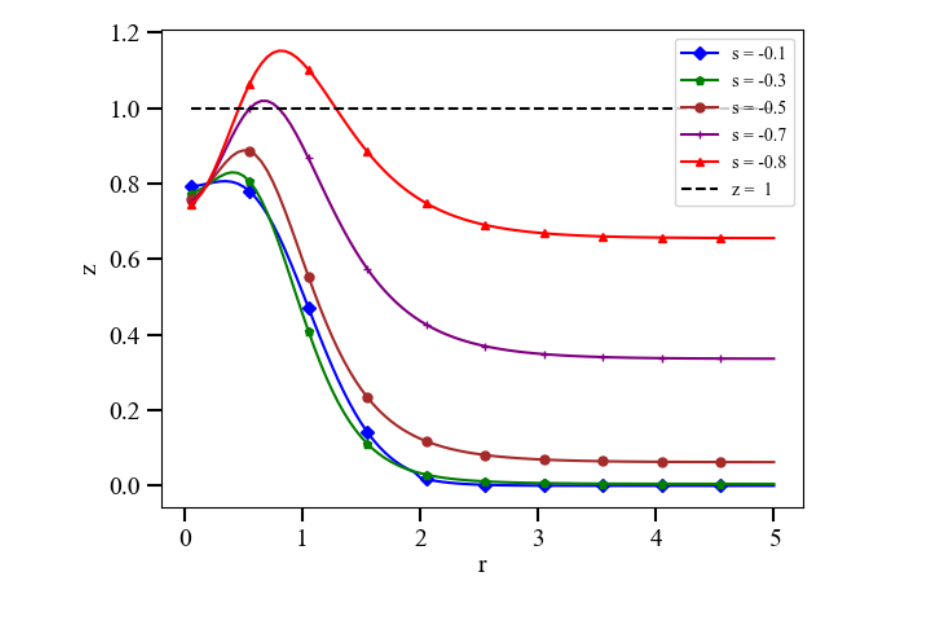}}
\subfigure [$s\leq-1$]
{\includegraphics[width=7cm,height=5.5cm]{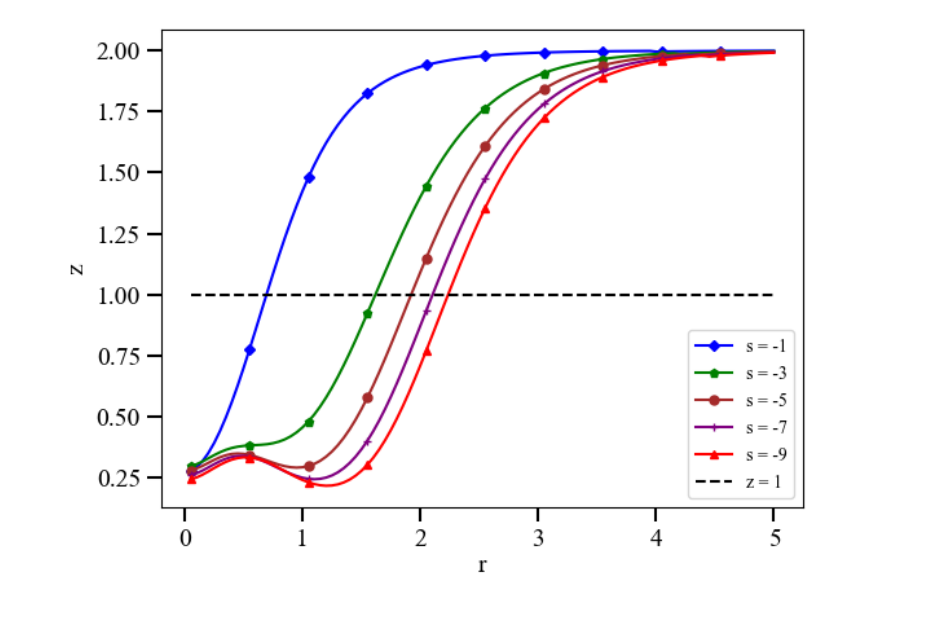}}
\caption{\small\quad Value of $z=B^{cy}(s,r)$ in $\mathcal{C}yc(5)$ for $(1+1)$-mode Gaussian state $\rho^{A_{j}A_{j+1}}=\rho^{A_{5}A_{1}}=\rho(1.2,1.2,r)(j=1,2,3,4)$, as a function of the parameter $r$ for fixed $s$. The figure (a) shows the case of $-1<s\leq0$, while the figure (b) shows the case of $s\leq-1$.}
\centering
\label{fig12}
\end{figure*}

{\bf Example 7.3.} Consider the cyclic network ${\mathcal C}yc(5)$ and assume that $\rho^{A_{j}A_{j+1}}=\rho^{A_{5}A_{1}}=\rho^{r}$ is an EPR state as in Eq.(\ref{eq3.7}) for each $j=1,2,3,4$. Let
$$  \begin{array}{rl}
\bm{\alpha}_{t}=&(\alpha_{x_{1}=t}^{\lambda_{1}},\alpha_{x_{2}=t}^{\lambda_{1}},\alpha_{x_{2}=t}^{\lambda_{2}},
\alpha_{x_{3}=t}^{\lambda_{2}},\alpha_{x_{3}=t}^{\lambda_{3}},\nonumber\\&\alpha_{x_{4}=t}^{\lambda_{3}},\alpha_{x_{4}=t}^{\lambda_{4}},
\alpha_{x_{5}=t}^{\lambda_{4}},\alpha_{x_{5}=t}^{\lambda_{5}},\alpha_{x_{1}=t}^{\lambda_{5}})\in\mathbb{C}^{10},
\end{array}
$$
 where $t\in\{0,1\}$. Then $$B^{cy}(s,r)=\sup_{\bm{\alpha}_{0},\bm{\alpha}_{1}}
 \mathcal{B}_{\rho}^{cy}(\bm{\alpha}_{0},\bm{\alpha}_{1};s)$$  is a function  of $s$ and $r$. By Theorem 7.1,   $B^{cy}(s,r)>1$ will demonstrate the network nonlocality in ${\mathcal C}yc(5)$. For some specific fixed values of $s$,  Figure  \ref{fig11} plots $B^{cy}(s,r)$ as a function of the parameter $r$.

In Figure \ref{fig11}(a), when $s$ is fixed within the interval $(-1,0]$, the inequality $B^{cy}(s,r)>1$ holds when $s\leq-0.5$ and this inequality is only satisfied when $r$ takes on relatively small values. Figure \ref{fig11}(b) shows that, when $s\leq-1$, the inequality $B^{cy}(s,r)>1$ holds when  $r$ is sufficiently large, and as $s$ increases, there are more values of $r$ that satisfy $B^{cy}(s,r)>1$. So we can see that using the generalized quasiprobability  function with $s=-1$ enables us to witness more network nonlocality.

Moreover, from Figure \ref{fig11}(a), the inequality $B^{cy}(-0.5,r)>1$ holds for $0<r\leq0.6$, while from Figure \ref{fig11}(b),  the inequality $B^{cy}(-1,r)>1$ holds for $r\geq0.55$. In conclusion, by taking $s=-0.5$ and $-1$, the inequality $B^{cy}(s,r)>1$ holds for any $r>0$. This reveals that if all five sources in cyclic network ${\mathcal C}yc(5)$ emit any identical entangled EPR states,  they generate a network nonlocality correlation in ${\mathcal C}yc(5)$. Therefore, we in fact obtain that {\it  if the five sources in ${\mathcal C}yc(5)$ emit any identical entangled Gaussian pure states, the cyclic network ${\mathcal C}yc(5)$ demonstrates the network nonlocality.}

{\bf Example 7.4.} Let us consider the scenario that all sources in the cyclic network ${\mathcal C}yc(5)$ emit any identical mixed Gaussian states. For simplicity, we assume that $\rho^{A_{j}A_{j+1}}=\rho^{A_{5}A_{1}}=\rho(1.2,1.2,r)$ $(j=1,2,3,4)$, where $\rho(1.2,1.2,r)$ is a symmetric STS as described in Eq.(\ref{eq3.10}). Similar to Example 7.3, $B^{cy}(s,r)=\sup_{\bm{\alpha}_{0},\bm{\alpha}_{1}}
\mathcal{B}_{\rho}^{cy}(\bm{\alpha}_{0},\bm{\alpha}_{1};s)$  is a function  of $s$ and $r$, and  by Theorem 7.1,  $B^{cy}(s,r)>1$  demonstrates the network nonlocality in ${\mathcal C}yc(5)$.
   Figure \ref{fig12} depicts the value of $B^{cy}(s,r)$ as a function of the parameter $r$ for several fixed $s$.

It is observed from Figures \ref{fig12}(a) and (b) that when $s=-0.8$ and $s\leq-1$,   $B^{cy}(s,r)>1$ for some $r$.  Furthermore, Figure \ref{fig12}(a) indicates that the inequality  $B^{cy}(-0.8,r)>1$ holds for $0.5\leq r\leq1.25$,  while Figure \ref{fig12}(b) reveals that the inequality $B^{cy}(-1,r)>1$ holds for  $r\geq0.7$. Combining these two facts reveals that, {\it the cyclic network ${\mathcal C}yc(5)$ exhibits network nonlocality if $r\geq0.5$}. \if false This suggests that using the Q function $(s=-1)$ allows for the witnessing of network nonlocality of a larger number of symmetric STSs in the cyclic network ${\mathcal C}yc(5)$.\fi

\section{Conclusion and discussion}\label{sec:8}

Gaussian states serve as a highly valuable resource in the realm of quantum information and quantum communication protocols, and the quantum networks that will be operated in the future are likely to be  hybrid networks. { However, so far,  there has been a significant limitation of research focusing on the nonlocality of networks for CV systems.} The challenge comes from the fact that Gaussian states can not recognize nonlocality when Gaussian measurements are performed. In this work, we aim to address the problem of detecting network nonlocality in CV systems by proposing an effective nonlinear Bell-type inequality for a general network of finite or infinite dimensional system, where each party has two inputs  and unrestricted outputs (which may be infinitely many). This inequality provides a general approach  to detect the network nonlocality of CV systems, where the quantum source states are   multimode Gaussian states  and the   measurement scheme is based on generalized quasiprobability functions. Notably, assessing the  nonlocality  of  networks in CV systems using this inequality is straightforward, as this inequality is solely related to the generalized quasiprobability functions of Gaussian states.  This allow us propose a general approach called the supremum strategy.   To illustrate how to utilize established nonlinear Bell-type inequality and the supremum strategy to detect the network nonlocality in the networks of CV systems, we establish several nonlinear  Bell-type inequalities for various network configurations of depth 2, including the chain, star, tree-shaped, and cyclic networks of CV systems with source states being  arbitrary $(1+1)$-mode Gaussian states.  In addition, the violation of these inequalities was demonstrated for identical entangled EPR states and certain symmetric STSs. Through these examples,  we infer that, for a general quantum network, if all source states are identical $(1+1)$-mode entangled Gaussian pure states, then the Bell-type inequalities of the network will be   violated, and this violation thereby naturally witnesses the network nonlocality. We also observe that,  using only special measurements based on the generalized quasiprobability functions, for instance with some choice of $s$, is sufficient to recognize the network nonlocality of networks with pure Gaussian source states. When source states are identical mixed Gaussian states such as the symmetric STSs $\rho(v,v,r)$,  the situation becomes more complex. However, by  utilizing the generalized quasiprobability function, one still can recognize  the  network nonlocality   for all $r> r_v$ for some $r_v>0$. For the entanglement swapping network, the scenarios where sources emit different pure Gaussian states or symmetric STSs are discussed in detail.

  Our scheme demonstrates that for Gaussian networks, measurements based on generalized quasiprobability functions can effectively detect network nonlocality.  This approach may  provide new insights into protocols including quantum sensing \cite{QJL} and device-independent quantum key distribution \cite{FCC}.

Moreover, our approach is physically feasible. As described in \cite{SHD}, when  all sources emit $(1+1)$-mode Gaussian states in the quantum network, the experimental
implementation based on this scheme is requires only a
beam-splitter coupled to the photodetectors. Notably, when the parameter $s=-1$, the measurement scheme
only requires a beam-splitter coupled to an intense coherent state, followed by a photodetector that distinguishes
between vacuum and non-vacuum states. Present day
photodetectors can achieve this with large efficiency in
the visible and near infrared. Thus, our scheme assumes
the significance of reality, especially in the construction
of practical CV quantum networks.

%In addition, using the Q-function$(s=-1)$ allows for the detection of a larger number of Gaussian states violating network local inequalities. We infer that for a general quantum network, if all resource states are identical $(1+1)$-mode entangled Gaussian pure states, it will violate the network local inequality and naturally demonstrate the network nonlocality. If the resource state contains Gaussian mixed states, the Q-function$(s=-1)$ test can be used by all parties. The realization of $s$-parametrized network local inequalities tests is expected along with the progress of photon detection technologies  in the near future.

It worth to point out, although our work is only focus on the network nonlocality of CV systems,  our scheme is well-suited for detecting network nonlocality within ``hybrid" quantum networks that are integrated with both DV and CV systems.
For example, consider the entanglement swapping network with one source emits a single-photon entangled state  $\rho=|\Psi\rangle\langle\Psi|$ with $|\Psi\rangle=\frac{1}{\sqrt{2}}(|01\rangle+|10\rangle)$  (its two-mode generalized quasiprobability function and its marginal single-mode distribution are given in \cite{SHD}) and the other source distributes a $(1+1)$-mode EPR state. \if false By substituting their generalized quasiprobability functions into the left-hand side of inequality (4.1), we analyze its maximum value to investigate whether this network exhibits network nonlocality.\fi Then, similar to the proof of Theorem 4.1,
one can establish a nonlinear Bell-type inequality by applying Theorem 2.1 to detect the network nonlocality in this hybrid entanglement swapping network.

{\bf Acknowledgments}
This work was supported by the National Natural Science Foundation
 of China (Grant Nos. 12071336, 12171290, 12271394)

{\bf InterestConflict} {The authors declare that they have no conflict of interest.}

\section*{appendix A: Proof of Theorem 2.1}

 Let $\Xi(y,z)$ be a general network with $k$ independent parties in system of any dimension and assume that it is network local. We prove that, for any dichotomic local measurements $\{M_{x_i}: x_i\in\{0,1\}\}_{i=1}^y$ of countable outcomes, the  nonlinear Bell-type inequality (\ref{eq2.2}) holds.

For any $i=1,2,\cdot\cdot\cdot,y$,  denote $\Upsilon_{x_{i}}=\{a_{h}: a_{h}\ \text{is the eigenvalue of} \ M_{x_{i}}\}$. Then $\langle M_{x_{1}}M_{x_{2}}\cdot\cdot\cdot M_{x_{y}}\rangle=\displaystyle\sum_{a_{1}\in\Upsilon_{x_{1}}}\displaystyle\sum_{a_{2}\in\Upsilon_{x_{2}}}\cdot\cdot\cdot\displaystyle\sum_{a_{y}\in\Upsilon_{x_{y}}}a_{1}a_{2}\cdot\cdot\cdot a_{y}P(a_{1},a_{2},\cdot\cdot\cdot, a_{y}|x_{1},x_{2},\cdot\cdot\cdot, x_{y})$ and the expectation of the outcomes of $M_{x_{i}}$ is
\begin{equation}\label{eqA1}\langle M_{x_{i}}\rangle=\displaystyle\sum_{a_{i}\in\Upsilon_{x_{i}}}a_{i}P(a_{i}|x_{i},\Lambda_{i}),\tag{A1}
\end{equation}
where $i=1,2,\cdots,y$ and $\Lambda _i=\{\lambda_{i_1}, \lambda_{i_2},\cdots, \lambda_{i_{e_i}}\}$  denotes the set of local variables associated with the sources that
connect to party $A_i$. Since $|a_{i}|\leq1$ for any $i$, we have $|\langle M_{x_{i}}\rangle|\leq1$.

Denote the index set of $k$ independent parties as $\mathcal{K}=\{i_{1},i_{2},\cdot\cdot\cdot, i_{k}\}$ and let $\bar{\mathcal{K}}=\{1,2,\cdot\cdot\cdot, y\}\backslash\mathcal{K}$.  By Eqs.(\ref{eq2.3}) and (\ref{eqA1}), we have
\begin{widetext}
\begin{equation}
\begin{aligned}
|\mathcal{I}|=&\frac{1}{2^{k}}|\int\int\cdot\cdot\cdot\int d\Lambda_{i_{1}}d\Lambda_{i_{2}}\cdot\cdot\cdot d\Lambda_{i_{k}}\mathop{\prod}\limits_{m=1}^{k}\mu_{i_{m}}(\Lambda_{i_{m}})\mathop{\prod}\limits_{i_{s}\in\mathcal{K}}(\langle M_{x_{i_{s}}=0}+M_{x_{i_{s}}=1}\rangle)\mathop{\prod}\limits_{j\in\bar{\mathcal{K}}}\langle M_{x_{j}=0}\rangle|\nonumber\\ \leq&
\frac{1}{2^{k}}\int\int\cdot\cdot\cdot\int d\Lambda_{i_{1}}d\Lambda_{i_{2}}\cdot\cdot\cdot d\Lambda_{i_{k}}\mathop{\prod}\limits_{m=1}^{k}\mu_{i_{m}}(\Lambda_{i_{m}})\mathop{\prod}\limits_{i_{s}\in\mathcal{K}}|\langle M_{x_{i_{s}}=0}+M_{x_{i_{s}}=1}\rangle|\mathop{\prod}\limits_{j\in\bar{\mathcal{K}}}|\langle M_{x_{j}=0}\rangle|\nonumber\\ \leq&\frac{1}{2^{k}}\int\int\cdot\cdot\cdot\int d\Lambda_{i_{1}}d\Lambda_{i_{2}}\cdot\cdot\cdot d\Lambda_{i_{k}}\mathop{\prod}\limits_{m=1}^{k}\mu_{i_{m}}(\Lambda_{i_{m}})\mathop{\prod}\limits_{i_{s}\in\mathcal{K}}|\langle M_{x_{i_{s}}=0}+M_{x_{i_{s}}=1}\rangle|\nonumber,
\end{aligned}
\end{equation}
where $\Lambda_{i_{m}}=\{\lambda_{(i_{m})_{1}},\lambda_{(i_{m})_{2}},\cdot\cdot\cdot,\lambda_{(i_{m})_{e_{i_{m}}}}\}$ denotes the set of local variables associated with the sources that connect to party $A_{i_{m}}$, $\mu_{i_{m}}(\Lambda_{i_{m}})=\mathop{\prod}\limits_{l=1}^{e_{i_{m}}}\mu_{(i_{m})_{l}}(\lambda_{(i_{m})_{l}})$ and $d\Lambda_{i_{m}}=d\lambda_{(i_{m})_{1}}d\lambda_{(i_{m})_{2}}\cdot\cdot\cdot d\lambda_{(i_{m})_{e_{i_{m}}}}$. Let $\langle\Delta^{\pm}M_{x_{i_{s}}}\rangle=\frac{1}{2}(\langle M_{x_{i_{s}}=0}\rangle\pm\langle M_{x_{i_{s}}=1}\rangle)$, then
\begin{equation}\tag{A2}\label{eqA2}
\begin{aligned}
|\mathcal{I}|&\leq\int\int\cdot\cdot\cdot\int d\Lambda_{i_{1}}d\Lambda_{i_{2}}\cdot\cdot\cdot d\Lambda_{i_{k}}\mathop{\prod}\limits_{m=1}^{k}\mu_{i_{m}}(\Lambda_{i_{m}})\mathop{\prod}\limits_{i_{s}\in\mathcal{K}}|\langle\Delta^{+}M_{x_{i_{s}}}\rangle|
\\&\leq\mathop{\prod}\limits_{i_{s}\in\mathcal{K}}\int d\Lambda_{i_{s}}\mu_{i_{s}}(\Lambda_{i_{s}})|\langle\Delta^{+}M_{x_{i_{s}}}\rangle|.
\end{aligned}
\end{equation}

Similarly, we have
\begin{equation}\tag{A3}\label{eqA3}
\begin{aligned}
|\mathcal{J}|&=|\int\int\cdot\cdot\cdot\int d\Lambda_{i_{1}}d\Lambda_{i_{2}}\cdot\cdot\cdot d\Lambda_{i_{k}}\mathop{\prod}\limits_{m=1}^{k}\mu_{i_{m}}(\Lambda_{i_{m}})\mathop{\prod}\limits_{i_{s}\in\mathcal{K}}\langle\Delta^{-}M_{x_{i_{s}}}\rangle\mathop{\prod}\limits_{j\in\bar{\mathcal{K}}}\langle M_{x_{j}=1}\rangle|
\\&\leq\mathop{\prod}\limits_{i_{s}\in\mathcal{K}}\int d\Lambda_{i_{s}}\mu_{i_{s}}(\Lambda_{i_{s}})|\langle\Delta^{-}M_{x_{i_{s}}}\rangle|.
\end{aligned}
\end{equation}

Using the Mahler inequality \cite{INE},  inequalities (\ref{eqA2}) and (\ref{eqA3}) together lead to
\begin{equation}\tag{A4}\label{eqA4}
\begin{aligned}
|\mathcal{I}|^{\frac{1}{k}}+|\mathcal{J}|^{\frac{1}{k}}\leq&[\mathop{\prod}\limits_{i_{s}\in\mathcal{K}}\int d\Lambda_{i_{s}}\mu_{i_{s}}(\Lambda_{i_{s}})(|\langle\Delta^{+}M_{x_{i_{s}}}\rangle|+|\langle\Delta^{-}M_{x_{i_{s}}}\rangle|)]^{\frac{1}{k}}
\\ \leq&[\mathop{\prod}\limits_{i_{s}\in\mathcal{K}}\int d\Lambda_{i_{s}}\mu_{i_{s}}(\Lambda_{i_{s}})]^{\frac{1}{k}}\\=&1.
\end{aligned}
\end{equation}
The second inequality in inequality (\ref{eqA4}) is from the inequality $|\langle\Delta^{+}M_{x_{i_{s}}}\rangle|+|\langle\Delta^{-}M_{x_{i_{s}}}\rangle|={\rm max}\{|\langle M_{x_{i_{s}}=0}\rangle|, |\langle M_{x_{i_{s}}=1}\rangle|\}\leq1$  for $s=1,2,\cdots,k$; and the final equality in inequality (\ref{eqA4}) is from the normalization condition of the probability distribution of hidden states. \hfill$\Box$
\end{widetext}

\section*{appendix B: Proof of Theorem 4.1}

 For arbitrary chain network scenario ${\mathcal C}ha(y)$ described in Figure \ref{fig1}, it is clear that (1) if $y$ is odd, then $k_{\rm max}=\frac{y+1}{2}$ with the index set of  maximal set of independent parties ${\mathcal K}={\mathcal K}_{\rm max}=\{1,3,\cdots, y\}$; (2) if $y$ is even, then $k_{\rm max}=\frac{y}{2}$ with the index set of  maximal set of independent parties ${\mathcal K}={\mathcal K}_{\rm max}=\{2,4,\cdots, y\}$. Denote by $\overline{\mathcal K}=\{1,2,\cdots, y\}\setminus {\mathcal K}$ and ${\mathcal K}'$  the set of indexes from ${\mathcal K}$ with the biggest index excluded. For example, when $y$ is odd, ${\mathcal K}'=\{1,3,\cdots, y-2\}$.  In Figure \ref{fig1}, if $y$ is odd, by Eqs. (\ref{eq2.3}) and (\ref{eq3.14}),
 \begin{widetext}
 \begin{equation}\tag{B1}\label{eqB1}
 \begin{aligned}
\mathcal{I}_{s}=&\frac{1}{2^{\frac{y+1}{2}}}\times\langle\mathop{\prod}\limits_{i\in\mathcal{K}}(M_{x_{i}=0}+M_{x_{i}=1})\mathop{\prod}\limits_{j\in\overline{\mathcal K}}M_{x_{j}=0}\rangle
\\=&\frac{1}{2^{\frac{y+1}{2}}}\times\langle(M_{x_{1}=0}+M_{x_{1}=1})M_{x_{2}=0}\cdot\cdot\cdot(M_{x_{y-2}=0}+M_{x_{y-2}=1}) M_{x_{y-1}=0}(M_{x_{y}=0}+M_{x_{y}=1})\rangle
\\=&\frac{1}{2^{\frac{y+1}{2}}}\times{\rm Tr}\{[(\hat{O}(\alpha_{x_{1}=0}^{\lambda_{1}};s)+\hat{O}(\alpha_{x_{1}=1}^{\lambda_{1}};s))\otimes\hat{O}(\alpha_{x_{2}=0}^{\lambda_{1}};s)\otimes\hat{O}(\alpha_{x_{2}=0}^{\lambda_{2}};s)\\&
\otimes((\hat{O}(\alpha_{x_{3}=0}^{\lambda_{2}};s)\otimes\hat{O}(\alpha_{x_{3}=0}^{\lambda_{3}};s))+(\hat{O}(\alpha_{x_{3}=1}^{\lambda_{2}};s)\otimes\hat{O}(\alpha_{x_{3}=1}^{\lambda_{3}};s)))
\\&\otimes\hat{O}(\alpha_{x_{4}=0}^{\lambda_{3}};s)\otimes\hat{O}(\alpha_{x_{4}=0}^{\lambda_{4}};s)\otimes\cdot\cdot\cdot\otimes((\hat{O}(\alpha_{x_{y-2}=0}^{\lambda_{y-3}};s)\otimes\hat{O}(\alpha_{x_{y-2}=0}^{\lambda_{y-2}};s))\\&+(\hat{O}(\alpha_{x_{y-2}=1}^{\lambda_{y-3}};s)\otimes\hat{O}(\alpha_{x_{y-2}=1}^{\lambda_{y-2}};s)))
\otimes\hat{O}(\alpha_{x_{y-1}=0}^{\lambda_{y-2}};s)\otimes\hat{O}(\alpha_{x_{y-1}=0}^{\lambda_{y-1}};s)\\&\otimes(\hat{O}(\alpha_{x_{y}=0}^{\lambda_{y-1}};s)+\hat{O}(\alpha_{x_{y}=1}^{\lambda_{y-1}};s))]\rho^{A_{1}A_{2}}\otimes\rho^{A_{2}A_{3}}\otimes\cdot\cdot\cdot\otimes\rho^{A_{y-1}A_{y}}\}
\\=&\frac{1}{2^{\frac{y+1}{2}}}\times{\rm Tr}\{[(\hat{O}(\alpha_{x_{1}=0}^{\lambda_{1}};s)+\hat{O}(\alpha_{x_{1}=1}^{\lambda_{1}};s))\otimes\hat{O}(\alpha_{x_{2}=0}^{\lambda_{1}};s)]\rho^{A_{1}A_{2}}\}
\\&\times{\rm Tr}\{[\hat{O}(\alpha_{x_{2}=0}^{\lambda_{2}};s)\otimes(\hat{O}(\alpha_{x_{3}=0}^{\lambda_{2}};s)+\hat{O}(\alpha_{x_{3}=1}^{\lambda_{2}};s))]\rho^{A_{2}A_{3}}\}
\\&\times\cdot\cdot\cdot\times{\rm Tr}\{[(\hat{O}(\alpha_{x_{y-2}=0}^{\lambda_{y-2}};s)+\hat{O}(\alpha_{x_{y-2}=1}^{\lambda_{y-2}};s))\otimes\hat{O}(\alpha_{x_{y-1}=0}^{\lambda_{y-2}};s)]\rho^{A_{y-2}A_{y-1}}\}
\\&\times{\rm Tr}\{[\hat{O}(\alpha_{x_{y-1}=0}^{\lambda_{y-1}};s)\otimes(\hat{O}(\alpha_{x_{y}=0}^{\lambda_{y-1}};s)+\hat{O}(\alpha_{x_{y}=1}^{\lambda_{y-1}};s))]\rho^{A_{y-1}A_{y}}\}.
\end{aligned}
\end{equation}

Similar to Eq.(\ref{eqB1}), when $y$ is even, we have
\begin{equation}\tag{B2}\label{eqB2}
\begin{aligned}
\mathcal{I}_{s}=&\frac{1}{2^{\frac{y}{2}}}\times\langle\mathop{\prod}\limits_{i\in\mathcal{K}}(M_{x_{i}=0}+M_{x_{i}=1})\mathop{\prod}\limits_{j\in\overline{\mathcal K}}M_{x_{j}=0}\rangle
\\=&\frac{1}{2^{\frac{y}{2}}}\times\langle M_{x_{1}=0}(M_{x_{2}=0}+M_{x_{2}=1})\cdot\cdot\cdot M_{x_{y-1}=0}(M_{x_{y}=0}+M_{x_{y}=1})\rangle
\\=&\frac{1}{2^{\frac{y}{2}}}\times{\rm Tr}\{[\hat{O}(\alpha_{x_{1}=0}^{\lambda_{1}};s)\otimes(\hat{O}(\alpha_{x_{2}=0}^{\lambda_{1}};s)+\hat{O}(\alpha_{x_{2}=1}^{\lambda_{1}};s))]\rho^{A_{1}A_{2}}\}
\\&\times{\rm Tr}\{[(\hat{O}(\alpha_{x_{2}=0}^{\lambda_{2}};s)+\hat{O}(\alpha_{x_{2}=1}^{\lambda_{2}};s))\otimes\hat{O}(\alpha_{x_{3}=0}^{\lambda_{2}};s)]\rho^{A_{2}A_{3}}\}
\\&\times\cdot\cdot\cdot\times{\rm Tr}\{[(\hat{O}(\alpha_{x_{y-2}=0}^{\lambda_{y-2}};s)+\hat{O}(\alpha_{x_{y-2}=1}^{\lambda_{y-2}};s))\otimes\hat{O}(\alpha_{x_{y-1}=0}^{\lambda_{y-2}};s)]\rho^{A_{y-2}A_{y-1}}\}
\\&\times{\rm Tr}\{[\hat{O}(\alpha_{x_{y-1}=0}^{\lambda_{y-1}};s)\otimes(\hat{O}(\alpha_{x_{y}=0}^{\lambda_{y-1}};s)+\hat{O}(\alpha_{x_{y}=1}^{\lambda_{y-1}};s))]\rho^{A_{y-1}A_{y}}\}.
\end{aligned}
\end{equation}

By combining Eqs.(\ref{eqB1}) and (\ref{eqB2}), for arbitrary chain network ${\mathcal C}ha(y)$,
\begin{equation}\tag{B3}\label{eqB3}
\begin{aligned}
\mathcal{I}_{s}=&\frac{1}{2^{k_{\rm max}}}\times\langle\mathop{\prod}\limits_{i\in\mathcal{K}}(M_{x_{i}=0}+M_{x_{i}=1})\mathop{\prod}\limits_{j\in\overline{\mathcal K}}M_{x_{j}=0}\rangle
\\=&\frac{1}{2^{k_{\rm max}}}\times\mathop{\prod}\limits_{i\in\mathcal{K}^{'}}{\rm Tr}\{[(\hat{O}(\alpha_{x_{i}=0}^{\lambda_{i}};s)+\hat{O}(\alpha_{x_{i}=1}^{\lambda_{i}};s))\otimes\hat{O}(\alpha_{x_{i+1}=0}^{\lambda_{i}};s)]\rho^{A_{i}A_{i+1}}\}
\\&\times\mathop{\prod}\limits_{j\in\overline{\mathcal{K}}}{\rm Tr}\{[\hat{O}(\alpha_{x_{j}=0}^{\lambda_{j}};s)\otimes(\hat{O}(\alpha_{x_{j+1}=0}^{\lambda_{j}};s)+\hat{O}(\alpha_{x_{j+1}=1}^{\lambda_{j}};s))]\rho^{A_{j}A_{j+1}}\}.
\end{aligned}
\end{equation}

Now, by Eqs.(\ref{eq3.3}), (\ref{eq3.15}), (\ref{eq3.16}) and (\ref{eq3.18}), for $-1<s\leq0$, we get
\begin{equation}\tag{B4}\label{eqB4}
\begin{aligned}
&{\rm Tr}\{[(\hat{O}(\alpha_{x_{i}=0}^{\lambda_{i}};s)+\hat{O}(\alpha_{x_{i}=1}^{\lambda_{i}};s))\otimes\hat{O}(\alpha_{x_{i+1}=0}^{\lambda_{i}};s)]\rho^{A_{i}A_{i+1}}\}\\=&{\rm Tr}\{[\hat{O}(\alpha_{x_{i}=0}^{\lambda_{i}};s)\otimes\hat{O}(\alpha_{x_{i+1}=0}^{\lambda_{i}};s)]\rho^{A_{i}A_{i+1}}\}+{\rm Tr}\{[\hat{O}(\alpha_{x_{i}=1}^{\lambda_{i}};s)\otimes\hat{O}(\alpha_{x_{i+1}=0}^{\lambda_{i}};s)]\rho^{A_{i}A_{i+1}}\}\\=&{\rm Tr}\{[((1-s)\hat{\Pi}(\alpha_{x_{i}=0}^{\lambda_{i}};s)+sI)\otimes((1-s)\hat{\Pi}(\alpha_{x_{i+1}=0}^{\lambda_{i}};s)+sI)]\rho^{A_{i}A_{i+1}}\}\\&+{\rm Tr}\{[((1-s)\hat{\Pi}(\alpha_{x_{i}=1}^{\lambda_{i}};s)+sI)\otimes((1-s)\hat{\Pi}(\alpha_{x_{i+1}=0}^{\lambda_{i}};s)+sI)]\rho^{A_{i}A_{i+1}}\}\\
=&(1-s)^{2}{\rm Tr}\{[\hat{\Pi}(\alpha_{x_{i}=0}^{\lambda_{i}};s)\otimes\hat{\Pi}(\alpha_{x_{i+1}=0}^{\lambda_{i}};s)]\rho^{A_{i}A_{i+1}}\}+s(1-s){\rm Tr}\{[\hat{\Pi}(\alpha_{x_{i}=0}^{\lambda_{i}};s)\otimes I]\rho^{A_{i}A_{i+1}}\}\\&+s(1-s){\rm Tr}\{[I\otimes\hat{\Pi}(\alpha_{x_{i+1}=0}^{\lambda_{i}};s)]\rho^{A_{i}A_{i+1}}\}+s^{2}+(1-s)^{2}{\rm Tr}\{[\hat{\Pi}(\alpha_{x_{i}=1}^{\lambda_{i}};s)\otimes\hat{\Pi}(\alpha_{x_{i+1}=0}^{\lambda_{i}};s)]\\&\rho^{A_{i}A_{i+1}}\}+s(1-s){\rm Tr}\{[\hat{\Pi}(\alpha_{x_{i}=1}^{\lambda_{i}};s)\otimes I]\rho^{A_{i}A_{i+1}}\}+s(1-s){\rm Tr}\{[I\otimes\hat{\Pi}(\alpha_{x_{i+1}=0}^{\lambda_{i}};s)]\rho^{A_{i}A_{i+1}}\}+s^{2}\\=&\frac{\pi^{2}(1-s)^{4}}{4}[Q_{\rho^{A_{i}A_{i+1}}}(\alpha_{x_{i}=0}^{\lambda_{i}},\alpha_{x_{i+1}=0}^{\lambda_{i}};s)+Q_{\rho^{A_{i}A_{i+1}}}(\alpha_{x_{i}=1}^{\lambda_{i}},\alpha_{x_{i+1}=0}^{\lambda_{i}};s)]
\\&+\frac{\pi s(1-s)^{2}}{2}[Q_{\rho^{A_{i}A_{i+1}}}(\alpha_{x_{i}=0}^{\lambda_{i}};s)+Q_{\rho^{A_{i}A_{i+1}}}(\alpha_{x_{i}=1}^{\lambda_{i}};s)+2Q_{\rho^{A_{i}A_{i+1}}}(\alpha_{x_{i+1}=0}^{\lambda_{i}};s)]+2s^{2}
\\=&\frac{\pi^{2}(1-s)^{4}}{4}C_{\rho^{A_{i}A_{i+1}}}^{+}(\alpha_{x_{i}=0}^{\lambda_{i}},\alpha_{x_{i+1}=0}^{\lambda_{i}},\alpha_{x_{i}=1}^{\lambda_{i}},\alpha_{x_{i+1}=0}^{\lambda_{i}};s)+\frac{\pi s(1-s)^{2}}{2}D_{\rho^{A_{i}A_{i+1}}}^{+}(\alpha_{x_{i}=0}^{\lambda_{i}},\alpha_{x_{i+1}=0}^{\lambda_{i}},
\\&\alpha_{x_{i}=1}^{\lambda_{i}},\alpha_{x_{i+1}=0}^{\lambda_{i}};s)+2s^{2}.
\end{aligned}
\end{equation}
Similarly,
\begin{equation}\tag{B5}\label{eqB5}
\begin{aligned}
&{\rm Tr}\{[\hat{O}(\alpha_{x_{j}=0}^{\lambda_{j}};s)\otimes(\hat{O}(\alpha_{x_{j+1}=0}^{\lambda_{j}};s)+\hat{O}(\alpha_{x_{j+1}=1}^{\lambda_{j}};s))]\rho^{A_{j}A_{j+1}}\}
\\=&\frac{\pi^{2}(1-s)^{4}}{4}C_{\rho^{A_{j}A_{j+1}}}^{+}(\alpha_{x_{j}=0}^{\lambda_{j}},\alpha_{x_{j+1}=0}^{\lambda_{j}},\alpha_{x_{j}=0}^{\lambda_{j}},\alpha_{x_{j+1}=1}^{\lambda_{j}};s)
\\&+\frac{\pi s(1-s)^{2}}{2}D_{\rho^{A_{j}A_{j+1}}}^{+}(\alpha_{x_{j}=0}^{\lambda_{j}},\alpha_{x_{j+1}=0}^{\lambda_{j}},\alpha_{x_{j}=0}^{\lambda_{j}},\alpha_{x_{j+1}=1}^{\lambda_{j}};s)+2s^{2}.
\end{aligned}
\end{equation}

While if $s\leq-1$, we have
\begin{equation}\tag{B6}\label{eqB6}
\begin{aligned}
&{\rm Tr}\{[(\hat{O}(\alpha_{x_{i}=0}^{\lambda_{i}};s)+\hat{O}(\alpha_{x_{i}=1}^{\lambda_{i}};s))\otimes\hat{O}(\alpha_{x_{i+1}=0}^{\lambda_{i}};s)]\rho^{A_{i}A_{i+1}}\}\\=&{\rm Tr}\{[(2\hat{\Pi}(\alpha_{x_{i}=0}^{\lambda_{i}};s)-I)\otimes(2\hat{\Pi}(\alpha_{x_{i+1}=0}^{\lambda_{i}};s)-I)]\rho^{A_{i}A_{i+1}}\}\\&+{\rm Tr}\{[(2\hat{\Pi}(\alpha_{x_{i}=1}^{\lambda_{i}};s)-I)\otimes(2\hat{\Pi}(\alpha_{x_{i+1}=0}^{\lambda_{i}};s)-I)]\rho^{A_{i}A_{i+1}}\}\\
=&\pi^{2}(1-s)^{2}[Q_{\rho^{A_{i}A_{i+1}}}(\alpha_{x_{i}=0}^{\lambda_{i}},\alpha_{x_{i+1}=0}^{\lambda_{i}};s)+Q_{\rho^{A_{i}A_{i+1}}}(\alpha_{x_{i}=1}^{\lambda_{i}},\alpha_{x_{i+1}=0}^{\lambda_{i}};s)]
\\&-\pi(1-s)[Q_{\rho^{A_{i}A_{i+1}}}(\alpha_{x_{i}=0}^{\lambda_{i}};s)+Q_{\rho^{A_{i}A_{i+1}}}(\alpha_{x_{i}=1}^{\lambda_{i}};s)+2Q_{\rho^{A_{i}A_{i+1}}}(\alpha_{x_{i+1}=0}^{\lambda_{i}};s)]+2
\\=&\pi^{2}(1-s)^{2}C_{\rho^{A_{i}A_{i+1}}}^{+}(\alpha_{x_{i}=0}^{\lambda_{i}},\alpha_{x_{i+1}=0}^{\lambda_{i}},\alpha_{x_{i}=1}^{\lambda_{i}},\alpha_{x_{i+1}=0}^{\lambda_{i}};s)
-\pi(1-s)\\&D_{\rho^{A_{i}A_{i+1}}}^{+}(\alpha_{x_{i}=0}^{\lambda_{i}},\alpha_{x_{i+1}=0}^{\lambda_{i}},\alpha_{x_{i}=1}^{\lambda_{i}},\alpha_{x_{i+1}=0}^{\lambda_{i}};s)+2
\end{aligned}
\end{equation}
and
\begin{equation}\tag{B7}\label{eqB7}
\begin{aligned}
&{\rm Tr}\{[\hat{O}(\alpha_{x_{j}=0}^{\lambda_{j}};s)\otimes(\hat{O}(\alpha_{x_{j+1}=0}^{\lambda_{j}};s)+\hat{O}(\alpha_{x_{j+1}=1}^{\lambda_{j}};s))]\rho^{A_{j}A_{j+1}}\}
\\=&\pi^{2}(1-s)^{2}C_{\rho^{A_{j}A_{j+1}}}^{+}(\alpha_{x_{j}=0}^{\lambda_{j}},\alpha_{x_{j+1}=0}^{\lambda_{j}},\alpha_{x_{j}=0}^{\lambda_{j}},\alpha_{x_{j+1}=1}^{\lambda_{j}};s)
\\&-\pi(1-s)D_{\rho^{A_{j}A_{j+1}}}^{+}(\alpha_{x_{j}=0}^{\lambda_{j}},\alpha_{x_{j+1}=0}^{\lambda_{j}},\alpha_{x_{j}=0}^{\lambda_{j}},\alpha_{x_{j+1}=1}^{\lambda_{j}};s)+2.
\end{aligned}
\end{equation}
Substitute Eqs.(\ref{eqB4})-(\ref{eqB7}) into  Eq.(\ref{eqB3}), we obtain the desired expression of $\mathcal{I}_{s}$ as in Theorem 4.1.

By a similar argument to Eq.(\ref{eqB3}), it can be derived from Eqs.(\ref{eq2.4}) and (\ref{eq3.14}) that
\begin{equation}\tag{B8}\label{eqB8}
\begin{aligned}
\mathcal{J}_{s}=&\frac{1}{2^{k_{\rm max}}}\times\langle\mathop{\prod}\limits_{i\in\mathcal{K}}(M_{x_{i}=0}-M_{x_{i}=1})\mathop{\prod}\limits_{j\in\overline{\mathcal{K}}}M_{x_{j}=1}\rangle
\\
=&\frac{1}{2^{k_{\rm max}}}\times\mathop{\prod}\limits_{i\in\mathcal{K}^{'}}{\rm Tr}\{[(\hat{O}(\alpha_{x_{i}=0}^{\lambda_{i}};s)-\hat{O}(\alpha_{x_{i}=1}^{\lambda_{i}};s))\otimes\hat{O}(\alpha_{x_{i+1}=1}^{\lambda_{i}};s)]\rho^{A_{i}A_{i+1}}\}
\\&\times\mathop{\prod}\limits_{j\in\overline{\mathcal{K}}}{\rm Tr}\{[\hat{O}(\alpha_{x_{j}=1}^{\lambda_{j}};s)\otimes(\hat{O}(\alpha_{x_{j+1}=0}^{\lambda_{j}};s)-\hat{O}(\alpha_{x_{j+1}=1}^{\lambda_{j}};s))]\rho^{A_{j}A_{j+1}}\}.
\end{aligned}
\end{equation}
Then by Eqs.(\ref{eq3.3}),  (\ref{eq3.15}), (\ref{eq3.17}) and (\ref{eq3.19}),  when $-1<s\leq0$, we have
\begin{equation}\tag{B9}\label{eqB9}
\begin{aligned}
&{\rm Tr}\{[(\hat{O}(\alpha_{x_{i}=0}^{\lambda_{i}};s)-\hat{O}(\alpha_{x_{i}=1}^{\lambda_{i}};s))\otimes\hat{O}(\alpha_{x_{i+1}=1}^{\lambda_{i}};s)]\rho^{A_{i}A_{i+1}}\}\\=&{\rm Tr}\{[((1-s)\hat{\Pi}(\alpha_{x_{i}=0}^{\lambda_{i}};s)+sI)\otimes((1-s)\hat{\Pi}(\alpha_{x_{i+1}=1}^{\lambda_{i}};s)+sI)]\rho^{A_{i}A_{i+1}}\}\\&-{\rm Tr}\{[((1-s)\hat{\Pi}(\alpha_{x_{i}=1}^{\lambda_{i}};s)+sI)\otimes((1-s)\hat{\Pi}(\alpha_{x_{i+1}=1}^{\lambda_{i}};s)+sI)]\rho^{A_{i}A_{i+1}}\}\\
=&(1-s)^{2}{\rm Tr}\{[\hat{\Pi}(\alpha_{x_{i}=0}^{\lambda_{i}};s)\otimes\hat{\Pi}(\alpha_{x_{i+1}=1}^{\lambda_{i}};s)]\rho^{A_{i}A_{i+1}}\}+s(1-s){\rm Tr}\{[\hat{\Pi}(\alpha_{x_{i}=0}^{\lambda_{i}};s)\otimes I]\rho^{A_{i}A_{i+1}}\}\nonumber\\&+s(1-s){\rm Tr}\{[I\otimes\hat{\Pi}(\alpha_{x_{i+1}=1}^{\lambda_{i}};s)]\rho^{A_{i}A_{i+1}}\}+s^{2}-(1-s)^{2}{\rm Tr}\{[\hat{\Pi}(\alpha_{x_{i}=1}^{\lambda_{i}};s)\otimes\hat{\Pi}(\alpha_{x_{i+1}=1}^{\lambda_{i}};s)]\\&\rho^{A_{i}A_{i+1}}\}-s(1-s){\rm Tr}\{[\hat{\Pi}(\alpha_{x_{i}=1}^{\lambda_{i}};s)\otimes I]\rho^{A_{i}A_{i+1}}\}-s(1-s){\rm Tr}\{[I\otimes\hat{\Pi}(\alpha_{x_{i+1}=1}^{\lambda_{i}};s)]\rho^{A_{i}A_{i+1}}\}-s^{2}\\=&\frac{\pi^{2}(1-s)^{4}}{4}[Q_{\rho^{A_{i}A_{i+1}}}(\alpha_{x_{i}=0}^{\lambda_{i}},\alpha_{x_{i+1}=1}^{\lambda_{i}};s)-Q_{\rho^{A_{i}A_{i+1}}}(\alpha_{x_{i}=1}^{\lambda_{i}},\alpha_{x_{i+1}=1}^{\lambda_{i}};s)]
\\&+\frac{\pi s(1-s)^{2}}{2}[Q_{\rho^{A_{i}A_{i+1}}}(\alpha_{x_{i}=0}^{\lambda_{i}};s)-Q_{\rho^{A_{i}A_{i+1}}}(\alpha_{x_{i}=1}^{\lambda_{i}};s)]
\\=&\frac{\pi^{2}(1-s)^{4}}{4}C_{\rho^{A_{i}A_{i+1}}}^{-}(\alpha_{x_{i}=0}^{\lambda_{i}},\alpha_{x_{i+1}=1}^{\lambda_{i}},\alpha_{x_{i}=1}^{\lambda_{i}},\alpha_{x_{i+1}=1}^{\lambda_{i}};s)\\&+\frac{\pi s(1-s)^{2}}{2}D_{\rho^{A_{i}A_{i+1}}}^{-}(\alpha_{x_{i}=0}^{\lambda_{i}},\alpha_{x_{i+1}=1}^{\lambda_{i}},
\alpha_{x_{i}=1}^{\lambda_{i}},\alpha_{x_{i+1}=1}^{\lambda_{i}};s),
\end{aligned}
\end{equation}
and similarly,
\begin{equation}\tag{B10}\label{eqB10}
\begin{aligned}
&{\rm Tr}\{[\hat{O}(\alpha_{x_{j}=1}^{\lambda_{j}};s)\otimes(\hat{O}(\alpha_{x_{j+1}=0}^{\lambda_{j}};s)-\hat{O}(\alpha_{x_{j+1}=1}^{\lambda_{j}};s))]\rho^{A_{j}A_{j+1}}\}
\\=&\frac{\pi^{2}(1-s)^{4}}{4}C_{\rho^{A_{j}A_{j+1}}}^{-}(\alpha_{x_{j}=1}^{\lambda_{j}},\alpha_{x_{j+1}=0}^{\lambda_{j}},\alpha_{x_{j}=1}^{\lambda_{j}},\alpha_{x_{j+1}=1}^{\lambda_{j}};s)
\\&+\frac{\pi s(1-s)^{2}}{2}D_{\rho^{A_{j}A_{j+1}}}^{-}(\alpha_{x_{j}=1}^{\lambda_{j}},\alpha_{x_{j+1}=0}^{\lambda_{j}},\alpha_{x_{j}=1}^{\lambda_{j}},\alpha_{x_{j+1}=1}^{\lambda_{j}};s).
\end{aligned}
\end{equation}

While when $s\leq-1$,
\begin{equation}\tag{B11}\label{eqB11}
\begin{aligned}
&{\rm Tr}\{[(\hat{O}(\alpha_{x_{i}=0}^{\lambda_{i}};s)-\hat{O}(\alpha_{x_{i}=1}^{\lambda_{i}};s))\otimes\hat{O}(\alpha_{x_{i+1}=1}^{\lambda_{i}};s)]\rho^{A_{i}A_{i+1}}\}\\=&{\rm Tr}\{[(2\hat{\Pi}(\alpha_{x_{i}=0}^{\lambda_{i}};s)-I)\otimes(2\hat{\Pi}(\alpha_{x_{i+1}=1}^{\lambda_{i}};s)-I)]\rho^{A_{i}A_{i+1}}\}\\&-{\rm Tr}\{[(2\hat{\Pi}(\alpha_{x_{i}=1}^{\lambda_{i}};s)-I)\otimes(2\hat{\Pi}(\alpha_{x_{i+1}=1}^{\lambda_{i}};s)-I)]\rho^{A_{i}A_{i+1}}\}\\
=&4{\rm Tr}\{[\hat{\Pi}(\alpha_{x_{i}=0}^{\lambda_{i}};s)\otimes\hat{\Pi}(\alpha_{x_{i+1}=1}^{\lambda_{i}};s)]\rho^{A_{i}A_{i+1}}\}-2{\rm Tr}\{[\hat{\Pi}(\alpha_{x_{i}=0}^{\lambda_{i}};s)\otimes I]\rho^{A_{i}A_{i+1}}\}\\&-2{\rm Tr}\{[I\otimes\hat{\Pi}(\alpha_{x_{i+1}=1}^{\lambda_{i}};s)]\rho^{A_{i}A_{i+1}}\}+1-4{\rm Tr}\{[\hat{\Pi}(\alpha_{x_{i}=1}^{\lambda_{i}};s)\otimes\hat{\Pi}(\alpha_{x_{i+1}=1}^{\lambda_{i}};s)]\rho^{A_{i}A_{i+1}}\}\\&+2{\rm Tr}\{[\hat{\Pi}(\alpha_{x_{i}=1}^{\lambda_{i}};s)\otimes I]\rho^{A_{i}A_{i+1}}\}+2{\rm Tr}\{[I\otimes\hat{\Pi}(\alpha_{x_{i+1}=1}^{\lambda_{i}};s)]\rho^{A_{i}A_{i+1}}\}-1 \\=&\pi^{2}(1-s)^{2}[Q_{\rho^{A_{i}A_{i+1}}}(\alpha_{x_{i}=0}^{\lambda_{i}},\alpha_{x_{i+1}=1}^{\lambda_{i}};s)-Q_{\rho^{A_{i}A_{i+1}}}(\alpha_{x_{i}=1}^{\lambda_{i}},\alpha_{x_{i+1}=1}^{\lambda_{i}};s)]
\\&-\pi(1-s)[Q_{\rho^{A_{i}A_{i+1}}}(\alpha_{x_{i}=0}^{\lambda_{i}};s)-Q_{\rho^{A_{i}A_{i+1}}}(\alpha_{x_{i}=1}^{\lambda_{i}};s)]
\\=&\pi^{2}(1-s)^{2}C_{\rho^{A_{i}A_{i+1}}}^{-}(\alpha_{x_{i}=0}^{\lambda_{i}},\alpha_{x_{i+1}=1}^{\lambda_{i}},\alpha_{x_{i}=1}^{\lambda_{i}},\alpha_{x_{i+1}=1}^{\lambda_{i}};s)
\\&-\pi(1-s)D_{\rho^{A_{i}A_{i+1}}}^{-}(\alpha_{x_{i}=0}^{\lambda_{i}},\alpha_{x_{i+1}=1}^{\lambda_{i}},
\alpha_{x_{i}=1}^{\lambda_{i}},\alpha_{x_{i+1}=1}^{\lambda_{i}};s)
\end{aligned}
\end{equation}
and
\begin{equation}\tag{B12}\label{eqB12}
\begin{aligned}
&{\rm Tr}\{[\hat{O}(\alpha_{x_{j}=1}^{\lambda_{j}};s)\otimes(\hat{O}(\alpha_{x_{j+1}=0}^{\lambda_{j}};s)-\hat{O}(\alpha_{x_{j+1}=1}^{\lambda_{j}};s))]\rho^{A_{j}A_{j+1}}\}
\\=&\pi^{2}(1-s)^{2}C_{\rho^{A_{j}A_{j+1}}}^{-}(\alpha_{x_{j}=1}^{\lambda_{j}},\alpha_{x_{j+1}=0}^{\lambda_{j}},\alpha_{x_{j}=1}^{\lambda_{j}},\alpha_{x_{j+1}=1}^{\lambda_{j}};s)
\\&-\pi(1-s)D_{\rho^{A_{j}A_{j+1}}}^{-}(\alpha_{x_{j}=1}^{\lambda_{j}},\alpha_{x_{j+1}=0}^{\lambda_{j}},\alpha_{x_{j}=1}^{\lambda_{j}},\alpha_{x_{j+1}=1}^{\lambda_{j}};s).
\end{aligned}
\end{equation}
Substituting Eqs.(\ref{eqB9})-(\ref{eqB12}) into $\mathcal{J}_{s}$ in Eq.(\ref{eqB8}) derives  the desired expression of $\mathcal{J}_{s}$ in Theorem 4.1, completing the proof.
\hfill$\Box$
\end{widetext}

\section*{appendix C: Proof of Theorem 5.1}

For a star network ${\mathcal S}(y)$  as in Figure \ref{fig4}, it is clear that the number of maximal independent parties  $k_{\rm max}=y-1$, and the corresponding index set of maximal set of  independent parties $\mathcal{K}=\mathcal{K}_{\rm max}=\{1,2,\cdots,y-1\}$. By Eqs.(\ref{eq2.3}) and (\ref{eq3.14}), we have
\begin{widetext}
\begin{equation}\tag{C1}\label{eqC1}
\begin{aligned}
\mathcal{I}_{s}=&\frac{1}{2^{y-1}}\langle(M_{x_{1}=0}+M_{x_{1}=1})(M_{x_{2}=0}+M_{x_{2}=1})\cdot\cdot\cdot (M_{x_{y-1}=0}+M_{x_{y-1}=1})M_{x_{y}=0}\rangle\\=&\frac{1}{2^{y-1}}{\rm Tr}\{[(\hat{O}(\alpha_{x_{1}=0}^{\lambda_{1}};s)+\hat{O}(\alpha_{x_{1}=1}^{\lambda_{1}};s))\otimes\cdot\cdot\cdot\otimes(\hat{O}(\alpha_{x_{y-1}=0}^{\lambda_{y-1}};s)+\hat{O}(\alpha_{x_{y-1}=1}^{\lambda_{y-1}};s))\\&\otimes\hat{O}(\alpha_{x_{y}=0}^{\lambda_{1}};s)\otimes\hat{O}(\alpha_{x_{y}=0}^{\lambda_{2}};s)\otimes\cdot\cdot\cdot\otimes\hat{O}(\alpha_{x_{y}=0}^{\lambda_{y-1}};s)](\rho^{A_{1}A_{y}}\otimes\cdot\cdot\cdot\otimes\rho^{A_{y-1}A_{y}})\}\\
=&\frac{1}{2^{y-1}}\times\mathop{\prod}\limits_{j\in\mathcal{K}}\{{\rm Tr}[((\hat{O}(\alpha_{x_{j}=0}^{\lambda_{j}};s)+\hat{O}(\alpha_{x_{j}=1}^{\lambda_{j}};s))\otimes\hat{O}(\alpha_{x_{y}=0}^{\lambda_{j}};s))\rho^{A_{j}A_{y}}]\}.
\end{aligned}
\end{equation}

Analogous to Eqs.(\ref{eqB4}) and \ref{eqB6}), when $-1<s\leq0$,
\begin{equation}\tag{C2}\label{eqC2}
\begin{aligned}
&{\rm Tr}\{[(\hat{O}(\alpha_{x_{j}=0}^{\lambda_{j}};s)+\hat{O}(\alpha_{x_{j}=1}^{\lambda_{j}};s))\otimes\hat{O}(\alpha_{x_{y}=0}^{\lambda_{j}};s)]\rho^{A_{j}A_{y}}\}
\\=&\frac{\pi^{2}(1-s)^{4}}{4}C_{\rho^{A_{j}A_{y}}}^{+}(\alpha_{x_{j}=0}^{\lambda_{j}},\alpha_{x_{y}=0}^{\lambda_{j}},\alpha_{x_{j}=1}^{\lambda_{j}},\alpha_{x_{y}=0}^{\lambda_{j}};s)+\frac{\pi s(1-s)^{2}}{2}D_{\rho^{A_{j}A_{y}}}^{+}\\&(\alpha_{x_{j}=0}^{\lambda_{j}},\alpha_{x_{y}=0}^{\lambda_{j}},\alpha_{x_{j}=1}^{\lambda_{j}},\alpha_{x_{y}=0}^{\lambda_{j}};s)+2s^{2}
\end{aligned}
\end{equation}
and  when $s\leq-1$,
\begin{equation}\tag{C3}\label{eqC3}
\begin{aligned}
&{\rm Tr}\{[(\hat{O}(\alpha_{x_{j}=0}^{\lambda_{j}};s)+\hat{O}(\alpha_{x_{j}=1}^{\lambda_{j}};s))\otimes\hat{O}(\alpha_{x_{y}=0}^{\lambda_{j}};s)]\rho^{A_{j}A_{y}}\}
\\=&\pi^{2}(1-s)^{2}C_{\rho^{A_{j}A_{y}}}^{+}(\alpha_{x_{j}=0}^{\lambda_{j}},\alpha_{x_{y}=0}^{\lambda_{j}},\alpha_{x_{j}=1}^{\lambda_{j}},\alpha_{x_{y}=0}^{\lambda_{j}};s)
-\pi(1-s)\\&D_{\rho^{A_{j}A_{y}}}^{+}(\alpha_{x_{j}=0}^{\lambda_{j}},\alpha_{x_{y}=0}^{\lambda_{j}},\alpha_{x_{j}=1}^{\lambda_{j}},\alpha_{x_{y}=0}^{\lambda_{j}};s)+2.
\end{aligned}
\end{equation}
Substitute Eqs.(\ref{eqC2}) and (\ref{eqC3}) into  Eq.(\ref{eqC1}) to obtain the desired expression of  $\mathcal{I}_{s}$ as in Theorem 5.1.

Additionally, by Eqs.(\ref{eq2.4}) and (\ref{eq3.14}),
\begin{equation}\tag{C4}\label{eqC4}
\begin{aligned}
\mathcal{J}_{s}=&\frac{1}{2^{y-1}}\langle(M_{x_{1}=0}-M_{x_{1}=1})(M_{x_{2}=0}-M_{x_{2}=1})\cdot\cdot\cdot (M_{x_{y-1}=0}-M_{x_{y-1}=1})M_{x_{y}=1}\rangle\\=&\frac{1}{2^{y-1}}{\rm Tr}\{[(\hat{O}(\alpha_{x_{1}=0}^{\lambda_{1}};s)-\hat{O}(\alpha_{x_{1}=1}^{\lambda_{1}};s))\otimes\cdot\cdot\cdot\otimes(\hat{O}(\alpha_{x_{y-1}=0}^{\lambda_{y-1}};s)-\hat{O}(\alpha_{x_{y-1}=1}^{\lambda_{y-1}};s))\\&\otimes\hat{O}(\alpha_{x_{y}=1}^{\lambda_{1}};s)\otimes\hat{O}(\alpha_{x_{y}=1}^{\lambda_{2}};s)\otimes\cdot\cdot\cdot\otimes\hat{O}(\alpha_{x_{y}=1}^{\lambda_{y-1}};s)](\rho^{A_{1}A_{y}}\otimes\cdot\cdot\cdot\otimes\rho^{A_{y-1}A_{y}})\}\\
=&\frac{1}{2^{y-1}}\times\mathop{\prod}\limits_{j\in\mathcal{K}}\{{\rm Tr}[((\hat{O}(\alpha_{x_{j}=0}^{\lambda_{j}};s)-\hat{O}(\alpha_{x_{j}=1}^{\lambda_{j}};s))\otimes\hat{O}(\alpha_{x_{y}=1}^{\lambda_{j}};s))\rho^{A_{j}A_{y}}]\}.
\end{aligned}
\end{equation}
Analogous to the cases in Eqs.(\ref{eqB9}) and (\ref{eqB11}), if $-1<s\leq0$, we get
\begin{equation}\tag{C5}\label{eqC5}
\begin{aligned}
&{\rm Tr}[((\hat{O}(\alpha_{x_{j}=0}^{\lambda_{j}};s)-\hat{O}(\alpha_{x_{j}=1}^{\lambda_{j}};s))\otimes\hat{O}(\alpha_{x_{y}=1}^{\lambda_{j}};s))\rho^{A_{j}A_{y}}]
\\=&\frac{\pi^{2}(1-s)^{4}}{4}C_{\rho^{A_{j}A_{y}}}^{-}(\alpha_{x_{j}=0}^{\lambda_{j}},\alpha_{x_{y}=1}^{\lambda_{j}},\alpha_{x_{j}=1}^{\lambda_{j}},\alpha_{x_{y}=1}^{\lambda_{j}};s)\\&+\frac{\pi s(1-s)^{2}}{2}D_{\rho^{A_{j}A_{y}}}^{-}(\alpha_{x_{j}=0}^{\lambda_{j}},\alpha_{x_{y}=1}^{\lambda_{j}},\alpha_{x_{j}=1}^{\lambda_{j}},\alpha_{x_{y}=1}^{\lambda_{j}};s)
\end{aligned}
\end{equation}
and if $s\leq-1$, we get
\begin{equation}\tag{C6}\label{eqC6}
\begin{aligned}
&{\rm Tr}[((\hat{O}(\alpha_{x_{j}=0}^{\lambda_{j}};s)-\hat{O}(\alpha_{x_{j}=1}^{\lambda_{j}};s))\otimes\hat{O}(\alpha_{x_{y}=1}^{\lambda_{j}};s))\rho^{A_{j}A_{y}}]
\\=&\pi^{2}(1-s)^{2}C_{\rho^{A_{j}A_{y}}}^{-}(\alpha_{x_{j}=0}^{\lambda_{j}},\alpha_{x_{y}=1}^{\lambda_{j}},\alpha_{x_{j}=1}^{\lambda_{j}},\alpha_{x_{y}=1}^{\lambda_{j}};s)
\\&-\pi(1-s)D_{\rho^{A_{j}A_{y}}}^{-}(\alpha_{x_{j}=0}^{\lambda_{j}},\alpha_{x_{y}=1}^{\lambda_{j}},\alpha_{x_{j}=1}^{\lambda_{j}},\alpha_{x_{y}=1}^{\lambda_{j}};s).
\end{aligned}
\end{equation}
Substitute Eqs.(\ref{eqC5}) and (\ref{eqC6}) into Eq.(\ref{eqC4}) gives the desired expression of  $\mathcal{J}_{s}$ in Theorem 5.1.
\hfill$\Box$
\end{widetext}

\section*{appendix D: Proof of Theorem 6.1}

In tree-shaped network $\mathcal{T}(m,f)$, it is clear that (1) if $m$ is odd, then $k_{\rm max}=\frac{f^{m+1}-1}{f^{2}-1}$ with the index set of  maximal set of independent parties ${\mathcal K}={\mathcal K}_{\rm max}=\{1,\frac{1-f^{t-1}}{1-f}+1, \frac{1-f^{t-1}}{1-f}+2, \cdots, \frac{1-f^{t}}{1-f}\}_{t=3,5,\cdot\cdot\cdot,m}$; (2) if $m$ is even, then $k_{\rm max}=\frac{f-f^{m+1}}{1-f^{2}}$ with the index set of  maximal set of independent parties ${\mathcal K}=\mathcal K_{\max}=\{\frac{1-f^{t-1}}{1-f}+1, \frac{1-f^{t-1}}{1-f}+2, \cdot\cdot\cdot, \frac{1-f^{t}}{1-f}\}_{t=2,4,\cdot\cdot\cdot,m}$. Denote by $\overline{\mathcal K}=\{1,2,\ldots \frac{1-f^{m}}{1-f}\}\setminus {\mathcal K}$ and ${\mathcal K}'$   the set of indexes from ${\mathcal K}$ with the last layer of indexes excluded. For example, when $m$ is odd, ${\mathcal K}'=\{1,\frac{1-f^{t-1}}{1-f}+1, \frac{1-f^{t-1}}{1-f}+2, \cdot\cdot\cdot, \frac{1-f^{t}}{1-f}\}_{t=3,5,\cdot\cdot\cdot,m-2}$. By Eqs.(\ref{eq2.3}) and (\ref{eq3.14}), we have
\begin{widetext}
\begin{equation}\tag{D1}\label{eqD1}
\begin{aligned}
\mathcal{I}_{s}=&\frac{1}{2^{k_{\rm max}}}\langle\mathop{\prod}\limits_{i\in\mathcal{K}}(M_{x_{i}=0}+M_{x_{i}=1})\mathop{\prod}\limits_{j\in\overline{\mathcal{K}}} M_{x_{j}=0}\rangle\\=&\frac{1}{2^{k_{\rm max}}}\times\mathop{\prod}\limits_{j=(i-1)f+2}^{if+1}\mathop{\prod}\limits_{i\in\mathcal{K}^{'}}{\rm Tr}\{[(\hat{O}(\alpha_{x_{i}=0}^{\lambda_{j-1}};s)+\hat{O}(\alpha_{x_{i}=1}^{\lambda_{j-1}};s))\otimes\hat{O}(\alpha_{x_{j}=0}^{\lambda_{j-1}};s)]\rho^{A_{i}A_{j}}\}\\
&\times\mathop{\prod}\limits_{q=(p-1)f+2}^{pf+1}\mathop{\prod}\limits_{p\in\overline{\mathcal{K}}}{\rm Tr}\{[\hat{O}(\alpha_{x_{p}=0}^{\lambda_{q-1}};s)\otimes(\hat{O}(\alpha_{x_{q}=0}^{\lambda_{q-1}};s)+\hat{O}(\alpha_{x_{q}=1}^{\lambda_{q-1}};s))]\rho^{A_{p}A_{q}}\}.
\end{aligned}
\end{equation}

Analogous to Eqs.(\ref{eqB4})-(\ref{eqB7}), when $-1<s\leq0$,
\begin{equation}\tag{D2}\label{eqD2}
\begin{aligned}
&{\rm Tr}\{[(\hat{O}(\alpha_{x_{i}=0}^{\lambda_{j-1}};s)+\hat{O}(\alpha_{x_{i}=1}^{\lambda_{j-1}};s))\otimes\hat{O}(\alpha_{x_{j}=0}^{\lambda_{j-1}};s)]\rho^{A_{i}A_{j}}\}
\\=&\frac{\pi^{2}(1-s)^{4}}{4}C_{\rho^{A_{i}A_{j}}}^{+}(\alpha_{x_{i}=0}^{\lambda_{j-1}},\alpha_{x_{j}=0}^{\lambda_{j-1}},\alpha_{x_{i}=1}^{\lambda_{j-1}},\alpha_{x_{j}=0}^{\lambda_{j-1}};s)+\frac{\pi s(1-s)^{2}}{2}\\&D_{\rho^{A_{i}A_{j}}}^{+}(\alpha_{x_{i}=0}^{\lambda_{j-1}},\alpha_{x_{j}=0}^{\lambda_{j-1}},\alpha_{x_{i}=1}^{\lambda_{j-1}},\alpha_{x_{j}=0}^{\lambda_{j-1}};s)+2s^{2}
\end{aligned}
\end{equation}
and
\begin{equation}\tag{D3}\label{eqD3}
\begin{aligned}
&{\rm Tr}\{[\hat{O}(\alpha_{x_{p}=0}^{\lambda_{q-1}};s)\otimes(\hat{O}(\alpha_{x_{q}=0}^{\lambda_{q-1}};s)+\hat{O}(\alpha_{x_{q}=1}^{\lambda_{q-1}};s))]\rho^{A_{p}A_{q}}\}
\\=&\frac{\pi^{2}(1-s)^{4}}{4}C_{\rho^{A_{p}A_{q}}}^{+}(\alpha_{x_{p}=0}^{\lambda_{q-1}},\alpha_{x_{q}=0}^{\lambda_{q-1}},\alpha_{x_{p}=0}^{\lambda_{q-1}},\alpha_{x_{q}=1}^{\lambda_{q-1}};s)
+\frac{\pi s(1-s)^{2}}{2}\\&D_{\rho^{A_{p}A_{q}}}^{+}(\alpha_{x_{p}=0}^{\lambda_{q-1}},\alpha_{x_{q}=0}^{\lambda_{q-1}},\alpha_{x_{p}=0}^{\lambda_{q-1}},\alpha_{x_{q}=1}^{\lambda_{q-1}};s)+2s^{2};
\end{aligned}
\end{equation}
while for $s\leq-1$,
\begin{equation}\tag{D4}\label{eqD4}
\begin{aligned}
&{\rm Tr}\{[(\hat{O}(\alpha_{x_{i}=0}^{\lambda_{j-1}};s)+\hat{O}(\alpha_{x_{i}=1}^{\lambda_{j-1}};s))\otimes\hat{O}(\alpha_{x_{j}=0}^{\lambda_{j-1}};s)]\rho^{A_{i}A_{j}}\}
\\=&\pi^{2}(1-s)^{2}C_{\rho^{A_{i}A_{j}}}^{+}(\alpha_{x_{i}=0}^{\lambda_{j-1}},\alpha_{x_{j}=0}^{\lambda_{j-1}},\alpha_{x_{i}=1}^{\lambda_{j-1}},\alpha_{x_{j}=0}^{\lambda_{j-1}};s)
-\pi(1-s)\\&D_{\rho^{A_{i}A_{j}}}^{+}(\alpha_{x_{i}=0}^{\lambda_{j-1}},\alpha_{x_{j}=0}^{\lambda_{j-1}},
\alpha_{x_{i}=1}^{\lambda_{j-1}},\alpha_{x_{j}=0}^{\lambda_{j-1}};s)+2
\end{aligned}
\end{equation}
and
\begin{equation}\tag{D5}\label{eqD5}
\begin{aligned}
&{\rm Tr}\{[\hat{O}(\alpha_{x_{p}=0}^{\lambda_{q-1}};s)\otimes(\hat{O}(\alpha_{x_{q}=0}^{\lambda_{q-1}};s)+\hat{O}(\alpha_{x_{q}=1}^{\lambda_{q-1}};s))]\rho^{A_{p}A_{q}}\}
\\=&\pi^{2}(1-s)^{2}C_{\rho^{A_{p}A_{q}}}^{+}(\alpha_{x_{p}=0}^{\lambda_{q-1}},\alpha_{x_{q}=0}^{\lambda_{q-1}},\alpha_{x_{p}=0}^{\lambda_{q-1}},\alpha_{x_{q}=1}^{\lambda_{q-1}};s)
-\pi(1-s)\\&D_{\rho^{A_{p}A_{q}}}^{+}(\alpha_{x_{p}=0}^{\lambda_{q-1}},\alpha_{x_{q}=0}^{\lambda_{q-1}},\alpha_{x_{p}=0}^{\lambda_{q-1}},\alpha_{x_{q}=1}^{\lambda_{q-1}};s)+2.
\end{aligned}
\end{equation}
Substitute Eqs.(\ref{eqD2})-(\ref{eqD5}) into Eq.(\ref{eqD1}) one obtains the desired expression of  $\mathcal{I}_{s}$ in Theorem 6.1.

Similarly,
\begin{equation}\tag{D6}\label{eqD6}
\begin{aligned}
\mathcal{J}_{s}=&\frac{1}{2^{k_{\rm max}}}\langle\mathop{\prod}\limits_{i\in\mathcal{K}}(M_{x_{i}=0}-M_{x_{i}=1})\mathop{\prod}\limits_{j\in\overline{\mathcal{K}}} M_{x_{j}=1}\rangle\\=&\frac{1}{2^{k_{\rm max}}}\times\mathop{\prod}\limits_{j=(i-1)f+2}^{if+1}\mathop{\prod}\limits_{i\in\mathcal{K}^{'}}{\rm Tr}\{[(\hat{O}(\alpha_{x_{i}=0}^{\lambda_{j-1}};s)-\hat{O}(\alpha_{x_{i}=1}^{\lambda_{j-1}};s))\otimes\hat{O}(\alpha_{x_{j}=1}^{\lambda_{j-1}};s)]\rho^{A_{i}A_{j}}\}\\
&\times\mathop{\prod}\limits_{q=(p-1)f+2}^{pf+1}\mathop{\prod}\limits_{p\in\overline{\mathcal{K}}}{\rm Tr}\{[\hat{O}(\alpha_{x_{p}=1}^{\lambda_{q-1}};s)\otimes(\hat{O}(\alpha_{x_{q}=0}^{\lambda_{q-1}};s)-\hat{O}(\alpha_{x_{q}=1}^{\lambda_{q-1}};s))]\rho^{A_{p}A_{q}}\}.
\end{aligned}
\end{equation}
Analogous to the cases in Eqs.(\ref{eqB9})-(\ref{eqB12}), when $-1<s\leq0$,
\begin{equation}\tag{D7}\label{eqD7}
\begin{aligned}
&{\rm Tr}\{[(\hat{O}(\alpha_{x_{i}=0}^{\lambda_{j-1}};s)-\hat{O}(\alpha_{x_{i}=1}^{\lambda_{j-1}};s))\otimes\hat{O}(\alpha_{x_{j}=1}^{\lambda_{j-1}};s)]\rho^{A_{i}A_{j}}\}
\\=&\frac{\pi^{2}(1-s)^{4}}{4}C_{\rho^{A_{i}A_{j}}}^{-}(\alpha_{x_{i}=0}^{\lambda_{j-1}},\alpha_{x_{j}=1}^{\lambda_{j-1}},\alpha_{x_{i}=1}^{\lambda_{j-1}},\alpha_{x_{j}=1}^{\lambda_{j-1}};s)\\&+\frac{\pi s(1-s)^{2}}{2}D_{\rho^{A_{i}A_{j}}}^{-}(\alpha_{x_{i}=0}^{\lambda_{j-1}},\alpha_{x_{j}=1}^{\lambda_{j-1}},\alpha_{x_{i}=1}^{\lambda_{j-1}},\alpha_{x_{j}=1}^{\lambda_{j-1}};s)
\end{aligned}
\end{equation}
and
\begin{equation}\tag{D8}\label{eqD8}
\begin{aligned}
&{\rm Tr}\{[\hat{O}(\alpha_{x_{p}=1}^{\lambda_{q-1}};s)\otimes(\hat{O}(\alpha_{x_{q}=0}^{\lambda_{q-1}};s)-\hat{O}(\alpha_{x_{q}=1}^{\lambda_{q-1}};s))]\rho^{A_{p}A_{q}}\}
\\=&\frac{\pi^{2}(1-s)^{4}}{4}C_{\rho^{A_{p}A_{q}}}^{-}(\alpha_{x_{p}=1}^{\lambda_{q-1}},\alpha_{x_{q}=0}^{\lambda_{q-1}},\alpha_{x_{p}=1}^{\lambda_{q-1}},\alpha_{x_{q}=1}^{\lambda_{q-1}};s)
\\&+\frac{\pi s(1-s)^{2}}{2}D_{\rho^{A_{p}A_{q}}}^{-}(\alpha_{x_{p}=1}^{\lambda_{q-1}},\alpha_{x_{q}=0}^{\lambda_{q-1}},\alpha_{x_{p}=1}^{\lambda_{q-1}},\alpha_{x_{q}=1}^{\lambda_{q-1}};s);
\end{aligned}
\end{equation}
while if $s\leq-1$,
\begin{equation}\tag{D9}\label{eqD9}
\begin{aligned}
&{\rm Tr}\{[(\hat{O}(\alpha_{x_{i}=0}^{\lambda_{j-1}};s)-\hat{O}(\alpha_{x_{i}=1}^{\lambda_{j-1}};s))\otimes\hat{O}(\alpha_{x_{j}=1}^{\lambda_{j-1}};s)]\rho^{A_{i}A_{j}}\}
\\=&\pi^{2}(1-s)^{2}C_{\rho^{A_{i}A_{j}}}^{-}(\alpha_{x_{i}=0}^{\lambda_{j-1}},\alpha_{x_{j}=1}^{\lambda_{j-1}},\alpha_{x_{i}=1}^{\lambda_{j-1}},\alpha_{x_{j}=1}^{\lambda_{j-1}};s)
\\&-\pi(1-s)D_{\rho^{A_{i}A_{j}}}^{-}(\alpha_{x_{i}=0}^{\lambda_{j-1}},\alpha_{x_{j}=1}^{\lambda_{j-1}},\alpha_{x_{i}=1}^{\lambda_{j-1}},\alpha_{x_{j}=1}^{\lambda_{j-1}};s)
\end{aligned}
\end{equation}
and
\begin{equation}\tag{D10}\label{eqD10}
\begin{aligned}
&{\rm Tr}\{[\hat{O}(\alpha_{x_{p}=1}^{\lambda_{q-1}};s)\otimes(\hat{O}(\alpha_{x_{q}=0}^{\lambda_{q-1}};s)-\hat{O}(\alpha_{x_{q}=1}^{\lambda_{q-1}};s))]\rho^{A_{p}A_{q}}\}
\\=&\pi^{2}(1-s)^{2}C_{\rho^{A_{p}A_{q}}}^{-}(\alpha_{x_{p}=1}^{\lambda_{q-1}},\alpha_{x_{q}=0}^{\lambda_{q-1}},\alpha_{x_{p}=1}^{\lambda_{q-1}},\alpha_{x_{q}=1}^{\lambda_{q-1}};s)
\\&-\pi(1-s)D_{\rho^{A_{p}A_{q}}}^{-}(\alpha_{x_{p}=1}^{\lambda_{q-1}},\alpha_{x_{q}=0}^{\lambda_{q-1}},\alpha_{x_{p}=1}^{\lambda_{q-1}},\alpha_{x_{q}=1}^{\lambda_{q-1}};s).
\end{aligned}
\end{equation}
Substituting Eqs.(\ref{eqD7})-(\ref{eqD10}) into Eq.(\ref{eqD6})  obtains the desired expression of $\mathcal{J}_{s}$ in Theorem 6.1.
\hfill$\Box$
\end{widetext}

\section*{appendix E: Proofs of Theorems 7.1 and 7.2}

{\it Proof of Theorem 7.1.} In a cyclic network $\mathcal{C}yc(y)$ (see Figure \ref{fig10}), there are $y$ parties
$A_{1}$, $A_{2}$, $\cdot\cdot\cdot$, $A_{y}$ with $A_{i}$ and $A_{i+1}$ shares a source $S_{i}$ for $i=1,2, \cdot\cdot\cdot, y-1$, $A_{y}$ and $A_{1}$ share a source $S_{y}$. It is clear that  if $y$ is odd, then $k_{\rm max}=\frac{y-1}{2}$ with the index set of  maximal set of independent parties ${\mathcal K}=\mathcal K_{\max}=\{1,3, 5, \cdot\cdot\cdot, y-2\}$. Let $\overline{\mathcal K}=\{1,2,\cdots, y\}\setminus {\mathcal K}$ and $\overline{{\mathcal K}}'=\overline{{\mathcal K}}\backslash \{y-1\}$, then, by Eqs.(\ref{eq2.3}) and (\ref{eq3.14}),
 \begin{widetext}
 \begin{equation}\tag{E1}\label{eqE1}
\begin{aligned}
\mathcal{I}_{s}=&\frac{1}{2^{\frac{y-1}{2}}}\langle(M_{x_{1}=0}+M_{x_{1}=1})M_{x_{2}=0}\cdot\cdot\cdot (M_{x_{y-2}=0}+M_{x_{y-2}=1})M_{x_{y-1}=0}M_{x_{y}=0}\rangle\\=&\frac{1}{2^{\frac{y-1}{2}}}{\rm Tr}\{[((\hat{O}(\alpha_{x_{1}=0}^{\lambda_{y}};s)\otimes\hat{O}(\alpha_{x_{1}=0}^{\lambda_{1}};s))
+(\hat{O}(\alpha_{x_{1}=1}^{\lambda_{y}};s)\otimes\hat{O}(\alpha_{x_{1}=1}^{\lambda_{1}};s)))\\&\otimes(\hat{O}(\alpha_{x_{2}=0}^{\lambda_{1}};s)\otimes\hat{O}(\alpha_{x_{2}=0}^{\lambda_{2}};s))\otimes\cdot\cdot\cdot\otimes((\hat{O}(\alpha_{x_{y-2}=0}^{\lambda_{y-3}};s)\otimes\hat{O}(\alpha_{x_{y-2}=0}^{\lambda_{y-2}};s))
\\&+(\hat{O}(\alpha_{x_{y-2}=1}^{\lambda_{y-3}};s)\otimes\hat{O}(\alpha_{x_{y-2}=1}^{\lambda_{y-2}};s)))\otimes(\hat{O}(\alpha_{x_{y-1}=0}^{\lambda_{y-2}};s)\otimes\hat{O}(\alpha_{x_{y-1}=0}^{\lambda_{y-1}};s))\\&\otimes(\hat{O}(\alpha_{x_{y}=0}^{\lambda_{y-1}};s)\otimes\hat{O}(\alpha_{x_{y}=0}^{\lambda_{y}};s))](\rho^{A_{1}A_{2}}\otimes\cdot\cdot\cdot\otimes\rho^{A_{y-1}A_{y}}\otimes\rho^{A_{y}A_{1}})\}\\=&\frac{1}{2^{\frac{y-1}{2}}}\times{\rm Tr}\{[(\hat{O}(\alpha_{x_{1}=0}^{\lambda_{1}};s)+\hat{O}(\alpha_{x_{1}=1}^{\lambda_{1}};s))\otimes\hat{O}(\alpha_{x_{2}=0}^{\lambda_{1}};s)]\rho^{A_{1}A_{2}}\}\\&\times{\rm Tr}\{[\hat{O}(\alpha_{x_{2}=0}^{\lambda_{2}};s)\otimes(\hat{O}(\alpha_{x_{3}=0}^{\lambda_{2}};s)+\hat{O}(\alpha_{x_{3}=1}^{\lambda_{2}};s))]\rho^{A_{2}A_{3}}\}\\&\times\cdot\cdot\cdot\times
{\rm Tr}\{[(\hat{O}(\alpha_{x_{y-2}=0}^{\lambda_{y-2}};s)+\hat{O}(\alpha_{x_{y-2}=1}^{\lambda_{y-2}};s))\otimes\hat{O}(\alpha_{x_{y-1}=0}^{\lambda_{y-2}};s)]\rho^{A_{y-2}A_{y-1}}\}\\&\times{\rm Tr}\{[\hat{O}(\alpha_{x_{y-1}=0}^{\lambda_{y-1}};s)\otimes\hat{O}(\alpha_{x_{y}=0}^{\lambda_{y-1}};s)]\rho^{A_{y-1}A_{y}}\}\\&\times{\rm Tr}\{[\hat{O}(\alpha_{x_{y}=0}^{\lambda_{y}};s)\otimes(\hat{O}(\alpha_{x_{1}=0}^{\lambda_{y}};s)+\hat{O}(\alpha_{x_{1}=1}^{\lambda_{y}};s))]\rho^{A_{y}A_{1}}\}\\
=&\frac{1}{2^{\frac{y-1}{2}}}\times\mathop{\prod}\limits_{i\in\mathcal{K}}{\rm Tr}\{[((\hat{O}(\alpha_{x_{i}=0}^{\lambda_{i}};s)+\hat{O}(\alpha_{x_{i}=1}^{\lambda_{i}};s))\otimes\hat{O}(\alpha_{x_{i+1}=0}^{\lambda_{i}};s)]\rho^{A_{i}A_{i+1}}\}\\&\times\mathop{\prod}\limits_{j\in\overline{\mathcal{K}}^{'}}{\rm Tr}\{[\hat{O}(\alpha_{x_{j}=0}^{\lambda_{j}};s)\otimes(\hat{O}(\alpha_{x_{j+1}=0}^{\lambda_{j}};s)+\hat{O}(\alpha_{x_{j+1}=1}^{\lambda_{j}};s))]\rho^{A_{j}A_{j+1}}\}\\&\times{\rm Tr}\{[\hat{O}(\alpha_{x_{y-1}=0}^{\lambda_{y-1}};s)\otimes\hat{O}(\alpha_{x_{y}=0}^{\lambda_{y-1}};s)]
\rho^{A_{y-1}A_{y}}\}.
\end{aligned}
\end{equation}
\if false where ${\mathcal K}=\{1,3, 5, \cdot\cdot\cdot, y-2\}$, $\overline{{\mathcal K}}=\{2,4, 6, \cdot\cdot\cdot, y-3,y-1,y\}$ and $\overline{{\mathcal K}}'=\overline{{\mathcal K}}\backslash \{y-1\}$. \fi Additionally, in Eq.(\ref{eqE1}), $y+1$ actually represents 1 due to the cyclic nature of the network.

 By Eqs.(\ref{eq3.3}), (\ref{eq3.15}), (\ref{eq3.16}) and (\ref{eq3.18}), if $-1<s\leq0$, then
\begin{equation}\tag{E2}\label{eqE2}
\begin{aligned}
&{\rm Tr}\{[\hat{O}(\alpha_{x_{y-1}=0}^{\lambda_{y-1}};s)\otimes\hat{O}(\alpha_{x_{y}=0}^{\lambda_{y-1}};s)]\rho^{A_{y-1}A_{y}}\}\\=&{\rm Tr}\{[((1-s)\hat{\Pi}(\alpha_{x_{y-1}=0}^{\lambda_{y-1}};s)+sI)\otimes((1-s)\hat{\Pi}(\alpha_{x_{y}=0}^{\lambda_{y-1}};s)+sI)]\rho^{A_{y-1}A_{y}}\}\\
=&(1-s)^{2}{\rm Tr}\{[\hat{\Pi}(\alpha_{x_{y-1}=0}^{\lambda_{y-1}};s)\otimes\hat{\Pi}(\alpha_{x_{y}=0}^{\lambda_{y-1}};s)]\rho^{A_{y-1}A_{y}}\}+s(1-s)\\&{\rm Tr}\{[\hat{\Pi}(\alpha_{x_{y-1}=0}^{\lambda_{y-1}};s)\otimes I]\rho^{A_{y-1}A_{y}}\}+s(1-s){\rm Tr}\{[I\otimes\hat{\Pi}(\alpha_{x_{y}=0}^{\lambda_{y-1}};s)]\rho^{A_{y-1}A_{y}}\}+s^{2}\\=&\frac{\pi^{2}(1-s)^{4}}{4}Q_{\rho^{A_{y-1}A_{y}}}(\alpha_{x_{y-1}=0}^{\lambda_{y-1}},\alpha_{x_{y}=0}^{\lambda_{y-1}};s)
+\frac{\pi s(1-s)^{2}}{2}[Q_{\rho^{A_{y-1}A_{y}}}(\alpha_{x_{y-1}=0}^{\lambda_{y-1}};s)\\&+Q_{\rho^{A_{y-1}A_{y}}}(\alpha_{x_{y}=0}^{\lambda_{y-1}};s)]+s^{2}
\\=&\frac{\pi^{2}(1-s)^{4}}{8}C_{\rho^{A_{y-1}A_{y}}}^{+}(\alpha_{x_{y-1}=0}^{\lambda_{y-1}},\alpha_{x_{y}=0}^{\lambda_{y-1}},\alpha_{x_{y-1}=0}^{\lambda_{y-1}},\alpha_{x_{y}=0}^{\lambda_{y-1}};s)
\\&+\frac{\pi s(1-s)^{2}}{4}D_{\rho^{A_{y-1}A_{y}}}^{+}(\alpha_{x_{y-1}=0}^{\lambda_{y-1}},\alpha_{x_{y}=0}^{\lambda_{y-1}},\alpha_{x_{y-1}=0}^{\lambda_{y-1}},\alpha_{x_{y}=0}^{\lambda_{y-1}};s)+s^{2};
\end{aligned}
\end{equation}
while if $s\leq-1$, then
\begin{equation}\tag{E3}\label{eqE3}
\begin{aligned}
&{\rm Tr}\{[\hat{O}(\alpha_{x_{y-1}=0}^{\lambda_{y-1}};s)\otimes\hat{O}(\alpha_{x_{y}=0}^{\lambda_{y-1}};s)]\rho^{A_{y-1}A_{y}}\}\\=&{\rm Tr}\{[(2\hat{\Pi}(\alpha_{x_{y-1}=0}^{\lambda_{y-1}};s)-I)\otimes(2\hat{\Pi}(\alpha_{x_{y}=0}^{\lambda_{y-1}};s)-I)]\rho^{A_{y-1}A_{y}}\}\\
=&4{\rm Tr}\{[\hat{\Pi}(\alpha_{x_{y-1}=0}^{\lambda_{y-1}};s)\otimes\hat{\Pi}(\alpha_{x_{y}=0}^{\lambda_{y-1}};s)]\rho^{A_{y-1}A_{y}}\}-2{\rm Tr}\{[\hat{\Pi}(\alpha_{x_{y-1}=0}^{\lambda_{y-1}};s)\otimes I]\rho^{A_{y-1}A_{y}}\}\\&-2{\rm Tr}\{[I\otimes\hat{\Pi}(\alpha_{x_{y}=0}^{\lambda_{y-1}};s)]\rho^{A_{y-1}A_{y}}\}+1\\=&\pi^{2}(1-s)^{2}Q_{\rho^{A_{y-1}A_{y}}}(\alpha_{x_{y-1}=0}^{\lambda_{y-1}},\alpha_{x_{y}=0}^{\lambda_{y-1}};s)
-\pi(1-s)[Q_{\rho^{A_{y-1}A_{y}}}(\alpha_{x_{y-1}=0}^{\lambda_{y-1}};s)\\&+Q_{\rho^{A_{y-1}A_{y}}}(\alpha_{x_{y}=0}^{\lambda_{y-1}};s)]+1
\\=&\frac{\pi^{2}(1-s)^{2}}{2}C_{\rho^{A_{y-1}A_{y}}}^{+}(\alpha_{x_{y-1}=0}^{\lambda_{y-1}},\alpha_{x_{y}=0}^{\lambda_{y-1}},\alpha_{x_{y-1}=0}^{\lambda_{y-1}},\alpha_{x_{y}=0}^{\lambda_{y-1}};s)
\\&-\frac{\pi(1-s)}{2}D_{\rho^{A_{y-1}A_{y}}}^{+}(\alpha_{x_{y-1}=0}^{\lambda_{y-1}},\alpha_{x_{y}=0}^{\lambda_{y-1}},\alpha_{x_{y-1}=0}^{\lambda_{y-1}},\alpha_{x_{y}=0}^{\lambda_{y-1}};s)+1.
\end{aligned}
\end{equation}
Now substituting  Eqs.(\ref{eqB4})-(\ref{eqB7}) and  Eqs.(\ref{eqE2})-(\ref{eqE3}) into  Eq.(\ref{eqE1}) gives the desired expressions of $\mathcal{I}_{s}$ in Theorem 7.1.

Similarly, by Eqs.(\ref{eq2.4}) and (\ref{eq3.14}),
\begin{equation}\tag{E4}\label{eqE4}
\begin{aligned}
\mathcal{J}_{s}=&\frac{1}{2^{\frac{y-1}{2}}}\langle(M_{x_{1}=0}-M_{x_{1}=1})M_{x_{2}=1}\cdot\cdot\cdot (M_{x_{y-2}=0}-M_{x_{y-2}=1})M_{x_{y-1}=1}M_{x_{y}=1}\rangle\\
=&\frac{1}{2^{\frac{y-1}{2}}}\times\mathop{\prod}\limits_{i\in\mathcal{K}}{\rm Tr}\{[(\hat{O}(\alpha_{x_{i}=0}^{\lambda_{i}};s)-\hat{O}(\alpha_{x_{i}=1}^{\lambda_{i}};s))\otimes\hat{O}(\alpha_{x_{i+1}=1}^{\lambda_{i}};s)]\rho^{A_{i}A_{i+1}}\}\\&\times\mathop{\prod}\limits_{j\in\overline{\mathcal{K}}'}{\rm Tr}\{[\hat{O}(\alpha_{x_{j}=1}^{\lambda_{j}};s)\otimes(\hat{O}(\alpha_{x_{j+1}=0}^{\lambda_{j}};s)-\hat{O}(\alpha_{x_{j+1}=1}^{\lambda_{j}};s))]\rho^{A_{j}A_{j+1}}\}\\&\times{\rm Tr}\{[\hat{O}(\alpha_{x_{y-1}=1}^{\lambda_{y-1}};s)\otimes\hat{O}(\alpha_{x_{y}=1}^{\lambda_{y-1}};s)]\rho^{A_{y-1}A_{y}}\}.
\end{aligned}
\end{equation}
\if false where ${\mathcal K}=\{1,3, 5, \cdot\cdot\cdot, y-2\}$, $\overline{{\mathcal K}}=\{2,4, 6, \cdot\cdot\cdot, y-3,y-1,y\}$ and $\overline{{\mathcal K}}'=\overline{{\mathcal K}}\backslash \{y-1\}$. \fi Likewise, in Eq.(\ref{eqE4}), $y+1$ actually represents 1 due to the cyclic nature of the network.

Analogous to Eqs.(\ref{eqE2}) and (\ref{eqE3}), when $-1<s\leq0$,
\begin{equation}\tag{E5}\label{eqE5}
\begin{aligned}
&{\rm Tr}\{[\hat{O}(\alpha_{x_{y-1}=1}^{\lambda_{y-1}};s)\otimes\hat{O}(\alpha_{x_{y}=1}^{\lambda_{y-1}};s)]\rho^{A_{y-1}A_{y}}\}
\\=&\frac{\pi^{2}(1-s)^{4}}{8}C_{\rho^{A_{y-1}A_{y}}}^{+}(\alpha_{x_{y-1}=1}^{\lambda_{y-1}},\alpha_{x_{y}=1}^{\lambda_{y-1}},\alpha_{x_{y-1}=1}^{\lambda_{y-1}},\alpha_{x_{y}=1}^{\lambda_{y-1}};s)
\\&+\frac{\pi s(1-s)^{2}}{4}D_{\rho^{A_{y-1}A_{y}}}^{+}(\alpha_{x_{y-1}=1}^{\lambda_{y-1}},\alpha_{x_{y}=1}^{\lambda_{y-1}},\alpha_{x_{y-1}=1}^{\lambda_{y-1}},\alpha_{x_{y}=1}^{\lambda_{y-1}};s)+s^{2};
\end{aligned}
\end{equation}
while if $s\leq-1$,
\begin{equation}\tag{E6}\label{eqE6}
\begin{aligned}
&{\rm Tr}\{[\hat{O}(\alpha_{x_{y-1}=1}^{\lambda_{y-1}};s)\otimes\hat{O}(\alpha_{x_{y}=1}^{\lambda_{y-1}};s)]\rho^{A_{y-1}A_{y}}\}
\\=&\frac{\pi^{2}(1-s)^{2}}{2}C_{\rho^{A_{y-1}A_{y}}}^{+}(\alpha_{x_{y-1}=1}^{\lambda_{y-1}},\alpha_{x_{y}=1}^{\lambda_{y-1}},\alpha_{x_{y-1}=1}^{\lambda_{y-1}},\alpha_{x_{y}=1}^{\lambda_{y-1}};s)
\\&-\frac{\pi(1-s)}{2}D_{\rho^{A_{y-1}A_{y}}}^{+}(\alpha_{x_{y-1}=1}^{\lambda_{y-1}},\alpha_{x_{y}=1}^{\lambda_{y-1}},\alpha_{x_{y-1}=1}^{\lambda_{y-1}},\alpha_{x_{y}=1}^{\lambda_{y-1}};s)+1.
\end{aligned}
\end{equation}
Substitute  Eqs.(\ref{eqB9})-(\ref{eqB12}) and  Eqs.(\ref{eqE5})-(\ref{eqE6}) into  Eq.(\ref{eqE4}) one obtains the desired expressions of $\mathcal{J}_{s}$ in Theorem 7.1, completing the proof.
\hfill$\Box$

 {\it Proof of Theorem 7.2.} For cyclic network $\mathcal{C}yc(y)$, if $y$ is even, then $k_{\rm max}=\frac{y}{2}$ with the index set of  maximal set of independent parties ${\mathcal K}=\mathcal K_{\max}= \{2,4,6,\cdot\cdot\cdot, y\}$. Denote by $\overline{\mathcal K}=\{1,2,\cdots, y\}\setminus {\mathcal K}$.  By Eqs.(\ref{eq2.3}) and (\ref{eq3.14}),
\begin{equation}\tag{E7}\label{eqE7}
\begin{aligned}
\mathcal{I}_{s}=&\frac{1}{2^{\frac{y}{2}}}\langle(M_{x_{1}=0}+M_{x_{1}=1})M_{x_{2}=0}\cdot\cdot\cdot (M_{x_{y-1}=0}+M_{x_{y-1}=1})M_{x_{y}=0}\rangle\\=&\frac{1}{2^{\frac{y}{2}}}{\rm Tr}\{[((\hat{O}(\alpha_{x_{1}=0}^{\lambda_{y}};s)\otimes\hat{O}(\alpha_{x_{1}=0}^{\lambda_{1}};s))
+(\hat{O}(\alpha_{x_{1}=1}^{\lambda_{y}};s)\otimes\hat{O}(\alpha_{x_{1}=1}^{\lambda_{1}};s)))\\&\otimes(\hat{O}(\alpha_{x_{2}=0}^{\lambda_{1}};s)\otimes\hat{O}(\alpha_{x_{2}=0}^{\lambda_{2}};s))\otimes\cdot\cdot\cdot\otimes((\hat{O}(\alpha_{x_{y-1}=0}^{\lambda_{y-2}};s)\otimes\hat{O}(\alpha_{x_{y-1}=0}^{\lambda_{y-1}};s))
\\&+(\hat{O}(\alpha_{x_{y-1}=1}^{\lambda_{y-2}};s)\otimes\hat{O}(\alpha_{x_{y-1}=1}^{\lambda_{y-1}};s)))\otimes(\hat{O}(\alpha_{x_{y}=0}^{\lambda_{y-1}};s)\otimes\hat{O}(\alpha_{x_{y}=0}^{\lambda_{y}};s))]\\&(\rho^{A_{1}A_{2}}\otimes\cdot\cdot\cdot\otimes\rho^{A_{y-1}A_{y}}\otimes\rho^{A_{y}A_{1}})\}\\
=&\frac{1}{2^{\frac{y}{2}}}\mathop{\prod}\limits_{i\in\mathcal{K}}{\rm Tr}\{[(\hat{O}(\alpha_{x_{i}=0}^{\lambda_{i}};s)+\hat{O}(\alpha_{x_{i}=1}^{\lambda_{i}};s))\otimes\hat{O}(\alpha_{x_{i+1}=0}^{\lambda_{i}};s)]\rho^{A_{i}A_{i+1}}\}\\&
\times\mathop{\prod}\limits_{j\in\overline{\mathcal{K}}}{\rm Tr}\{[\hat{O}(\alpha_{x_{j}=0}^{\lambda_{j}};s)\otimes(\hat{O}(\alpha_{x_{j+1}=0}^{\lambda_{j}};s)+\hat{O}(\alpha_{x_{j+1}=1}^{\lambda_{j}};s))]\rho^{A_{j}A_{j+1}}\},
\end{aligned}
\end{equation}
\if false where ${\mathcal K}=\{2,4,\cdot\cdot\cdot, y\}$ and $\overline{{\mathcal K}}=\{1,3, \cdot\cdot\cdot, y-1\}$. Additionally, in Eq.(F9),\fi with $y+1$ actually represents 1 (mod $y$). Substituting Eqs.(\ref{eqB4})-(\ref{eqB7}) into   Eq.(\ref{eqE7})   obtains the expression of $\mathcal{I}_{s}$ in Theorem 7.2.

Similarly, by Eqs.(\ref{eq2.4}) and (\ref{eq3.14}),
\begin{equation}\tag{E8}\label{eqE8}
\begin{aligned}
\mathcal{J}_{s}=&\frac{1}{2^{\frac{y}{2}}}\langle(M_{x_{1}=0}-M_{x_{1}=1})M_{x_{2}=1}\cdot\cdot\cdot (M_{x_{y-1}=0}-M_{x_{y-1}=1})M_{x_{y}=1}\rangle\\=&\frac{1}{2^{\frac{y}{2}}}{\rm Tr}\{[((\hat{O}(\alpha_{x_{1}=0}^{\lambda_{y}};s)\otimes\hat{O}(\alpha_{x_{1}=0}^{\lambda_{1}};s))
-(\hat{O}(\alpha_{x_{1}=1}^{\lambda_{y}};s)\otimes\hat{O}(\alpha_{x_{1}=1}^{\lambda_{1}};s)))\\&\otimes(\hat{O}(\alpha_{x_{2}=1}^{\lambda_{1}};s)\otimes\hat{O}(\alpha_{x_{2}=1}^{\lambda_{2}};s))\otimes\cdot\cdot\cdot\otimes((\hat{O}(\alpha_{x_{y-1}=0}^{\lambda_{y-2}};s)\otimes\hat{O}(\alpha_{x_{y-1}=0}^{\lambda_{y-1}};s))
\\&-(\hat{O}(\alpha_{x_{y-1}=1}^{\lambda_{y-2}};s)\otimes\hat{O}(\alpha_{x_{y-1}=1}^{\lambda_{y-1}};s)))\otimes(\hat{O}(\alpha_{x_{y}=1}^{\lambda_{y-1}};s)\otimes\hat{O}(\alpha_{x_{y}=1}^{\lambda_{y}};s))]\\&(\rho^{A_{1}A_{2}}\otimes\cdot\cdot\cdot\otimes\rho^{A_{y-1}A_{y}}\otimes\rho^{A_{y}A_{1}})\}\\
=&\frac{1}{2^{\frac{y}{2}}}\times\mathop{\prod}\limits_{i\in\mathcal{K}}{\rm Tr}\{[(\hat{O}(\alpha_{x_{i}=0}^{\lambda_{i}};s)-\hat{O}(\alpha_{x_{i}=1}^{\lambda_{i}};s))\otimes\hat{O}(\alpha_{x_{i+1}=1}^{\lambda_{i}};s)]\rho^{A_{i}A_{i+1}}\}\\&\times\mathop{\prod}\limits_{j\in\overline{\mathcal{K}}}{\rm Tr}[(\hat{O}(\alpha_{x_{j}=1}^{\lambda_{j}};s)\otimes(\hat{O}(\alpha_{x_{j+1}=0}^{\lambda_{j}};s)
-\hat{O}(\alpha_{x_{j+1}=1}^{\lambda_{j}};s)))\rho^{A_{j}A_{j+1}}]\}.
\end{aligned}
\end{equation}
\if false where ${\mathcal K}=\{2,4,\cdot\cdot\cdot, y\}$ and $\overline{{\mathcal K}}=\{1,3, \cdot\cdot\cdot, y-1\}$. Additionally, in Eq.(F10), $y+1$ actually represents 1 due to the cyclic nature of the network. \fi Substituting Eqs.(\ref{eqB9})-(\ref{eqB12}) into Eq.(\ref{eqE8}) yields the desired expressions of $\mathcal{J}_{s}$ in Theorem 7.2.
The proof is completed.\hfill$\Box$
\end{widetext}

\end{document}